\documentclass{aa}
\usepackage[varg]{txfonts}
\usepackage{graphicx}
\usepackage{amssymb}
\usepackage{lscape}
\newcommand{\HII}{\ion{H}{II}\,\,}

\usepackage{natbib,twoopt}
\makeatletter
\newcommandtwoopt{\citeads}[3][][]{\href{http://adsabs.harvard.edu/abs/#3}%
{\def\hyper@linkstart##1##2{}%
\let\hyper@linkend\@empty\citealp[#1][#2]{#3}}}
\newcommandtwoopt{\citepads}[3][][]{\href{http://adsabs.harvard.edu/abs/#3}%
{\def\hyper@linkstart##1##2{}%
\let\hyper@linkend\@empty\citep[#1][#2]{#3}}}
\newcommandtwoopt{\citetads}[3][][]{\href{http://adsabs.harvard.edu/abs/#3}%
{\def\hyper@linkstart##1##2{}%
\let\hyper@linkend\@empty\citet[#1][#2]{#3}}}
\newcommandtwoopt{\citeyearads}[3][][]%
{\href{http://adsabs.harvard.edu/abs/#3}
{\def\hyper@linkstart##1##2{}%
\let\hyper@linkend\@empty\citeyear[#1][#2]{#3}}}
\makeatother
\begin{document}

\title{Relationship between the column density distribution and
evolutionary class of molecular
clouds as viewed by ATLASGAL}


\author{Abreu-Vicente, J.\inst{1}\thanks{Member of the
International Max Planck Research School (IMPRS) at the University
of Heidelberg}, Kainulainen, J.\inst{1}, Stutz,
A.\inst{1}, Henning, Th.\inst{1}, Beuther, H.\inst{1}}

\authorrunning{Abreu-Vicente, J. et al.}
\titlerunning{Relationship between density structure and evolutionary
class of molecular
clouds}

\offprints{J. Abreu-Vicente, \email{abreu@mpia-hd.mpg.de}}

\institute{Max-Planck-Institut f\"{u}r Astronomie (MPIA),
K\"{o}nigstuhl 17, 69117, Heidelberg, Germany}

\date{\today}


\abstract{We present the first study of the relationship between the
  column density distribution of molecular clouds within nearby
  Galactic spiral arms and their evolutionary status as measured from
  their stellar content.  We analyze a sample of 195 molecular clouds
  located at distances below $5.5$\,kpc, identified from the ATLASGAL
  870\,$\mu$m data.  We define three evolutionary classes within this
  sample: starless clumps, star-forming clouds with associated young
  stellar objects, and clouds associated with \HII regions.  We find
  that the $N(H_{2})$ probability density functions
  (\textit{N}-PDFs) of these three classes of objects are clearly
  different: the $N$-PDFs of starless clumps are narrowest and close
to log-normal in shape, while star-forming clouds and \HII
regions exhibit a power-law shape over a wide range of column
densities and log-normal-like components only at low column densities.
We use the $N$-PDFs to estimate the evolutionary time-scales of the
three classes of objects based on a simple analytic model
from literature.  Finally, we show that the integral of the
\textit{N}-PDFs, the dense gas mass fraction, depends on the total
mass of the regions as measured by ATLASGAL: more massive clouds
contain greater relative amounts of dense gas across all evolutionary
classes. }

\maketitle

\section{Introduction}\label{intro}

Molecular clouds (MCs) are the densest regions of the interstellar
medium and the birth sites of stars. Nevertheless, despite this
important role in star formation, key aspects of MC evolution remain
unclear: What are the key parameters in determining the star-forming
activity of MCs? How do these parameters change with MC evolution? The
column density distribution of MCs has been found to be sensitive to
the relevant physical processes ~\citep{2012A&ARv..20...55H}.  The
study of the density structure of clouds that are at different
evolutionary stages can therefore help to understand which physical
processes are dominating the cloud structure at those stages.

Column density probability density functions (\textit{N}-PDFs) are
useful tools for inferring the role of different physical processes in
shaping the structure of molecular clouds. Observations have shown
that non-star-forming molecular clouds show
\emph{bottom-heavy}\footnote{Most of their mass is in low-column
density material.}  \textit{N}-PDFs, while the star-forming
molecular clouds show \emph{top-heavy}\footnote{They have a
significant amount of mass enclosed in high-column density regions.}
\textit{N}-PDFs~\citep{2009A&A...508L..35K,2011A&A...530A..64K,2014Sci...344..183K,2013A&A...549A..53K,2013ApJ...766L..17S}.
It is generally accepted that the \emph{top-heavy} \textit{N}-PDFs
are well described by a power-law function in their high-column
density regimes.  The description of the shapes of the low-column
density regimes of both kinds of \textit{N}-PDFs is still a matter
of debate.  The papers cited above describe the low-column density
regimes as log-normal functions.  In contrast,~\citet{alves-2014}
and~\citet{lombardi-15} argue that a power-law function fits the
observed \textit{N}-PDFs throughout their range. The origin of
these differences is currently unclear.

Simulations predict that turbulence-dominated gas develops
a log-normal \textit{N}-PDF~\citep{2013ApJ...763...51F}; such a form is predicted
for the volume density PDF (hereafter \textit{$\rho$}-PDF) of isothermal,
supersonic turbulent, and non-self-gravitating
gas~\citep{1994ApJ...423..681V,1997ApJ...474..730P,1998ApJ...504..835S,2001ApJ...546..980O,2011ApJ...730...40P,2011MNRAS.416.1436B,2013ApJ...763...51F}.
Log-normal \textit{$\rho$}-PDFs can, however, result also
from processes other than supersonic turbulence such as 
gravity opposed only by thermal-pressure forces or
gravitationally-driven ambipolar diffusion~\citep{2010MNRAS.408.1089T}.

The log-normal $N$-PDF is defined as:
\begin{equation}\label{eq:log-normal}
p(s; \mu, \sigma_{s}) = \frac{1}{\sigma_{s}\sqrt{2\pi}}exp \left(\frac{-(s-\mu)^{2}}{2\sigma_{s}^{2}}\right),
\end{equation}
where $s=\mathrm{ln\,}(A_{V}/\overline{A_{V}})$
is the mean-normalized visual extinction
(tracer of column density, see Section~\ref{sec:NH}), and $\mu$
and $\sigma_{s}$ are respectively the mean and standard
deviation of the distribution. The log-normal
component that is used to describe low column densities
has typically the width of
$\sigma_{s}=0.3-0.4$~\citep{2009A&A...508L..35K}. 
It has been suggested that the determination 
of the width can be affected by issues such as unrelated 
dust emission along the line of sight to the cloud~\citep{schneider-15}.
Practically all
star-forming clouds in the Solar neighborhood
show an excess to this component at higher column densities,
following a power-law, or a wider log-normal
function~\citep{2009A&A...508L..35K,2013A&A...549A..53K,2013ApJ...766L..17S},
especially reflecting their ongoing
star formation activity~\citep{2014Sci...344..183K,2014ApJ...787L..18S,2015arXiv150405188S}.
Such behavior is suggested by the predictions that
develope \textit{top-heavy} \textit{$\rho$}-PDFs for self-gravitating
systems~\citep{2000ApJ...535..869K,2000ApJS..128..287K,2011ApJ...727L..20K,2014ApJ...781...91G}.

Another interesting measure of the density structure of
molecular clouds is the dense gas mass fraction (DGMF) that
describes the mass enclosed by regions with
$M(A_{V} \geq A_{V}\arcmin)$, relative to the total mass
of the cloud, $M_{\mathrm{tot}}$.
\begin{equation}\label{dgmf}
dM\arcmin=\frac{M(A_{V} \geq A_{V}\arcmin)}{M_{\mathrm{tot}}}.
\end{equation}
The DGMF has been recently linked to the star-forming
rates of molecular clouds:~\citet{2010ApJ...723.1019H}
and~\citet{2010ApJ...724..687L,2012ApJ...745..190L} showed,
using samples of nearby molecular clouds and external galaxies,
that there is a relation between the mean star-forming rate (SFR)
surface density ($\Sigma_{\mathrm{SFR}}$) and the mean
mass surface density ($\Sigma_{\mathrm{mass}}$) of MCs:
$\overline{\Sigma}_{\mathrm{SFR}}\propto f_{\mathrm{DG}}\overline{\Sigma}_{mass}$, where
$f_{\mathrm{DG}}=\frac{M(A_{V}>7.0\,\mathrm{mag})}{M_{\mathrm{tot}}}$.
Furthermore, in a sample of eight molecular clouds within 1\,kpc,
a correlation $\Sigma_{\mathrm{SFR}}\propto \Sigma_{\mathrm{mass}}^{2}$ was reported
by~\citet{2011ApJ...739...84G}.
Combining these two results suggests $f_{\mathrm{DG}}\propto\Sigma_{\mathrm{mass}}$.

Despite their utility, a complete, global understanding of the
\textit{N}-PDFs and DGMFs of
molecular clouds is still missing. One of the main
problems arises from the fact that the dynamic ranges
of different observational techniques sample the \textit{N}-PDFs
and DGMFs differently.
Previous works have employed various methods:
CO line emission only samples \textit{N}-PDFs between 
$A_{V}\approx3-8$\,mag~\citep{2009ApJ...692...91G}.
NIR extinction traces column density at wider, but still narrow,
dynamic range, $A_{V}\approx1-25$\,mag~\citep{2001A&A...377.1023L}.
~\citet{2013A&A...549A..53K} and~\citet{2013A&A...553L...8K} used a novel extinction technique
that combines NIR and MIR data, considerably increasing the observable dynamic
range, $A_{V}=3 - 100$\,mag.~\citet{2013ApJ...766L..17S} and~\citet{lombardi-15} used \textit{Herschel}
FIR data to sample \textit{N}-PDFs at $A_{V}<100$\,mag.

Another observational hindrance 
arises from the limited spatial resolution of observations. 
The high-column densities typically correspond to small spatial 
scales in molecular clouds; to probe the \textit{N}-PDFs at high-column 
densities requires spatial resolution that approaches the scale 
of dense cores in the clouds ($\sim$0.1\,pc).~\textit{Herschel} reaches resolution 
of $\sim$36$\arcsec$ that corresponds to 0.17\,pc at 1\,kpc distance. 
Extinction mapping using both NIR and MIR wavelengths can 
reach arcsecond-scale resolution, but only about ten clouds 
have been studied so far with that technique~\citep{2013A&A...549A..53K,2013A&A...557A.120K}.

Born out of the observational limitations above, 
the most important weakness in previous studies is that they only
analyze relatively nearby molecular clouds ($d\lesssim1.5\,\mathrm{kpc}$).
Thus, they probe only a very limited range of Galatic
environments, which prohibits the development of a global
picture of the factors that control \textit{N}-PDFs
across different Galactic environments.
Extending \textit{N}-PDF studies to
larger distances is imperative for three principal reasons.
First, studying the more massive and distant MCs will allow us to sample the
entire MC mass range present in the Galaxy. Second, larger numbers of
MCs over all masses provide statistically meaningful samples.  Finally,
extending to larger distances is necessary to study the possible
effect of the Galactic structure on the mass distribution statistics.

In this paper, we employ the ATLASGAL~\citep{2009A&A...504..415S,csengeri-2014} survey to study a large
sample of molecular clouds in the Galaxy. 
The ATLASGAL survey traces submillimeter dust emission at 870\,$\mu$m.
Submillimeter dust emission is an optically thin tracer of interstellar dust, 
and hence a direct tracer of gas if a canonical
dust-gas mass ratio is assumed. The submillimeter observing 
technique employed in the ATLASGAL survey filters out diffuse emission
on spatial scales greater than 2.5$\arcmin$, 
hence making the survey most sensitive to the densest material of the 
interstellar medium in which star formation occurs.
With this data set we can observe the cold dense interiors of 
molecular clouds in both the near and far sides of the Galactic plane.
With an angular resolution of $19.2\arcsec$, ATLASGAL improves 
by almost a factor of two the resolution of $Herschel$
observations, thus providing a more detailed view of the
dense material inside molecular clouds. 
We will use this data sample to study the 
\textit{N}-PDFs and DGMFs of molecular clouds at different evolutionary classes.

\section{Data and methods}\label{sec:data}

We used continuum maps at 870\,$\mu$m from
ATLASGAL to identify MCs in
the Galactic plane region between
$l\in[9\degr,21\degr]$ and $|b|\leq1\degr$, where the
rms of the survey is 50\,mJy/beam.
We selected this area, because extensive auxiliary data
sets were available for it, and specifically, starless clumps
have already been identified by~\citet{2012A&A...540A.113T}.
We classified the identified molecular cloud regions in
three groups based on their evolutionary classes:
starless clumps (SLCs), star-forming clouds (SFCs), and \HII regions.
In the following, we describe how each
class was defined and how we
estimated the distance to each region.

\subsection{Source selection}\label{sec:def}

\begin{figure*}
\centering
\includegraphics[width=0.9\textwidth]{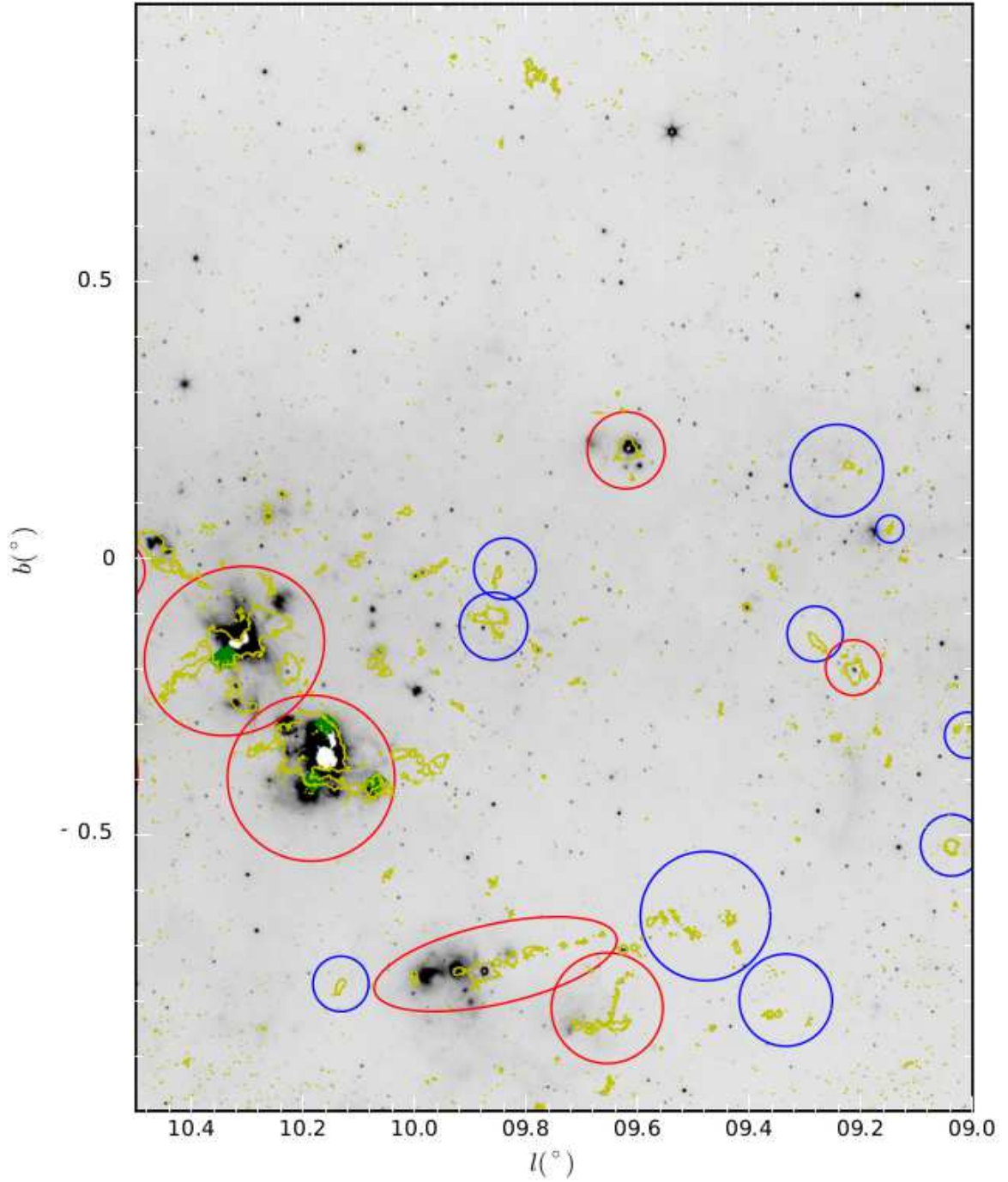}
\caption{MIPSGAL 24\,$\mu$m map of the Galactic plane
between $9\deg < l < 10.5\deg$. Yellow
contours indicate the 3$\sigma$ (0.15\,Jy/beam) emission level of
the ATLASGAL data. Red and blue ellipses show
the \HII regions and SFCs, respectively.
SLCs are shown with green filled diamonds.
Similar maps for the Galactic plane between
$10.5\deg < l < 21\deg$ and $|b|\leq1\deg$ are shown in
Appendix~\ref{ap:maps}.}
\label{fig:galPlane}
\end{figure*}

We identified molecular cloud regions based primarily on ATLASGAL
dust emission data. As a first step, we defined objects from ATLASGAL data
simply by using $3\sigma$ emission contours (0.15\,Jy/beam) to define the region boundaries.
Then, we used distances available in literature (see Sect.~\ref{sec-dist})
to group together neighbouring objects 
located at similar distances (within the assumed distance
uncertainty of 0.5\,kpc), i.e., those that are likely
associated with the same molecular cloud. As a next step, 
we expanded the boundaries of the regions down to their 1$\sigma$ level
in the cases in which they show close contours at 1$\sigma$.
Finally, each region created in this manner was classified either as a SLC,
SFC, or \HII region using information about their stellar content
available in literature.
An example of the region definition is shown in Fig.~\ref{fig:galPlane} (see also Appendix~\ref{ap:maps}).
We identify a total of 615 regions, 330 of them with known distances and classified
either as SLC, SFC, or \HII regions (Fig.~\ref{fig:dist-hist}).
Throughout this paper we refer to each of the ellipses shown in Fig.~\ref{fig:galPlane} with the
term \emph{region}. In the following we explain the definition of the three evolutionary classes in detail.

\HII regions are defined as regions hosting previously cataloged \HII
regions. We used the
catalogues~\citet{1989ApJS...69..831W},~\citet{1989ApJS...71..469L},~\citet{1993ApJ...418..368G},~\citet{1996A&AS..115...81B},
~\citet{1996ApJ...472..173L},~\citet{2000ApJ...530..371F},
and~\citet{2013MNRAS.435..400U}. We identified 114 \HII regions in the
considered area. Distances are known for 84 of them (74\%). Two thirds
(57) of the \HII regions with distance estimates lie at near distances
($d<5.5$\,kpc).  If we assume the same distribution for the 30 \HII
regions with unknown distances, 20 of them would be located at near
distances.  Nevertheless, we exclude these regions from our analysis.
We summarize the number of regions with and without known distances in
Table~\ref{tab:number} (see Sect.~\ref{sec-dist}).

The star-forming clouds (SFCs) are defined as the subset of regions
devoid of \HII regions but containing young stellar objects (YSOs) and
protostars. Here the presence of YSOs and protostars is assumed to be a clear indication
of ongoing star formation.
For this purpose, we used the YSO catalogues of~\citet{2011ApJ...731...90D} and~\citet{2014PASJ...66...17T}. 
The former search signs of active star formation in the Bolocam Galactic Plane Survey~\citep[BGPS]{bgps}
using the GLIMPSE Red Source catalogue~\citep{robitaille-2008},
the EGO catalogue~\citep{cyganowski-2008}, and the RMS catalogue~\citep{lumsden-2013}.
They found 1341 YSOs in the area $l\in[9\degr,21\degr]$ and $|b|\leq0.5\degr$
and it is $>$98\% complete at the  0.4\,Jy level~\citep{2011ApJ...731...90D}.
~\citet{2014PASJ...66...17T} present a catalog of 44001 YSO candidates,
2138 in the area $l\in[9\degr,21\degr]$ and $|b|\leq1\degr$, with a reliability of 90\%
in the YSO classification. 
All the regions showing spatially coincident YSOs were classified as SFCs.   
We only require one YSO to classify a region as SFC,
but our SFCs have more than one.
The probability of classifying a SFC as a 
region without YSOs due to completeness issues in the 
YSOs catalogues is therefore very low.
We identified 184 SFCs, 126 of them with
known distances. The 80\% (99) of the SFCs with known
distances lie at $d<5.5$\,kpc and are therefore studied in 
this paper. Assuming the same SFC distribution
for the SFCs with unknown distances, we estimate that the 80\% (46) of the SFCs with
no distance estimates would be located at near distances.

Finally, we adopted the starless clump catalog
from~\citet{2012A&A...540A.113T} to define our sample of SLCs.
They present a SLC sample with peak
column densities $N>10^{23}\mathrm{\,cm^{-2}}$. The properties of this SLC sample
were specifically chosen in order to detect potential high-mass star progenitors. 
~\citet{2012A&A...540A.113T} used uniform criteria to classify their
SLC sample: absence of GLIMPSE and/or 24$\,\mu$m MIPSGAL sources.
~\citet{2012A&A...540A.113T} identified 120 SLCs with
known distances\footnote{We adopt only
regions with solved kinematic distance ambiguity (KDA) as
sources with known distances.}
in the Galactic plane area studied.  
All SLCs are located inside our previously defined
\HII regions or SFCs (see Fig.~\ref{fig:galPlane}). 

We note a caveat in the above evolutionary class definition scheme.
Our scheme makes an effort to capture the dominant
evolutionary phase of the region, but it is clear that not all the regions
are straightforward to classify. In principle, the distinction between
\HII regions and SFCs is well defined; it depends on whether the regions
host an \HII region or not. However, eight regions harbor only UC\HII regions
whose extent is tiny compared to the full extent of
those regions (\#34, \#54, \#192, \#195, \#233, \#246, \#247 and \#390).
Since our aim is to capture the dominant evolutionary phase, we classified
these regions as SFCs. 

We also note that our evolutionary class
definition is based only on the stellar content of the regions.
The SLCs exhibit no indications of star-forming activity, SFCs have star-forming sources,
\HII region have formed massive stars.
However, we emphasize that we cannot assume that all the SLCs will definitely
form stars.  Similarly, we cannot assume that all the star-forming content
within SFCs will become  massive enough to create \HII regions, although some 
of them will. Therefore
we do not aim to draw a \textit{sequential} evolutionary link between these three
classes of regions. Instead, the estimated time-scales
for each class
instead aim to identify independent evolutionary time-scales for each observational class.

\subsection{Distance estimates and convolution to a common spatial resolution}
\label{sec-dist}

We adopted distances to each region from literature.
The two main literature sources used were~\citet{2013ApJ...770...39E} 
and~\citet{2012A&A...544A.146W}.
The former catalog measures kinematic distances of molecular clumps
identified with sub-mm dust emission.
They solve the kinematic distance ambiguity (KDA) 
using Bayesian distance probability density functions.
They use previous data sets to establish the prior
distance probabilities to be used in the Bayesian analysis.
This method has a 92\% agreement with Galactic Ring Survey based distances. 
In total, 68 out of 330 regions have counterparts in~\citet{2013ApJ...770...39E}.
~\citet{2012A&A...544A.146W} measured the kinematic distances to dense
clumps in the ATLASGAL survey using ammonia observations. We obtained
distance estimates for 80 regions from this catalog. We also used
other catalogs based on kinematic distances~\citep{1997MNRAS.291..261W,2006ApJ...639..227S,2006ApJS..162..346R,2009ApJ...699.1153R,2013MNRAS.435..400U,2012A&A...540A.113T}, 
and in a three-dimensional model of interstellar extinction~\citep{2009ApJ...706..727M}.
A detailed discussion on the methods for distance estimates is
beyond the scope of this paper. We therefore refer to the
cited papers for a detailed discussion on them. Table~\ref{tab:dist-ref} shows the
number of distance estimates adopted from each literature source.
In regions with more than one distance estimate, 
we estimated the distance averaging the different values. 
For all but six of the studied regions ($\sim96$\%) 
the distance ambiguity was solved in at least one of the cited papers. 
Regions with different KDA solutions in literature 
(i.e. with several clouds along the same line-of-sight) were 
removed from our sample to avoid line-of-sight contamination.
For the remaining six regions we used maps from the
GLIMPSE and MIPS surveys to search for dark shadows against
background emission (e.g. Stutz et al. 2009; Ragan et al. 2012).
The near distance was adopted for regions associated with IRDCs.

Since all the SLCs of our sample are
embedded in \HII regions or SFCs (see Section~\ref{sec:def}), we
compared the distance estimates for the SLCs
and for their hosting regions. In every but one case, the  
distance estimates of the SLCs and their hosting SFCs or \HII regions
were in good agreement. In the only inconsistent case, the SLC was 
located at the far distance in~\citep{2012A&A...540A.113T} and its hosting
SFC was located at the near distance. Since the KDA solutions of
the SLC and its hosting SFC differ, we removed out both regions
from the final sample (see also previous paragraph). 

\begin{figure}
\centering
\resizebox{\hsize}{!}{\includegraphics{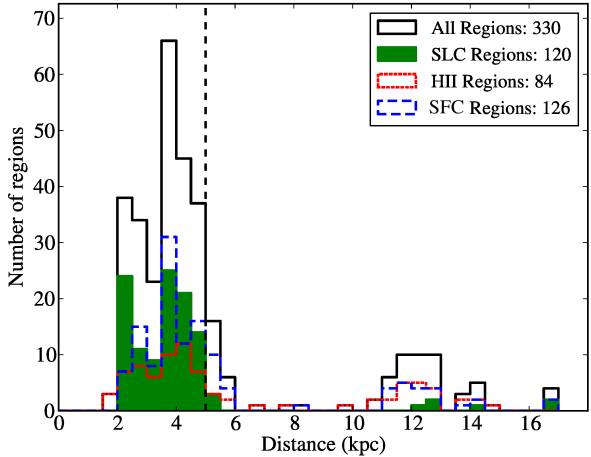}}
\caption{Distance distribution of the molecular
cloud regions. Black solid line shows the total number of regions.
Red dotted line shows the HII regions,
the blue dashed line the SFCs, and green filled area the SLCs.
Black dashed vertical line at 5\,kpc shows the common distance to
which we have smoothed the data.}
\label{fig:dist-hist}
\end{figure}

\begin{figure}[h]
\centering
\resizebox{\hsize}{!}{\includegraphics{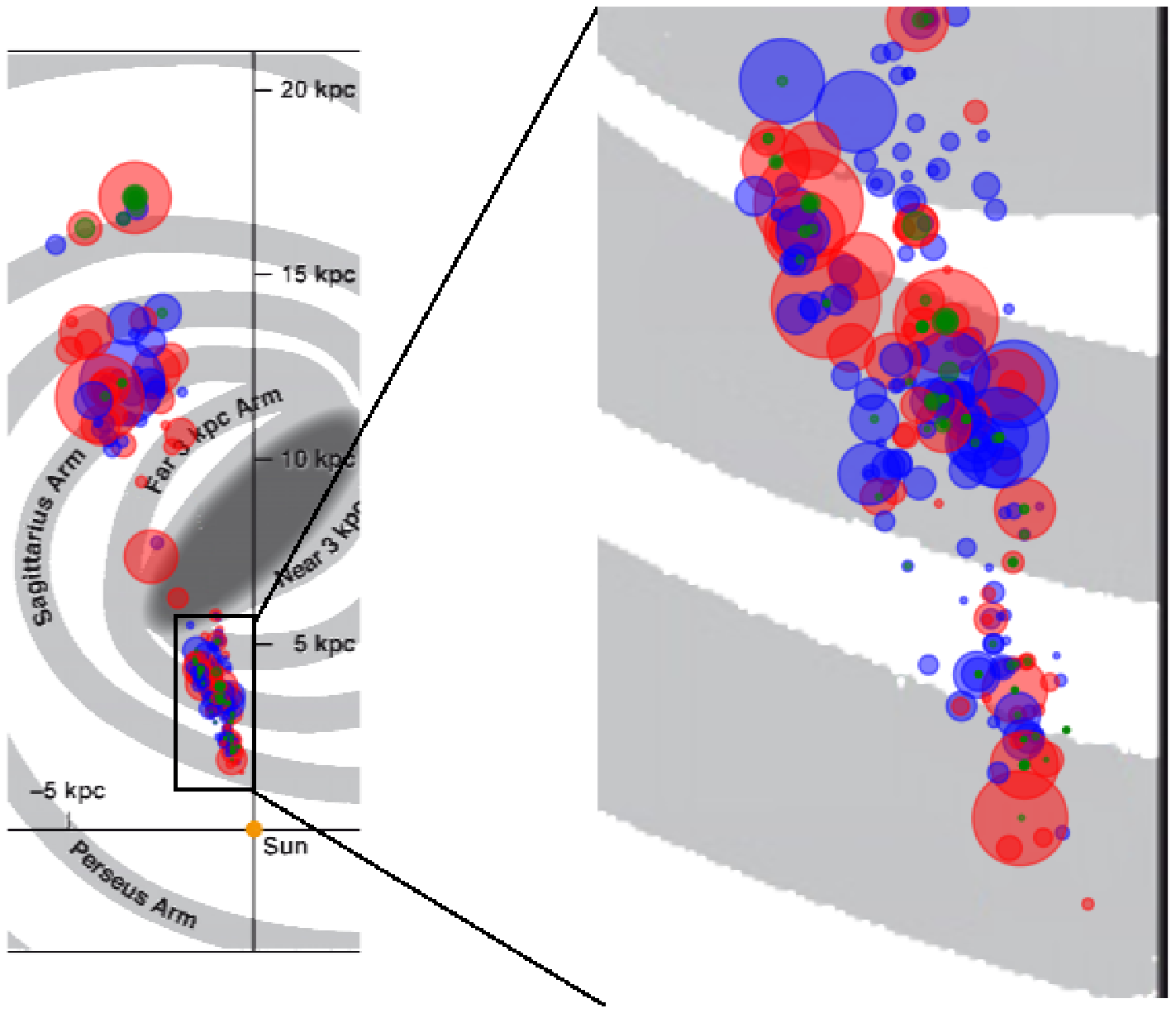}}
\caption{Artist impression of face-on view of the Milky Way
(R. Hurt, SSC-Caltech, MPIA graphic, Tackenberg et al. 2012).
\HII regions are shown as red circles, star-forming clouds as blue circles and
starless clumps as green circles.
Circle sizes are proportional to region sizes. The right panel shows a zoom
to the region enclosed by the black rectangle in the left panel,
where the source density is highest.}
\label{fig:MW-faceon}
\end{figure}

Figure~\ref{fig:dist-hist} shows the distance distribution of our
sample. A vast majority ($\sim$ 80\%) of our regions is located
within 5\,kpc distance.
There is a gap between 6 and 10\,kpc,
coinciding with the central hole of the Galactic molecular
ring~\citep{1989ApJ...339..919S}\footnote{We
note that the existence of the Galactic Ring has recently been questioned
by~\citet{2012MNRAS.421.2940D}, who proposed a two symmetric spiral
arm pattern for the Milky Way as an explanation of observations.}.
At the far side of the Galaxy, there are three 
density enhancements that coincide
with the Sagittarius, Norma, and Perseus spiral arms
(Fig.~\ref{fig:MW-faceon}).

We study only regions within 5\,kpc since the highest
source density of our sample is located there.
Assuming an error in distance determination of about 0.5\,kpc,
we also included regions located between 5\,kpc and 5.5\,kpc.
We convolved the ATLASGAL data of all closer regions to a
common 5\,kpc distance resolution using
a Gaussian kernel of
$FWHM=\sqrt{(19.2\arcsec)^{2}(\frac{5\mathrm{\,kpc}}{d\mathrm{\,kpc}})^{2}-(19.2\arcsec)^{2}}$.
This convolution was done for each region individually.
At the distance of 5\,kpc, the $19.2\arcsec$ resolution of the ATLASGAL
translates to about 0.5\,pc.
We therefore do not resolve the dense cores ultimately
linked to star formation that have typically a size of
$\sim$0.1\,pc~\citep{2009EAS....34..195M,2013A&A...559A..79R}.

When smoothing maps to a common distance,
some of the smaller SLCs
were washed out by strong emission gradients
likely associated with nearby strong sources. This artificially
increases the SLCs column densities.
To minimize the effect, we inspected each SLC by eye,
discarding those that were significantly affected by
strong gradients. Appendix~\ref{app:TSL} shows the SLCs
included in the final sample.

The total number of regions studied in this paper,
and the number of regions in each evolutionary class,
are listed in Table~\ref{tab:number}.

\begin{table}
\caption{Completeness of each evolutionary class} %
\centering 
\begin{tabular}{c c c c c c} 
\hline\hline 
					&\HII	&SFCs	&SLCs			&No class	\\\hline 
Total					&114	&184	&210			&107		\\
Known $d$\tablefootmark{a}		&84	&126	&120			&---		\\
$d<5.5$\,kpc				&57	&99	&111			&---		\\
$d>5.5$\,kpc				&27	&27	&9			&---		\\
miss. $d<5.5$\,kpc\tablefootmark{b}	&20	&18	&102			&87		\\
miss. $d>5.5$\,kpc\tablefootmark{b}	&10	&8	&8			&20		\\
Studied					&57	&99	&31\tablefootmark{c}	&---		\\
\hline
\label{tab:number}
\end{tabular}
\tablefoot{
\tablefoottext{a}{Only SLCs with KDA solved and, if more than one distance estimate,
		  agreement between different literature sources.}
\tablefoottext{b}{Number of regions lost due to 
		  lack of distance estimates. We assume homogeneous distribution of the sources
		  along the Galactic plane area studied.}
\tablefoottext{c}{We only studied isolated SLCs (see Sect.~\ref{sec-dist})}}
\end{table}

\begin{table}
\caption{Literature sources from which distances were obtained} %
\centering 
\begin{tabular}{c c c c} 
\hline\hline
	Reference			&\HII	&SFCs	&SLCs		\\\hline 
1	&7	&50	&11		\\
2	&19	&6	&18		\\
3	&23	&1	&7		\\
4	&39	&23	&18		\\
5	&7	&11	&5		\\
6	&14	&26	&5		\\
7&---	&13	&---		\\
8&17	&---	&5		\\
\hline
\label{tab:dist-ref}
\end{tabular}
\tablebib{(1)~\citet{2013ApJ...770...39E};
(2)~\citet{2012A&A...540A.113T}; (3)~\citet{2013MNRAS.435..400U};
(4)~\citet{2012A&A...544A.146W}; (5)~\citet{2009ApJ...699.1153R};
(6)~\citet{2009ApJ...706..727M}; (7)~\citet{2006ApJ...639..227S}; 
(8)~\citet{1997MNRAS.291..261W}.}
\end{table}

\onllongtab{
\begin{landscape}
\begin{longtable}{ccccccccccc}
\caption{Regions studied in this paper}\\
\hline
\hline
ID        &        Type        &        RA (J2000)        &        DE (J2000)        &        Radius ($\arcsec$)        &        Maj. Ax
($\arcsec$)        &        Min. Ax ($\arcsec$)        &        Angle ($\degr$)        &        Distance
(kpc)        &   Dist. dispersion (kpc) &   References        \\
\hline
\endfirsthead
\caption{Continued.} \\
\hline
ID        &        Type        &        RA (J2000)        &        DE (J2000)        &        Radius ($\arcsec$)        &        Maj. Ax
($\arcsec$)        &        Min. Ax ($\arcsec$)        &        Angle ($\degr$)        &        Distance\tablefootmark{b}
(kpc)        &  Dist. dispersion\tablefootmark{b} (kpc)  &   References        \\
\hline
\endhead
\hline
\endfoot
\hline
\endlastfoot
\label{tab:regions}
2	&	SFC	&	18:05:35.91	&	-20:52:28.00	&	90	&	---	&	---	&	---	&	3.2	&	---	&	1	\\
3	&	SFC	&	18:05:47.64	&	-21:00:32.00	&	90	&	---	&	---	&	---	&	4	&	---	&	1	\\
4	&	HII	&	18:06:15.12	&	-20:31:36.50	&	150	&	---	&	---	&	---	&	5.5	&	0.4	&	3	\\
5	&	SFC	&	18:06:46.98	&	-20:59:04.10	&	180	&	---	&	---	&	---	&	4.5	&	---	&	1	\\
7	&	HII	&	18:06:52.03	&	-21:04:27.35	&	180	&	---	&	---	&	---	&	4.8	&	<0.1	&	4,6	\\
8	&	SFC	&	18:06:53.75	&	-21:18:46.50	&	150	&	---	&	---	&	---	&	4.4	&	---	&	1	\\
11	&	SFC	&	18:07:34.17	&	-20:26:03.30	&	150	&	---	&	---	&	---	&	2.6	&	---	&	1	\\
13	&	SFC	&	18:07:42.18	&	-21:23:02.20	&	200	&	---	&	---	&	---	&	4.7	&	0.3	&	1	\\
16	&	SFC	&	18:07:55.85	&	-20:28:26.60	&	220	&	---	&	---	&	---	&	2.5	&	---	&	1	\\
27	&	HII	&	18:08:53.62	&	-18:16:08.10	&	350	&	---	&	---	&	---	&	3	&	0.1	&	4,2	\\
28	&	SFC	&	18:08:59.71	&	-20:11:29.40	&	150	&	---	&	---	&	---	&	2	&	---	&	4	\\
29	&	HII	&	18:09:23.23	&	-20:08:38.60	&	150	&	---	&	---	&	---	&	3.6	&	0.1	&	4,3	\\
31	&	SFC	&	18:09:24.38	&	-20:01:35.70	&	220	&	---	&	---	&	---	&	2.3	&	0.4	&	1,4	\\
30	&	HII	&	18:09:01.33	&	-19:48:33.30	&	---	&	290	&	250	&	-60	&	5.2	&	---	&	2	\\
33	&	HII	&	18:09:05.33	&	-19:28:00.80	&	150	&	---	&	---	&	---	&	2.5	&	---	&	3	\\
36	&	HII	&	18:09:37.96	&	-20:19:16.10	&	540	&	---	&	---	&	---	&	1.9	&	0.2	&	4	\\
37	&	SFC	&	18:09:06.10	&	-21:03:37.20	&	420	&	---	&	---	&	---	&	3.7	&	0.3	&	4	\\
38	&	SFC	&	18:09:22.35	&	-21:15:35.70	&	300	&	---	&	---	&	---	&	5.2	&	---	&	6	\\
44	&	HII	&	18:09:29.30	&	-19:16:19.30	&	---	&	620	&	500	&	-60	&	3.4	&	---	&	8	\\
47	&	SFC	&	18:09:54.44	&	-19:44:46.90	&	---	&	450	&	360	&	-10	&	3.5	&	0.2	&	1,4,2	\\
49	&	SFC	&	18:10:02.27	&	-18:50:12.00	&	280	&	---	&	---	&	---	&	3.4	&	---	&	1	\\
50	&	HII	&	18:10:05.43	&	-20:59:11.90	&	360	&	---	&	---	&	---	&	3.7\tablefootmark{a}	&	---	&	4,8	\\
51	&	HII	&	18:10:12.40	&	-20:46:22.90	&	---	&	800	&	260	&	-50	&	3.7	&	0.2	&	4,8	\\
54	&	SFC	&	18:10:26.70	&	-19:20:56.20	&	---	&	920	&	380	&	-65	&	3.5	&	0.2	&	1,4,3,2,6	\\
55	&	HII	&	18:10:54.13	&	-19:52:35.30	&	---	&	860	&	630	&	-30	&	5.2	&	---	&	6	\\
60	&	HII	&	18:10:51.72	&	-17:55:57.40	&	240	&	---	&	---	&	---	&	2.4	&	0.3	&	1,4,6	\\
61	&	SFC	&	18:10:54.72	&	-20:32:50.30	&	180	&	---	&	---	&	---	&	5.4	&	---	&	6	\\
65	&	SFC	&	18:11:05.69	&	-19:37:05.40	&	---	&	300	&	180	&	-30	&	5.2	&	<0.1	&	6,2	\\
72	&	HII	&	18:11:32.30	&	-19:30:42.10	&	300	&	---	&	---	&	---	&	5.2	&	<0.1	&	8,6	\\
73	&	SFC	&	18:11:46.63	&	-18:17:40.40	&	---	&	240	&	120	&	90	&	3.8	&	---	&	1	\\
74	&	HII	&	18:11:43.26	&	-18:16:54.00	&	---	&	480	&	300	&	-10	&	3.6	&	---	&	1	\\
75	&	SFC	&	18:11:56.19	&	-18:48:14.50	&	200	&	---	&	---	&	---	&	3.2	&	---	&	1	\\
78	&	SFC	&	18:11:59.63	&	-19:07:48.50	&	400	&	---	&	---	&	---	&	4.6	&	---	&	1	\\
79	&	SFC	&	18:12:00.81	&	-19:36:00.90	&	350	&	---	&	---	&	---	&	3.2	&	---	&	1	\\
81	&	SFC	&	18:12:04.07	&	-17:52:43.80	&	210	&	---	&	---	&	---	&	4.3	&	---	&	6	\\
82	&	SFC	&	18:12:08.63	&	-16:42:41.60	&	360	&	---	&	---	&	---	&	3.4	&	---	&	6	\\
83	&	HII	&	18:12:12.42	&	-17:40:52.80	&	---	&	450	&	360	&	-63	&	2.2	&	0.4	&	4,2,6	\\
86	&	SFC	&	18:12:14.29	&	-18:26:52.00	&	130	&	---	&	---	&	---	&	4.8	&	---	&	6	\\
88	&	SFC	&	18:12:26.51	&	-17:32:46.30	&	---	&	240	&	90	&	-10	&	3.9	&	---	&	6	\\
90	&	HII	&	18:12:32.90	&	-18:30:13.00	&	80	&	---	&	---	&	---	&	4.2	&	<0.1	&	4,3	\\
91	&	SFC	&	18:12:32.62	&	-17:29:16.80	&	---	&	120	&	100	&	80	&	3.6	&	---	&	1	\\
94	&	SFC	&	18:12:41.69	&	-18:42:58.80	&	120	&	---	&	---	&	---	&	5	&	---	&	4	\\
98	&	SFC	&	18:12:56.03	&	-19:04:23.40	&	---	&	700	&	250	&	-80	&	3.9	&	---	&	6	\\
100	&	HII	&	18:13:09.20	&	-18:07:49.70	&	---	&	450	&	250	&	-40	&	2.6	&	<0.1	&	4,2	\\
102	&	SFC	&	18:12:51.72	&	-18:48:13.50	&	---	&	500	&	120	&	-50	&	5	&	---	&	-6	\\
103	&	HII	&	18:13:12.52	&	-18:00:07.20	&	150	&	---	&	---	&	---	&	2.6	&	0.1	&	4,3	\\
104	&	SFC	&	18:13:12.15	&	-16:41:11.40	&	---	&	210	&	100	&	30	&	3	&	---	&	6	\\
105	&	SFC	&	18:13:09.52	&	-18:15:57.60	&	---	&	370	&	240	&	-45	&	3.5	&	---	&	1	\\
106	&	SFC	&	18:13:30.90	&	-17:18:06.70	&	165	&	---	&	---	&	---	&	2.3	&	---	&	1	\\
108	&	SFC	&	18:13:38 	&	-18:12:15.20	&	---	&	180	&	120	&	35	&	3.6	&	---	&	1	\\
110	&	SFC	&	18:13:35.59	&	-17:23:47.20	&	---	&	270	&	210	&	-10	&	4.5	&	---	&	1	\\
111	&	SFC	&	18:13:54.20	&	-17:16:28.50	&	---	&	330	&	210	&	90	&	2.4	&	---	&	1	\\
114	&	HII	&	18:13:52.96	&	-18:56:44.90	&	---	&	450	&	300	&	100	&	3.9\tablefootmark{a}	&	0.3	&	4,3,8,6	\\
116	&	HII	&	18:14:09.82	&	-17:27:24.20	&	---	&	570	&	220	&	-45	&	4.4	&	0.1	&	4,3,2	\\
120	&	SFC	&	18:14:05.27	&	-18:14:49.00	&	300	&	---	&	---	&	---	&	3.5	&	---	&	1	\\
121	&	SFC	&	18:14:13.00	&	-18:20:41.80	&	---	&	150	&	120	&	-30	&	5.1	&	---	&	6	\\
122	&	HII	&	18:14:16.00	&	-17:57:18.60	&	---	&	780	&	550	&	-35	&	4	&	0.6	&	4,8,3,2	\\
123	&	SFC	&	18:14:16.28	&	-17:14:57.00	&	---	&	210	&	120	&	-60	&	2.8	&	---	&	1	\\
134	&	HII	&	18:14:35.87	&	-16:45:31.30	&	---	&	960	&	560	&	-60	&	4.4	&	0.3	&	4	\\
136	&	HII	&	18:14:30.50	&	-17:38:34.00	&	---	&	750	&	400	&	-65	&	4.4	&	0.6	&	1,4,2,6	\\
140	&	SFC	&	18:14:50.37	&	-17:20:31.50	&	---	&	420	&	210	&	-75	&	4.5	&	---	&	4	\\
146	&	SFC	&	18:14:55.69	&	-17:46:58.50	&	---	&	210	&	120	&	80	&	5.1	&	---	&	6	\\
149	&	SFC	&	18:14:57.28	&	-18:27:01.10	&	270	&	---	&	---	&	---	&	5	&	---	&	6	\\
152	&	SFC	&	18:15:01.51	&	-17:42:22.80	&	100	&	---	&	---	&	---	&	4.6	&	---	&	6	\\
154	&	SFC	&	18:15:11.03	&	-17:47:21.20	&	---	&	360	&	130	&	95	&	3.7	&	---	&	1	\\
159	&	SFC	&	18:15:46.15	&	-17:34:20.50	&	---	&	750	&	540	&	80	&	3.8	&	0.4	&	1,4,2,6	\\
160	&	HII	&	18:15:37.51	&	-17:04:23.50	&	400	&	---	&	---	&	---	&	4.1	&	<0.1	&	4,3	\\
162	&	SFC	&	18:15:38.97	&	-18:10:42.50	&	760	&	---	&	---	&	---	&	3.5	&	---	&	2,6	\\
163	&	HII	&	18:15:40.22	&	-17:19:46.70	&	260	&	---	&	---	&	---	&	2.6	&	---	&	1	\\
165	&	HII	&	18:15:45.18	&	-16:38:58.20	&	240	&	---	&	---	&	---	&	5.4	&	---	&	3	\\
170	&	SFC	&	18:16:00.58	&	-16:04:53.20	&	440	&	---	&	---	&	---	&	2.8	&	---	&	4	\\
175	&	SFC	&	18:16:12.68	&	-16:43:53.70	&	---	&	240	&	150	&	-10	&	4.3	&	---	&	6	\\
177	&	SFC	&	18:16:20.00	&	-17:05:00.90	&	---	&	120	&	70	&	80	&	3.7	&	---	&	1	\\
178	&	HII	&	18:16:31.62	&	-16:50:59.60	&	---	&	600	&	240	&	240	&	3.7	&	<0.1	&	4,2	\\
179	&	HII	&	18:16:04.39	&	-19:42:05.20	&	460	&	---	&	---	&	---	&	1.6	&	---	&	8	\\
183	&	SFC	&	18:16:45.65	&	-17:02:26.90	&	300	&	---	&	---	&	---	&	3.7	&	---	&	1	\\
185	&	HII	&	18:16:57.86	&	-16:15:33.90	&	---	&	600	&	360	&	-60	&	2.4	&	0.3	&	4,3,2,8	\\
186	&	HII	&	18:16:51.62	&	-18:41:28.30	&	300	&	---	&	---	&	---	&	4	&	---	&	4	\\
187	&	HII	&	18:16:58.96	&	-16:42:08.70	&	---	&	380	&	240	&	30	&	3.6	&	0.2	&	4,2,6	\\
190	&	HII	&	18:17:11.87	&	-16:27:00.60	&	---	&	630	&	300	&	-25	&	3.7\tablefootmark{a}	&	0.1	&	4,3,2	\\
192	&	SFC	&	18:16:18.76	&	-14:18:57.50	&	1100	&	---	&	---	&	---	&	2.2	&	0.2	&	4	\\
195	&	SFC	&	18:17:27.20	&	-17:08:58.50	&	---	&	480	&	220	&	-45	&	2.7	&	0.3	&	1	\\
196	&	SFC	&	18:17:21.85	&	-17:00:48.80	&	---	&	330	&	220	&	-60	&	2.3	&	---	&	6	\\
200	&	SFC	&	18:17:36.02	&	-15:56:27.60	&	---	&	360	&	200	&	-30	&	2.7	&	0.1	&	4,7	\\
202	&	SFC	&	18:17:31.86	&	-16:16:59.20	&	---	&	330	&	140	&	-80	&	3.7	&	---	&	1	\\
207	&	SFC	&	18:17:35.93	&	-15:47:45.70	&	240	&	---	&	---	&	---	&	3.1	&	0.2	&	1,7	\\
209	&	SFC	&	18:17:52.47	&	-16:29:02.00	&	---	&	370	&	210	&	-45	&	4.6	&	---	&	1	\\
210	&	SFC	&	18:17:55.42	&	-17:13:05.50	&	250	&	---	&	---	&	---	&	2.9	&	---	&	6	\\
214	&	HII	&	18:17:45.73	&	-16:01:22.20	&	---	&	400	&	200	&	-30	&	2.8	&	<0.1	&	7,6	\\
217	&	SFC	&	18:17:59.25	&	-16:15:15.20	&	---	&	480	&	270	&	100	&	3.7	&	0.2	&	1,2,4	\\
218	&	HII	&	18:18:09.85	&	-16:52:16.80	&	---	&	750	&	500	&	-70	&	2.5	&	0.4	&	1,2,4	\\
219	&	SFC	&	18:18:15.15	&	-16:04:42.50	&	---	&	360	&	200	&	-10	&	3.9	&	---	&	7a	\\
220	&	HII	&	18:18:16.55	&	-15:59:13.50	&	---	&	180	&	135	&	80	&	4	&	---	&	4,3	\\
227	&	SFC	&	18:18:46.76	&	-16:22:29.30	&	---	&	170	&	150	&	80	&	4.6	&	---	&	1	\\
228	&	HII	&	18:18:47.08	&	-13:32:56.90	&	---	&	2000	&	1200	&	-60	&	2	&	0.4	&	4,3,8,2,6	\\
230	&	HII	&	18:18:54.20	&	-16:48:55.70	&	---	&	260	&	160	&	90	&	2.7	&	0.3	&	4,2,8	\\
233	&	SFC	&	18:19:06.85	&	-16:33:09.50	&	---	&	570	&	280	&	100	&	2.3	&	0.2	&	1,4,2	\\
234	&	SFC	&	18:19:06.97	&	-16:11:20.30	&	300	&	---	&	---	&	---	&	4.7	&	<0.1	&	4	\\
236	&	SFC	&	18:19:11.92	&	-16:18:25.60	&	---	&	420	&	150	&	---	&	2.6	&	---	&	1	\\
238	&	SFC	&	18:19:38.17	&	-16:42:43.10	&	---	&	650	&	330	&	-45	&	2.4	&	---	&	1	\\
242	&	HII	&	18:19:51.20	&	-15:55:05.60	&	200	&	---	&	---	&	---	&	2.3	&	---	&	4	\\
246	&	SFC	&	18:20:12.44	&	-13:51:05.10	&	---	&	450	&	360	&	70	&	2.6	&	<0.1	&	7,6	\\
247	&	SFC	&	18:20:37.42	&	-15:37:55.10	&	100	&	---	&	---	&	---	&	3.6	&	---	&	1	\\
248	&	HII	&	18:20:27.49	&	-16:07:40.60	&	850	&	---	&	---	&	---	&	2.2	&	0.2	&	4,8,3,2	\\
251	&	SFC	&	18:20:39.33	&	-14:02:41.30	&	450	&	---	&	---	&	---	&	2.6	&	---	&	1	\\
253	&	SFC	&	18:20:56.57	&	-15:24:41.80	&	180	&	---	&	---	&	---	&	3.8	&	<0.1	&	1,4,6	\\
256	&	HII	&	18:21:05.67	&	-14:17:49.10	&	---	&	580	&	340	&	90	&	2.5	&	---	&	7	\\
259	&	HII	&	18:21:09.17	&	-14:31:46.20	&	---	&	700	&	350	&	-40	&	4.3\tablefootmark{a}	&	0.5	&	4,3,8	\\
260	&	HII	&	18:21:10.92	&	-15:02:42.00	&	---	&	300	&	150	&	80	&	3.3	&	---	&	3	\\
261	&	SFC	&	18:21:12.27	&	-16:30:08.20	&	210	&	---	&	---	&	---	&	2.6	&	---	&	1	\\
263	&	HII	&	18:21:04.64	&	-14:46:39.40	&	---	&	600	&	450	&	-10	&	3.9	&	<0.1	&	4	\\
265	&	SFC	&	18:21:44.83	&	-14:56:52.90	&	---	&	380	&	250	&	-60	&	3.9	&	---	&	6	\\
270	&	HII	&	18:21:59.64	&	-14:16:11.60	&	---	&	360	&	260	&	-40	&	3.6\tablefootmark{a}	&	0.2	&	4,7	\\
271	&	HII	&	18:22:04.97	&	-14:08:55.40	&	---	&	210	&	150	&	-30	&	3.6	&	---	&	4,3	\\
272	&	HII	&	18:22:05.14	&	-14:48:42.90	&	100	&	---	&	---	&	---	&	3.8	&	---	&	1	\\
279	&	HII	&	18:22:22.95	&	-14:35:35.30	&	---	&	190	&	120	&	-60	&	3.8	&	---	&	1	\\
287	&	SFC	&	18:22:41.01	&	-14:27:54.20	&	330	&	---	&	---	&	---	&	2.4	&	---	&	7	\\
292	&	HII	&	18:22:52.78	&	-13:59:26.60	&	550	&	---	&	---	&	---	&	4.8	&	---	&	6	\\
297	&	SFC	&	18:23:18.85	&	-13:15:50.60	&	240	&	---	&	---	&	---	&	2.2	&	<0.1	&	5,4,7	\\
298	&	SFC	&	18:23:26.37	&	-13:49:53.30	&	240	&	---	&	---	&	---	&	4.3	&	---	&	7	\\
301	&	HII	&	18:23:35.90	&	-13:09:28.50	&	240	&	---	&	---	&	---	&	6.6	&	---	&	5	\\
302	&	SFC	&	18:23:37.57	&	-13:18:52.90	&	210	&	---	&	---	&	---	&	4.2	&	---	&	5	\\
313	&	HII	&	18:24:34.34	&	-12:52:02.70	&	300	&	---	&	---	&	---	&	4	&	0.3	&	4,3,8	\\
320	&	SFC	&	18:25:07.20	&	-14:35:12.50	&	400	&	---	&	---	&	---	&	3.4	&	---	&	4	\\
322	&	SFC	&	18:25:09.19	&	-12:44:38.70	&	---	&	300	&	120	&	100	&	3.5	&	<0.1	&	1,4	\\
323	&	HII	&	18:24:58.83	&	-12:01:05.50	&	650	&	---	&	---	&	---	&	4.5	&	---	&	6	\\
325	&	HII	&	18:25:16.95	&	-13:14:13.00	&	850	&	---	&	---	&	---	&	4.6	&	0.5	&	4,3,8,2,5	\\
326	&	SFC	&	18:25:19.61	&	-12:53:00.50	&	---	&	500	&	230	&	95	&	3.9	&	0.2	&	1,5,7	\\
327	&	SFC	&	18:25:22.32	&	-13:34:38.30	&	320	&	---	&	---	&	---	&	4	&	0.5	&	1,7	\\
328	&	HII	&	18:25:35.01	&	-12:20:58.10	&	---	&	660	&	360	&	65	&	4.2	&	0.1	&	4,2,5	\\
330	&	SFC	&	18:25:28.88	&	-13:58:26.00	&	420	&	---	&	---	&	---	&	5.1	&	---	&	6	\\
332	&	SFC	&	18:25:53.36	&	-12:06:08.30	&	360	&	---	&	---	&	---	&	2.5	&	<0.1	&	1	\\
336	&	HII	&	18:26:00.56	&	-11:52:26.90	&	---	&	720	&	440	&	-20	&	1.9	&	---	&	4,8,3,5	\\
340	&	HII	&	18:26:20.45	&	-12:40:52.20	&	240	&	---	&	---	&	---	&	4.6	&	0.3	&	4,7,6	\\
341	&	SFC	&	18:26:22.16	&	-12:57:16.60	&	180	&	---	&	---	&	---	&	3.6	&	---	&	1	\\
342	&	HII	&	18:26:24.02	&	-12:03:09.10	&	---	&	200	&	160	&	80	&	2.5	&	0.2	&	8,3	\\
343	&	SFC	&	18:26:30.07	&	-12:32:42.00	&	---	&	360	&	300	&	10	&	4.7	&	0.3	&	7,6	\\
348	&	SFC	&	18:26:46.14	&	-12:03:25.70	&	230	&	---	&	---	&	---	&	4.6	&	---	&	6	\\
351	&	HII	&	18:26:44.64	&	-12:24:05.30	&	---	&	390	&	330	&	200	&	4.5	&	0.2	&	4,8,5	\\
359	&	HII	&	18:27:03.75	&	-12:43:48.30	&	480	&	---	&	---	&	---	&	4.5	&	0.3	&	5,4,2,3	\\
363	&	SFC	&	18:27:17.69	&	-10:36:28.60	&	300	&	---	&	---	&	---	&	2.6	&	0.1	&	1,7	\\
365	&	SFC	&	18:27:09.45	&	-13:19:49.80	&	480	&	---	&	---	&	---	&	4.2	&	---	&	5	\\
366	&	SFC	&	18:27:35.55	&	-12:19:14.70	&	290	&	---	&	---	&	---	&	4	&	0.1	&	1	\\
367	&	SFC	&	18:27:45.09	&	-12:46:41.10	&	320	&	---	&	---	&	---	&	4.4	&	0.2	&	5,4	\\
370	&	SFC	&	18:27:42.08	&	-11:36:29.60	&	---	&	300	&	180	&	-60	&	4.4	&	---	&	1	\\
372	&	HII	&	18:27:52.09	&	-12:36:17.50	&	440	&	---	&	---	&	---	&	4.9	&	---	&	2,5	\\
373	&	SFC	&	18:28:06.14	&	-11:39:09.90	&	---	&	100	&	70	&	-60	&	4.2	&	---	&	1	\\
376	&	SFC	&	18:28:20.95	&	-11:35:51.60	&	---	&	180	&	110	&	-50	&	3.5	&	<0.1	&	1,4	\\
380	&	SFC	&	18:28:23.18	&	-11:41:19.00	&	---	&	260	&	220	&	-30	&	4.4	&	0.3	&	1,4,7	\\
381	&	SFC	&	18:28:23.59	&	-11:47:37.90	&	240	&	---	&	---	&	---	&	3.5	&	0.2	&	4,5	\\
386	&	SFC	&	18:28:48.60	&	-11:48:40.00	&	---	&	420	&	230	&	-60	&	4.6	&	---	&	1	\\
389	&	SFC	&	18:29:18.34	&	-12:10:38.60	&	---	&	820	&	430	&	-10	&	4.2	&	---	&	5	\\
390	&	SFC	&	18:29:14.36	&	-11:50:21.70	&	250	&	---	&	---	&	---	&	3.4	&	0.2	&	4,2,5	\\
396	&	SFC	&	18:29:59.97	&	-11:00:30.50	&	330	&	---	&	---	&	---	&	4.2	&	0.2	&	1	\\
399	&	SFC	&	18:30:30.89	&	-11:12:11.90	&	---	&	630	&	330	&	-30	&	4.7	&	---	&	5	\\
400	&	SFC	&	18:30:30.81	&	-11:52:50.40	&	730	&	---	&	---	&	---	&	3.5	&	<0.1	&	5,7	\\
406	&	HII	&	18:18:47.30	&	-15:48:58.30	&	220	&	---	&	---	&	---	&	1.9	&	---	&	7	\\
407	&	HII	&	18:19:59.95	&	-15:27:06.90	&	560	&	---	&	---	&	---	&	3.7	&	---	&	6	\\
26a	&	SLC	&	18:08:49.40	&	-19:40:15.00	&	80	&	---	&	---	&	---	&	3.1	&	<0.1	&	2,1	\\
30a	&	SLC	&	18:09:08.56	&	-19:45:56.20	&	60	&	---	&	---	&	---	&	5.2	&	---	&	2	\\
54c	&	SLC	&	18:10:40.07	&	-19:10:40.90	&	100	&	---	&	---	&	---	&	3.5	&	0.2	&	1,3,2,4	\\
65a	&	SLC	&	18:11:09.60	&	-19:36:32.00	&	96	&	---	&	---	&	---	&	5.2	&	<0.1	&	2,6	\\
83c	&	SLC	&	18:12:19.40	&	-17:40:12.20	&	70	&	---	&	---	&	---	&	2.2	&	0.4	&	2,4,6	\\
106a	&	SLC	&	18:13:35.58	&	-17:18:39.60	&	---	&	60	&	35	&	-50	&	2.3	&	---	&	1	\\
116b	&	SLC	&	18:14:13.22	&	-17:25:15.80	&	50	&	---	&	---	&	---	&	4.4	&	0.1	&	3,2,4	\\
122b	&	SLC	&	18:13:43.80	&	-17:56:34.00	&	---	&	80	&	60	&	---	&	4	&	0.6	&	3,2,4,8	\\
185a	&	SLC	&	18:16:50.35	&	-16:21:44.80	&	---	&	150	&	80	&	60	&	2.4	&	0.3	&	3,2,4,8	\\
203a	&	SLC	&	18:17:32.59	&	-17:06:41.60	&	120	&	---	&	---	&	---	&	2.6	&	---	&	2	\\
217a	&	SLC	&	18:18:02.99	&	-16:17:23.10	&	80	&	---	&	---	&	---	&	3.7	&	0.2	&	1,2,4	\\
218b	&	SLC	&	18:18:03.23	&	-16:54:35.50	&	---	&	95	&	60	&	-80	&	2.5	&	0.4	&	1,2,4	\\
233a	&	SLC	&	18:19:02.22	&	-16:39:59.00	&	125	&	---	&	---	&	---	&	2.3	&	0.2	&	1,2,4	\\
233b	&	SLC	&	18:19:13.55	&	-16:34:57.10	&	85	&	---	&	---	&	---	&	2.3	&	0.2	&	1,2,4	\\
233e	&	SLC	&	18:19:13.36	&	-16:35:04.20	&	100	&	---	&	---	&	---	&	2.3	&	0.2	&	1,2,4	\\
247a	&	SLC	&	18:20:23.82	&	-15:38:31.80	&	---	&	170	&	80	&	95	&	3.6	&	---	&	1	\\
325g	&	SLC	&	18:25:23.77	&	-13:19:04.50	&	---	&	125	&	80	&	-10	&	4.6	&	0.5	&	2,3,4,5,8	\\
325i	&	SLC	&	18:25:22.56	&	-13:17:13.40	&	40	&	---	&	---	&	---	&	4.6	&	0.5	&	2,3,4,5,8	\\
328c	&	SLC	&	18:25:22.76	&	-12:23:26.90	&	100	&	---	&	---	&	---	&	4.2	&	0.1	&	2,4,5	\\
379a	&	SLC	&	18:28:18.30	&	-12:06:26.00	&	170	&	---	&	---	&	---	&	4.4	&	---	&	2	\\
390b	&	SLC	&	18:29:26.67	&	-11:50:45.60	&	50	&	---	&	---	&	---	&	3.4	&	0.2	&	2,4,5	\\
159a	&	SLC	&	18:16:06.90	&	-17:36:27.10	&	65	&	---	&	---	&	---	&	3.8	&	0.4	&	1,2,4,8	\\
162a	&	SLC	&	18:15:40.50	&	-18:13:08.00	&	75	&	---	&	---	&	---	&	3.5	&	<0.1	&	2,6	\\
160a	&	SLC	&	18:15:25.30	&	-17:05:43.70	&	90	&	---	&	---	&	---	&	4.1	&	<0.1	&	3,2,4	\\
187f	&	SLC	&	18:16:59.68	&	-16:39:26.80	&	---	&	90	&	70	&	-40	&	4	&	0.2	&	2,4,6	\\
212a	&	SLC	&	18:17:50.60	&	-15:53:34.00	&	50	&	---	&	---	&	---	&	2.7	&	---	&	2	\\
212e	&	SLC	&	18:17:52.84	&	-15:55:16.90	&	50	&	---	&	---	&	---	&	2.7	&	---	&	2	\\
253a	&	SLC	&	18:20:55.50	&	-15:24:05.00	&	50	&	---	&	---	&	---	&	3.8	&	<0.1	&	1,4,6	\\
329a	&	SLC	&	18:25:32.88	&	-13:01:51.70	&	120	&	---	&	---	&	---	&	4.5	&	---	&	2	\\
372a	&	SLC	&	18:27:43.01	&	-12:35:42.70	&	50	&	---	&	---	&	---	&	4.9	&	<0.1	&	2,5	\\
397a	&	SLC	&	18:30:02.90	&	-12:15:51.00	&	120	&	---	&	---	&	---	&	3.1	&	---	&	2	\\
\end{longtable}
\tablebib{(1)~\citet{2013ApJ...770...39E};
(2)~\citet{2012A&A...540A.113T}; (3)~\citet{2013MNRAS.435..400U};
(4)~\citet{2012A&A...544A.146W}; (5)~\citet{2009ApJ...699.1153R};
(6)~\citet{2009ApJ...706..727M}; (7)~\citet{2006ApJ...639..227S}; 
(8)~\citet{1997MNRAS.291..261W}.}
\tablefoot{
\tablefoottext{a}{These regions have not previous solution for the KDA, but they show shadows against
the NIR background radiation, so we located them at the near distance solution.}
\tablefoottext{b}{The distance shown in this table is the result of averaging all the distance
estimates available for each source.}
\tablefoottext{c}{The dispersion of all the distance
estimates available for each region, when more than one
distance estimates are available.}}
\end{landscape}}

\subsection{Column density and mass estimation}\label{sec:NH}
Column densities of molecular gas were calculated via
\begin{equation}\label{eq:NH2}
N_{\mathrm{H_{2}}} [\mathrm{cm^{-2}}] =
\frac{RF_{\lambda}}{B_{\lambda}(T_{Dust})\mu
m_{\mathrm{H}}\kappa\Omega},
\end{equation}
where $F_{\lambda}$ and $B_{\lambda}(T)$ are respectively the flux and
the blackbody radiation as a function of temperature, $T$, at
870\,$\mu$m.  The quantity $\mu$ is the mean molecular weight (assumed
to be 2.8) of the interstellar medium per hydrogen molecule,
$m_{\mathrm{H}}$ is the mass of the hydrogen atom, $\Omega$ is the
beam solid angle, and $R=154$ is the gas-to-dust
ratio~\citep{2011piim.book.....D}.  We used a dust absorption
coefficient $\kappa=1.85\mathrm{\,cm^{2}g^{-1}}$ at $870\,\mathrm{\mu}$m, which was calculated by interpolation of
the~\citet{1994A&A...291..943O} dust model of grains with thin ice
mantles and a mean density of $n=10^{6}\mathrm{\,cm^{-3}}$.  We
assumed $T=15$\,K for SLCs and SFCs~\citep{2012A&A...544A.146W}, in
agreement with previous dust temperature estimations within infrared
dark clouds~\citep[IRDCs]{2010ApJ...723..555P} and in envelopes of
star-forming cores~\citep{2010A&A...518L..87S,2013A&A...551A..98L}.
For \HII regions we assumed $T=25$\,K.  This dust temperature is in
agreement with the average dust temperatures in PDR regions
surrounding \HII regions ($T=26$\,K), where most of the FIR-submm dust
emission of these objects comes from~\citep{anderson-12}. It also
agrees with the mean temperature found in the central region of
NGC6334 ($T\sim24$\,K), that is an expanding \HII
region~\citep{2013A&A...554A..42R}.  For a better comparison with
previous works, we present the column density data also in units of
visual extinction using a conversion: $N_{\mathrm{H_{2}}} =
0.94\times10^{21}\,A_{V}\,\mathrm{cm^{-2}}\,\mathrm{mag^{-1}}$
~\citep{1978ApJ...224..132B}.  The rms noise of the ATLASGAL data
(50\,mJy) corresponds to $A_{V}=4.5$\,mag for both the SFCs and SLCs
and 2.2\,mag\footnote{The difference in the rms values in terms of
  $A_{V}$ is due to the temperatures assumed for each evolutionary
  class.} for \HII regions.  No saturation problems were found in the
ATLASGAL survey. The optical depth is $<<1$, therefore our
measurements do not suffer from optical depth effects in the
high-column density regime~\citep{2009A&A...504..415S}.

We estimated the total gas mass of each region from
dust continuum emission, assuming that
emission is optically thin:
\begin{equation}\label{mass}
M_{g} = \frac{Rd^{2}F_{\lambda}}{B_{\lambda}(T_{Dust})\kappa},
\end{equation}
where $d$ is the distance to the region. We assume the same values 
for the other listed quantities as we assume for
the column density determination (Eq.~\ref{eq:NH2}).
Masses of the regions cover three orders of magnitude
(Fig.~\ref{fig:mass-hist}).
The masses of the SLCs span $0.2-4\times10^{3}\,\mathrm{M_{\mathrm{\sun}}}$,
the SFCs $0.3-15\times10^{3}\,\mathrm{M_{\mathrm{\sun}}}$,
and the \HII regions $0.2-200\times10^{3}\,\mathrm{M_{\mathrm{\sun}}}$.
Larger masses for \HII regions and SFCs are expected since both
have much larger sizes than SLCs (Fig.~\ref{fig:galPlane}).

The derived mass and column density values depend on the assumed dust
properties,
specifically on $\kappa_{870\,\mathrm{\mu m}}$, $R$ and $T_{Dust}$.
Both $\kappa_{870\,\mathrm{\mu m}}$ and $R$ are subject to 
uncertainties:
$\kappa_{870\,\mathrm{\mu m}}$ values differ by $\sim1$\,dex in different
dust models~\citep{2005ApJ...632..982S,2011ApJ...728..143S}.
Eq.~\ref{eq:NH2} and Eq.~\ref{mass} assume isothermal clouds.
This is clearly an oversimplification,
increasing the uncertainty in the derived masses.
Mass depends also on $d^{2}$, making
uncertainties in distance a major contributor
to the absolute uncertainties.
If we adopt $\Delta d\sim 0.5$\,kpc, nearby regions
will be more affected by distance uncertainties (50\% at 1\,kpc)
than the most distant regions (10\% at 5\,kpc).
This assumption agrees with the distance uncertainties 
reported by~\citep{2009ApJ...699.1153R}.
We note that the absolute uncertainty in our derived column densities is
very large, potentially larger than a factor 10. 
The relative uncertainties between the evolutionary classes 
can be influenced by the different temperature assumptions or
intrinsic differences in the dust properties. 
The isothermal assumption introduces differences in the low-column
density regime of the $N$-PDFs, but it has negligible effect in 
shaping the column density distribution at high-column densities (see App.~\ref{app:tcomp}).
In the case of dust properties, we have no knowledge about any observational-based study suggesting changes
in them in molecular clouds at different evolutionary
phases. We therefore assume that the dust properties do not
introduce relative uncertainties between the three molecular cloud
classes defined.

\subsection{Physical properties of the evolutionary classes}

We define and analyze in this work three distinctive evolutionary 
classes of objects: SLCs, SFCs and \HII regions. The objects in these 
classes are different in their physical characteristics. 
These differences originate dominantly from the fact that the \HII 
regions and SFCs are typically extended regions 
(i.e., molecular clouds or even cloud complexes), while SLCs are smaller, 
"clump-like" structures. We quantify here the basic physical properties 
of the objects in our three evolutionary classes. 
The properties are also listed in Table ~\ref{tab:ph-prop}.

\begin{figure}
\centering
\resizebox{\hsize}{!}{\includegraphics{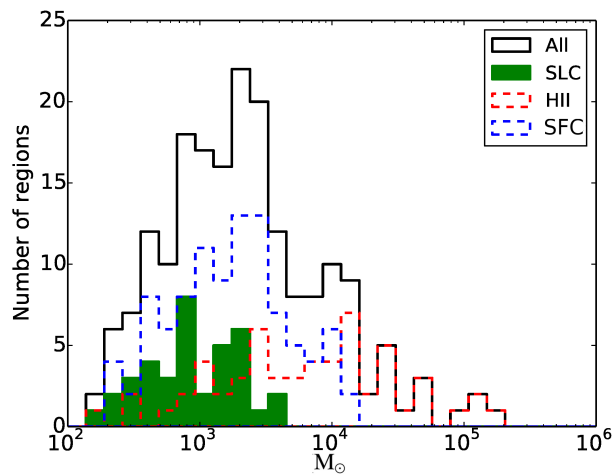}}
\caption{Mass distribution of the molecular cloud regions. Filled green area
shows mass for starless clumps, dashed blue line shows star-forming clouds
and red dashed line shows \HII regions. Masses are given in solar mass
units.}
\label{fig:mass-hist}
\end{figure}

Figure~\ref{fig:mass-hist} 
shows the mass distribution of our regions. The mass distribution of \HII regions spans
3\,dex from  $10^{2}-10^{5}\mathrm{\,M_{\sun}}$. SFCs have
masses of $10^{2}-10^{4}\mathrm{\,M_{\sun}}$, and SLCs show the most narrow
mass range, $10^{2}-10^{3}\mathrm{M_{\,\sun}}$.

\begin{figure}
\centering
\resizebox{\hsize}{!}{\includegraphics{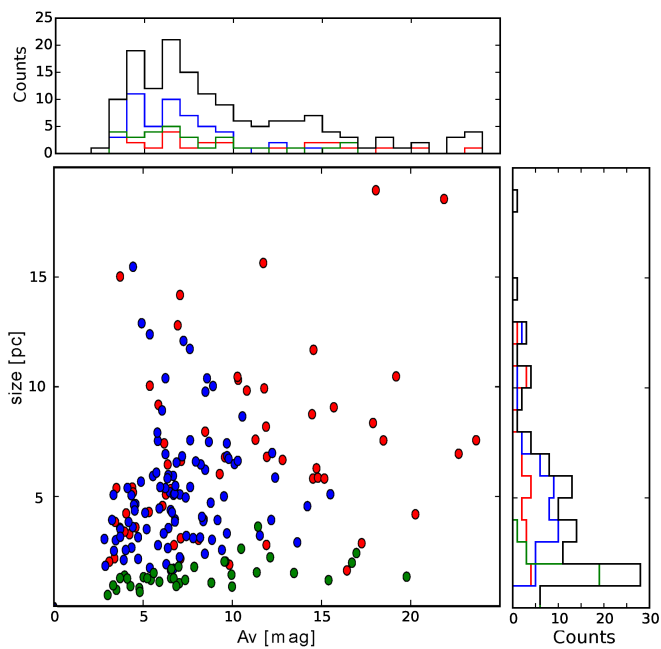}}
\caption{Mean column density, $\overline{A_{V}}$, and size distribution of
all the regions. In the scatter plot we show
the \HII regions in red, the SFCs in blue and 
the SLCs in green. The histograms show the $\overline{A_{V}}$ and size distributions
of each evolutionary class (same colors) and the whole sample (black).}
\label{fig:mean-av}
\end{figure}

The spread of the distribution of mean 
column densities is $\overline{A_{V}}=3-25$\,mag
and it peaks at $\overline{A_{V}}=7$\,mag (see Fig.~\ref{fig:mean-av}).
The $\overline{A_{V}}$ distribution differs in each evolutionary class.
While most SFCs and SLCs have $\overline{A_{V}}\lesssim10$\,mag,
a considerable number of \HII regions show $\overline{A_{V}}>10$\,mag. 
We note that the mean column densities of our sample are overestimated
due to the spatial filtering of ATLASGAL,
and so is the peak of the $\overline{A_{V}}$ distribution.
This effect is more important in \HII regions and SFCs since 
they have larger areas and hence larger fraction of diffuse 
material that is filtered out than the SLCs.

Figure~\ref{fig:mean-av} also shows the size
distribution of each class and of the total sample.
The SLCs have the smallest sizes of the sample with 
a mean size of 1.4\,pc and a range of sizes between
1\,pc and 2.5\,pc. The range of sizes of the SFCs 
is 2-15\,pc, with a mean of 5.3\,pc.
The \HII regions have the largest mean size of the three evolutionary
classes, 7\,pc, and also the largest spread, 2-18\,pc.

\begin{table}
\caption{Mean physical properties of the evolutionary classes} %
\centering 
\begin{tabular}{c c c c} 
\hline\hline 
					&\HII			&SFCs			&SLCs		\\\hline 
Mass [M$_{\sun}$]\tablefootmark{a}	&18$\times10^{3}$ 	&2.7$\times10^{3}$	&1.2$\times10^{3}$\\
$\overline{A_{V}}$ [mag]		&10.6 $\pm$ 6.2		&7.0 $\pm$ 2.5		&8.2 $\pm$ 4.2	\\
Size [pc]				&7.0 $\pm$ 4.0		&5.3 $\pm$ 2.7		&1.4 $\pm$ 0.6	\\\hline
\label{tab:ph-prop}
\end{tabular}
\end{table}

\section{Results}\label{results}

\subsection{Column density distribution}

We use the column density data to study the column density
distributions of the regions.
In the following, we first analyze the \textit{N}-PDFs and DGMFs.
We then examine the relationship between the
total mass and the column density distribution of the regions.

\subsubsection{\textit{N}-PDFs}\label{sec:pdf}

\begin{figure*}
\centering
\includegraphics[width=0.33\textwidth]{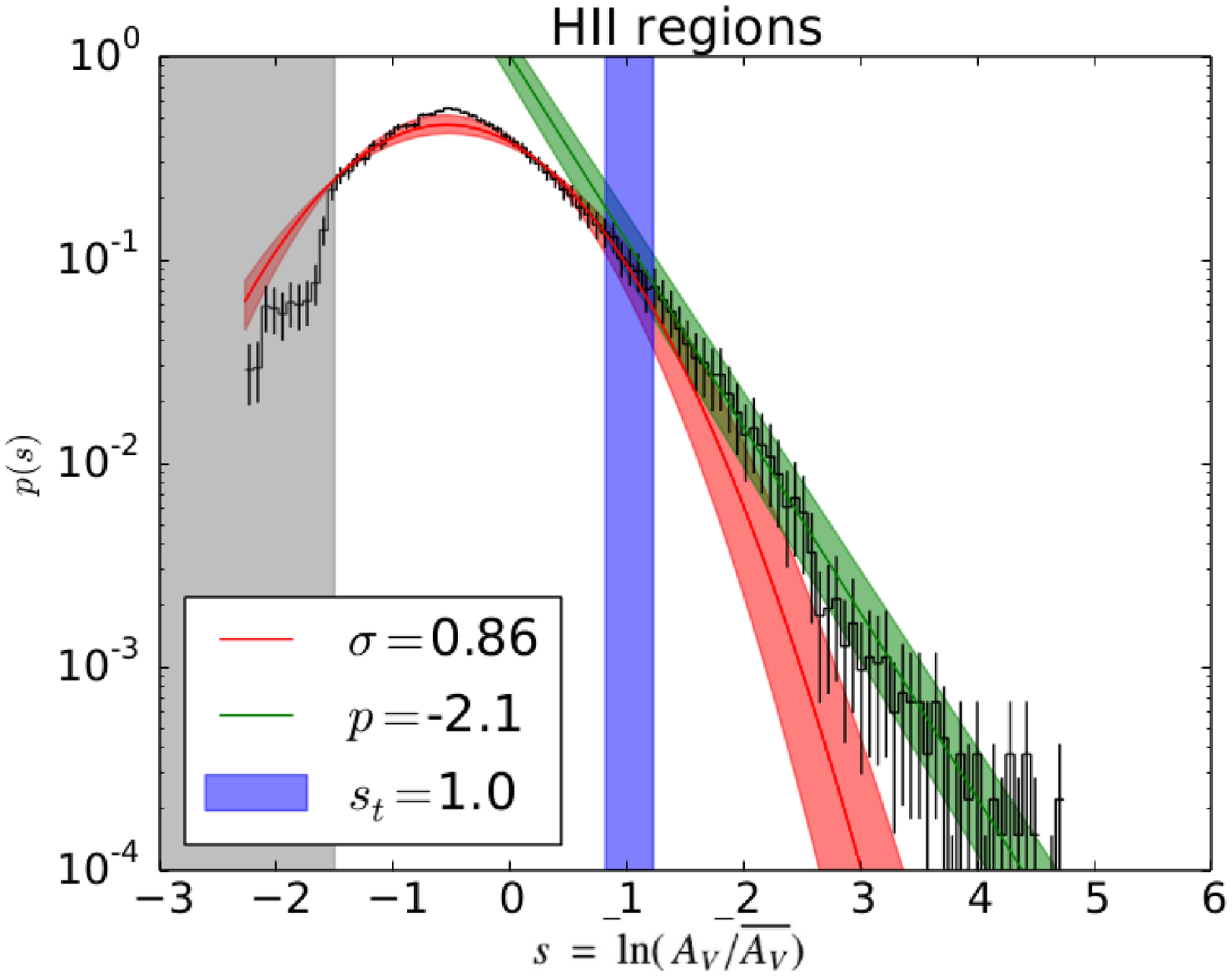}
\includegraphics[width=0.33\textwidth]{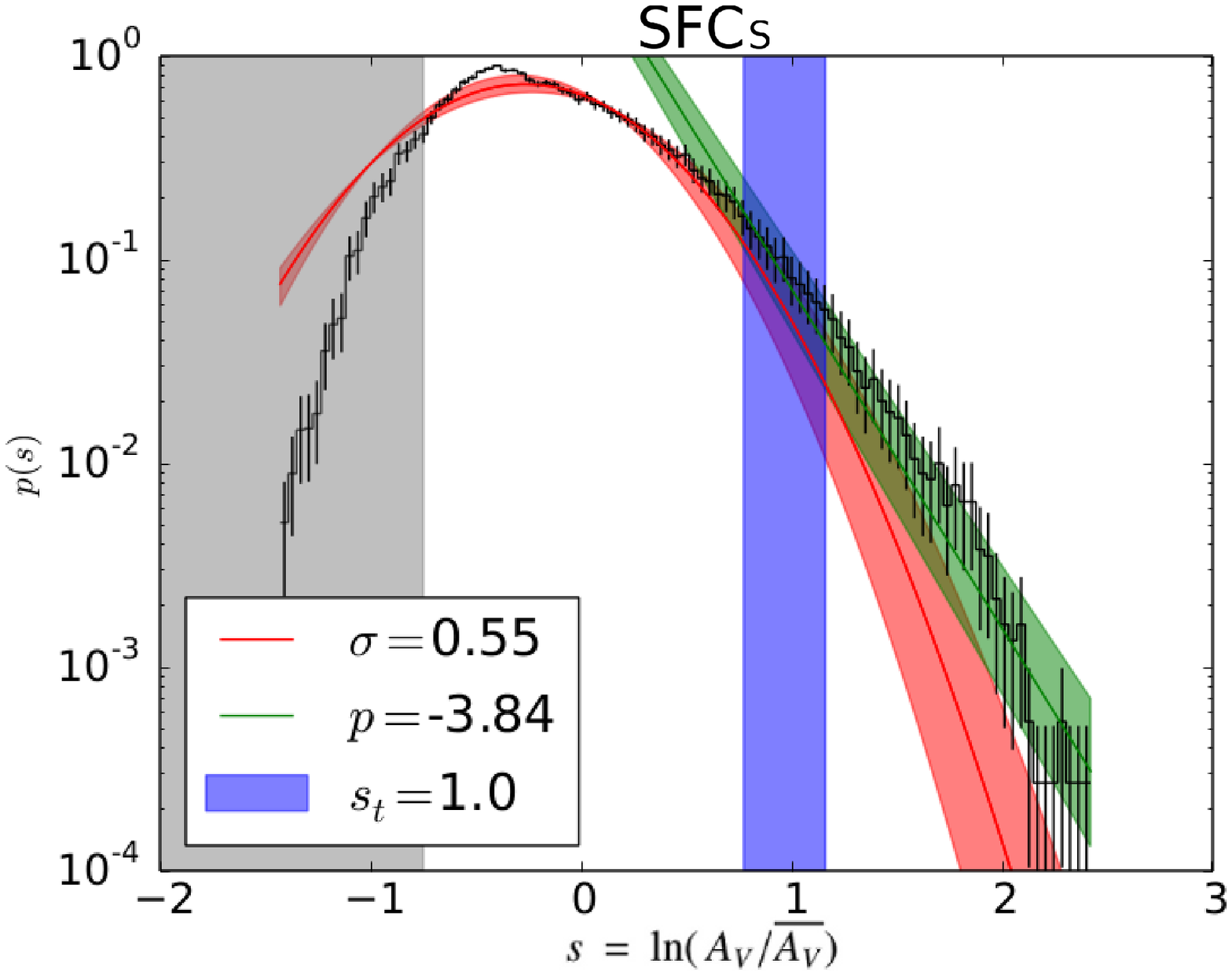}
\includegraphics[width=0.33\textwidth]{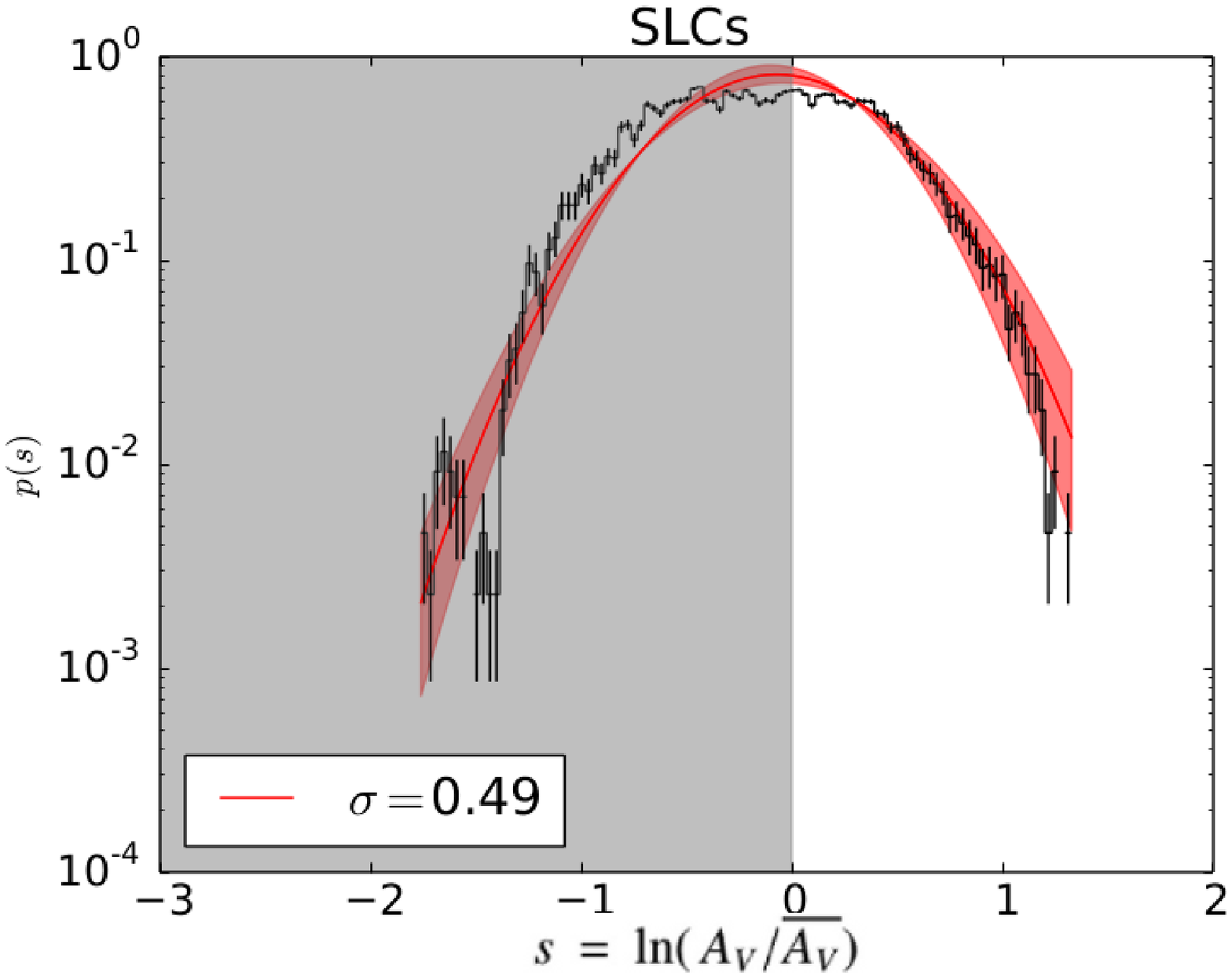}
\caption{Total mean normalized column density PDFs of \HII regions (left),
SFCs (center) and SLCs (right).
All panels: horizontal axis show mean normalized column densities,
$s=\mathrm{ln}(\,A_{V}/\overline{A_{V}})$.
Vertical error bars
show Poisson standard deviation, $\sigma_{poisson} \propto \sqrt{N}$.
The best-fit curves assuming a combination of log-normal and
power-law functions are indicated, respectively, by red and green solid lines,
with the fit errors indicated as shaded regions of the same colors.
The gray shaded regions indicate data below the reliability limit.
These data were excluded from the fit. Blue shaded regions
show the range of values obtained for
the mean-normalized column density value at which \textit{N}-PDFs
deviate from a log-normal to a power-law like function, $s_{t}$.
}
\label{fig:total-pdfs}
\end{figure*}

We first analyze the total \textit{N}-PDFs of the three evolutionary classes.
To construct the \textit{N}-PDFs, we used the mean
normalized column densities $s=\mathrm{ln}\,(A_{V}/\overline{A_{V}})$ (see
Eq.~\ref{eq:log-normal}) of each region. We calculated $\overline{A_{V}}$ as the
mean column density of all the pixels of each region.
The resulting \textit{N}-PDFs were then stacked together
to form the total \textit{N}-PDFs, shown in Fig.~\ref{fig:total-pdfs}
as the black histogram. The three classes show clearly different \textit{N}-PDFs: the \HII regions
have the widest (or shallowest) \textit{N}-PDF, followed by a slightly narrower (or steeper)
\textit{N}-PDF of SFCs. The SLCs have the narrowest \textit{N}-PDF.

Interpreting the low-column density shape of the \textit{N}-PDFs requires
taking into account two issues.
First, ideally the \textit{N}-PDF should not be affected by how exactly
the field-of-view
toward an individual region is cropped, i.e., it must include all column
density values above a given level.
Second, one must ascertain that the pixels are not dominated by noise or
contamination
from neighbouring regions. To fold these two limitations into one, we
define a
''reliability limit'' of the \textit{N}-PDFs as the minimum column density
value above which all regions of the evolutionary class are well defined by a
closed emission iso-contour (see Appendix~\ref{app:TSL},~\ref{app:HII}
and~\ref{app:SFC}).
These levels are $s=-1.5, -0.75,$ and $0$ for \HII regions, SFCs, and
SLCs, respectively.
These levels correspond typically to $A_{V} = 2, 4,$ and $9$\,mag,
all at least 1-$\sigma$ above the noise level (50\,mJy; see
Section~\ref{sec:NH}).
The larger reliability limit in SLCs originates from the
fact that they are embedded in \HII regions and SFCs, being surrounded
by emission levels higher than the map noise. The total number of pixels
above these
limits are $20\times10^{4}$, $9\times10^{4}$, and $10^{4}$
for \HII regions, SFCs, and SLCs, respectively.
We note that this definition of the reliability
limit is very conservative; it is set by the lowest iso-contour above
which \emph{all} regions of the evolutionary class show a closed contour.
Most regions, however, have this limit at lower $s$ values. We also
note that systematic uncertainties such as the dust opacity uncertainty do
not affect
(or, are unlikely to affect) the relative shapes of the three classes with
respect to each others.

To quantify the shapes of the \textit{N}-PDFs shown in
Fig.~\ref{fig:total-pdfs},
we fit them, equally sampled in the log space, with
a combination of log-normal (see Eq.~\ref{eq:log-normal})
and power-law ($p(s) \propto cs^{p}$) functions.
We used five free parameters in the fit: the width ($\sigma_{s}$)
and mean ($\mu$) of the log-normal function,
the slope ($p$) and constant ($c$) of the power-law, and
the break point between both functions ($s_{t}$).
Furthermore, molecular cloud masses
should be recovered when integrating the fitted function, representing
an extra boundary condition to the fit.
The fitting range was defined as all $s$
values larger than the reliability limit. We
weighted the data points by their Poisson noise.
We obtained the uncertainties of the fitted parameters
by fitting the \textit{N}-PDFs using different bin
sizes~\citep{2014ApJ...787L..18S}.
Results are summarized in Table~\ref{tab:fit-results}.

SLCs are well described by a log-normal \textit{N}-PDF
($\sigma_{s,SLC}=0.5\pm0.1$).
Even though the peak of the \textit{N}-PDFs is below the
reliability limit, it is well constrained by the fit
because of the normalization factor in Eq.~\ref{eq:log-normal}.
The \textit{N}-PDFs of \HII regions and SFCs
are inconsistent with a single log-normal function;
they are better described by a
combination of a log-normal function at low column densities
and a power-law function at high column densities.
The low-column density log-normal portion of \HII
regions is wider ($\sigma_{s,\HII}=0.9\pm0.09$)
than that of SFCs ($\sigma_{s,SFC}=0.5\pm0.05$).
The mean-normalized column densities at which the
\textit{N}-PDFs transition from log-normal to power-law
is similar in both classes, \HII regions and SFCs: $s_{t}=1.0\pm0.2$.
We also find differences in the power-law slopes of the \textit{N}-PDFs.
The power-law slope is clearly shallower for \HII regions ($p=-2.1\pm0.1$)
than for SFCs ($p=-3.8\pm0.3$).

\subsubsection{Dense Gas Mass Fraction}\label{sec:dgmf}

\begin{figure*}
\centering
\includegraphics[width=0.33\textwidth]{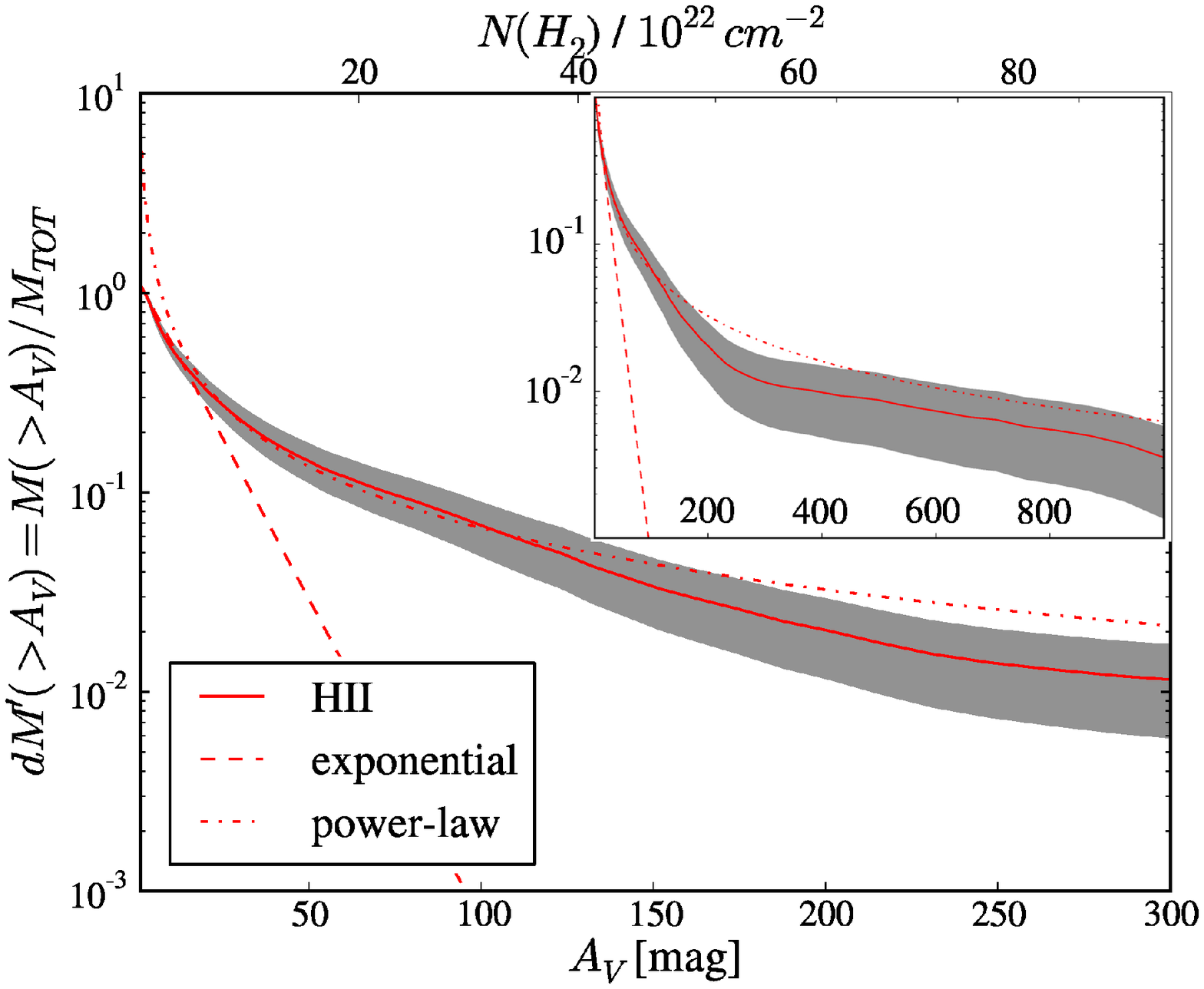}
\includegraphics[width=0.33\textwidth]{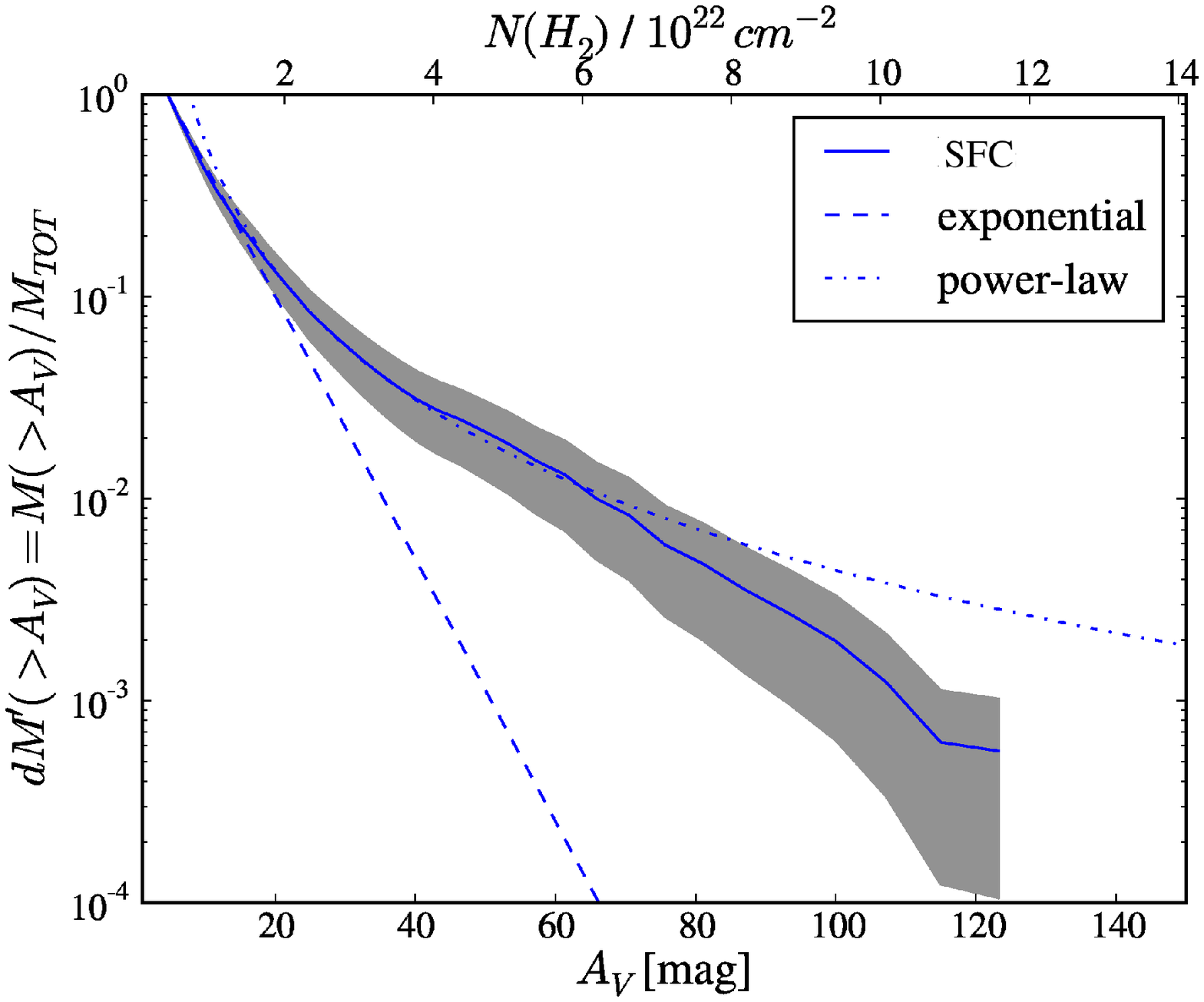}
\includegraphics[width=0.33\textwidth]{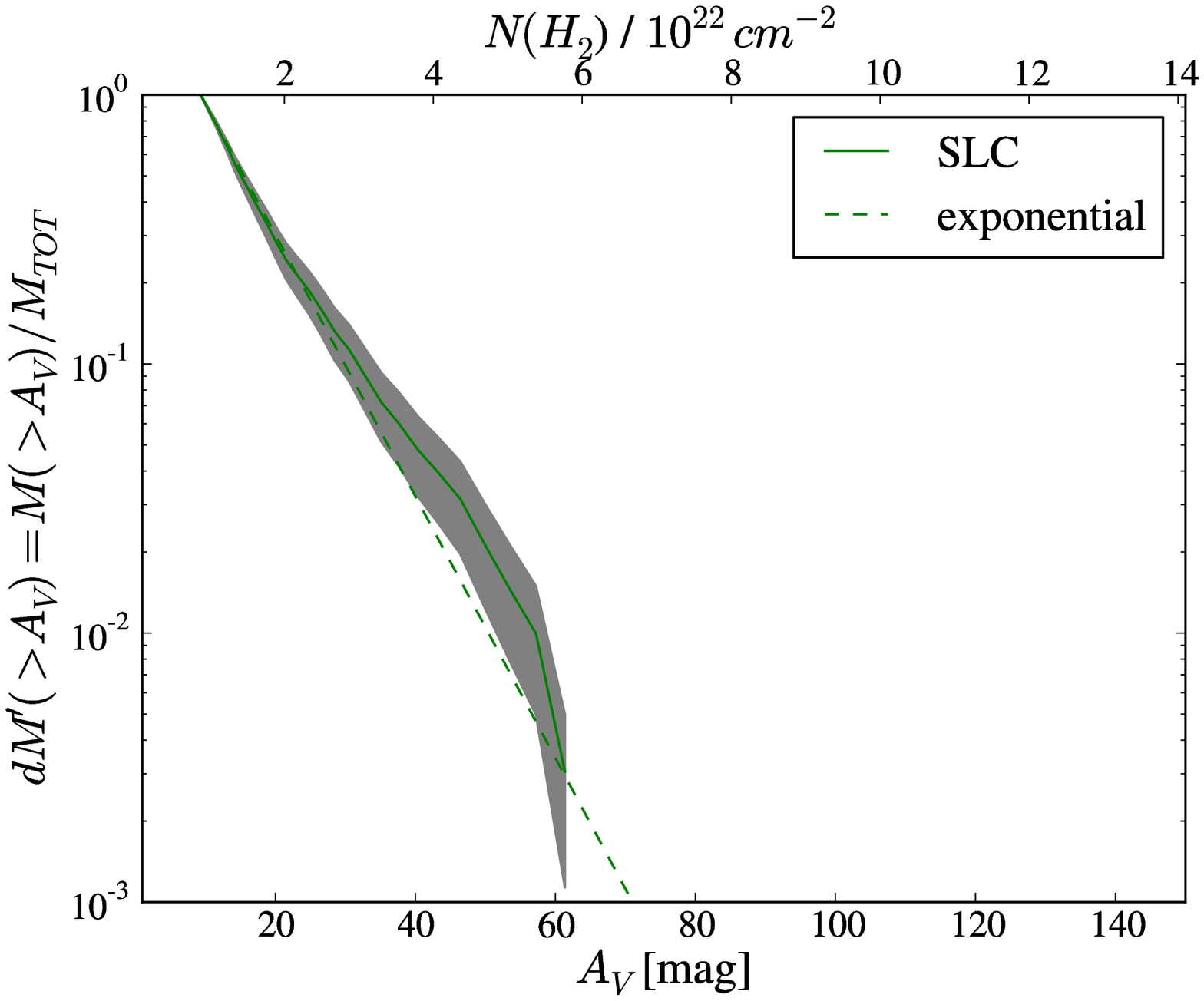}
\caption{Mean DGMFs of \HII regions (left),
SFCs (center) and SLCs (right).
Solid colored lines show mean normalized
DGMFs. DGMFs were normalized to the reliability limit
of each evolutionary class: $A_{V}=2,\,4,\,9$\,mag for
\HII regions, SFCs and SLCs, respectively.
Colored dashed lines show the
fit of the DGMFs with exponential functions. Dashed-dotted colored
lines show fits with power-law tails in the higher $A_{V}$ range.
Grey shaded regions show
statistical poisson errors of the DGMFs.
Small box in left panel shows the whole mean DGMF of \HII regions
up to $A_{V}=1000$\,mag.}
\label{fig:dgmf-mean}
\end{figure*}

In Section~\ref{intro} we defined the DGMFs as the fraction of gas mass
enclosed by regions with $M(A_{V} \geq A_{V}\arcmin)$, relative to the
total mass of the cloud (see Eq.~\ref{dgmf}).
Fig.~\ref{fig:dgmf-mean} shows the mean DGMFs of each evolutionary class.
Generally, \HII regions exhibit larger reservoirs of
high-column density gas than the SLC and SFC regions.

We quantified the shapes of the mean DGMFs by fitting
them with a combination of exponential
($\propto e^{\alpha A_{V}}$)
and power-law ($\propto A_{V}^{\beta}$) functions,
leaving both exponents and the breaking point as
free parameters and weighting each point by the
Poisson standard deviation.
Fit errors were calculated as in Section~\ref{sec:pdf}, resulting in
parameter value uncertainties of 10\%-15\%.
While the mean DGMF of the SLCs is well fitted by an exponential,
\HII regions and SFCs transition from an exponential to a power-law
shape at $A_{V}\geq20$\,mag.
This change is evidently linked to the change from log-normal to
power-law shape in the \textit{N}-PDF because the DGMFs are an integral of
the \textit{N}-PDF.
\HII regions show the shallowest mean DGMF ($\alpha = -0.06$), followed by
SLCs ($\alpha =
-0.11$) and SFCs ($\alpha = -0.14$).
In the power-law portion of the DGMFs, \HII regions
are also shallower ($\beta=-1.0$) than SFCs ($\beta=-2.1$).
The amount of mass enclosed by the power-law DGMF is 30\% of the
total mass in \HII regions, almost a factor of three lower, 10\%, 
for the SFCs.

The mean DGMF of \HII regions above $A_{V}=300$\,mag is
dominated by regions \#4, \#55 and \#122 (see Table~\ref{tab:regions}).
This flat tail is built up by less than 1\% of the pixels in
each of the mentioned regions, and hence, is not representative
of the whole \HII sample.

\begin{table}
\caption{Results of the best-fit parameters to the total \textit{N}-PDFs
and DGMFs.}
\centering 
\begin{tabular}{c | c c c | c c } 
\hline\hline 
               &  & \textit{N}-PDFs &  &  & DGMFs \\
\hline 
         & $\sigma_{s}$\tablefootmark{a} & $p$\tablefootmark{b}           
& $s_{t}$\tablefootmark{c}              &
$\alpha$\tablefootmark{d}  & $\beta$\tablefootmark{e} \\\hline
  \HII   & 0.9$\pm$0.09                   & -2.1$\pm$0.1    & 1.0$\pm$0.2 
         & -0.06\tablefoottext{f}     & -1.0   \\
  SFCs   & 0.5$\pm$0.05                   & -3.8$\pm$0.3    & 1.0$\pm$0.2 
         & -0.14     & -2.1   \\
  SLCs   & 0.5$\pm$0.1                    &  ---            & ---         
         & -0.11     & ---
\label{tab:fit-results}
\end{tabular}
\tablefoot{
\tablefoottext{a}{Standard deviation of the log-normal portion of the
$N$-PDFs.}
\tablefoottext{b}{Slope of the power-law portion of the $N$-PDFs.}
\tablefoottext{c}{Transition from log-normal to power-law portion of the
\textit{N}-PDFs in mean-normalized column densities.}
\tablefoottext{d}{Slope of the exponential portion of the DGMFs.}
\tablefoottext{e}{Slope of the power-law portion of the DGMFs.}
\tablefoottext{f}{Relative errors of DGMFs account for 10\%.}
}
\end{table}

\subsubsection{Relationship between the region's mass and column density
distribution}\label{sec:massDgmf}

Does the dense gas mass fraction of a region depend on its mass
and from therein, affect the SFR - cloud mass relation
presented by~\citet{2012ApJ...745..190L}?
We analyze the DGMFs of each evolutionary 
class divided in five mass intervals (listed in Table~\ref{tab:dgmf-mass}) 
that have at least 9 regions each to answer this question.

\begin{table}
\caption{Mass intervals of each evolutionary class in
Fig.~\ref{fig:dgmf-mass} }%
\centering 
\begin{tabular}{c c c c} 
\hline\hline 
[$\mathrm{M_{\sun}}$]&  \HII & SFCs & SLC \\
\hline 
$M_{i.\,1}$ & ---		     & $<10^{3}$            &$<10^{3}$         \\
$M_{i.\,2}$ & $1-2\times10^{3}$      & $1-2\times10^{3}$    &$1-2\times10^{3}$ \\
$M_{i.\,3}$ & $2-5\times10^{3}$      & $2-5\times10^{3}$    &$2-5\times10^{3}$ \\
$M_{i.\,4}$ & $0.5-1\times10^{4}$    & $0.5-1\times10^{4}$  & ---              \\
$M_{i.\,5}$ & $>10^{4}$              & ---		    & ---              \\

\hline
\label{tab:dgmf-mass}
\end{tabular}
\end{table}


\begin{figure*}
\centering
\includegraphics[width=0.33\textwidth]{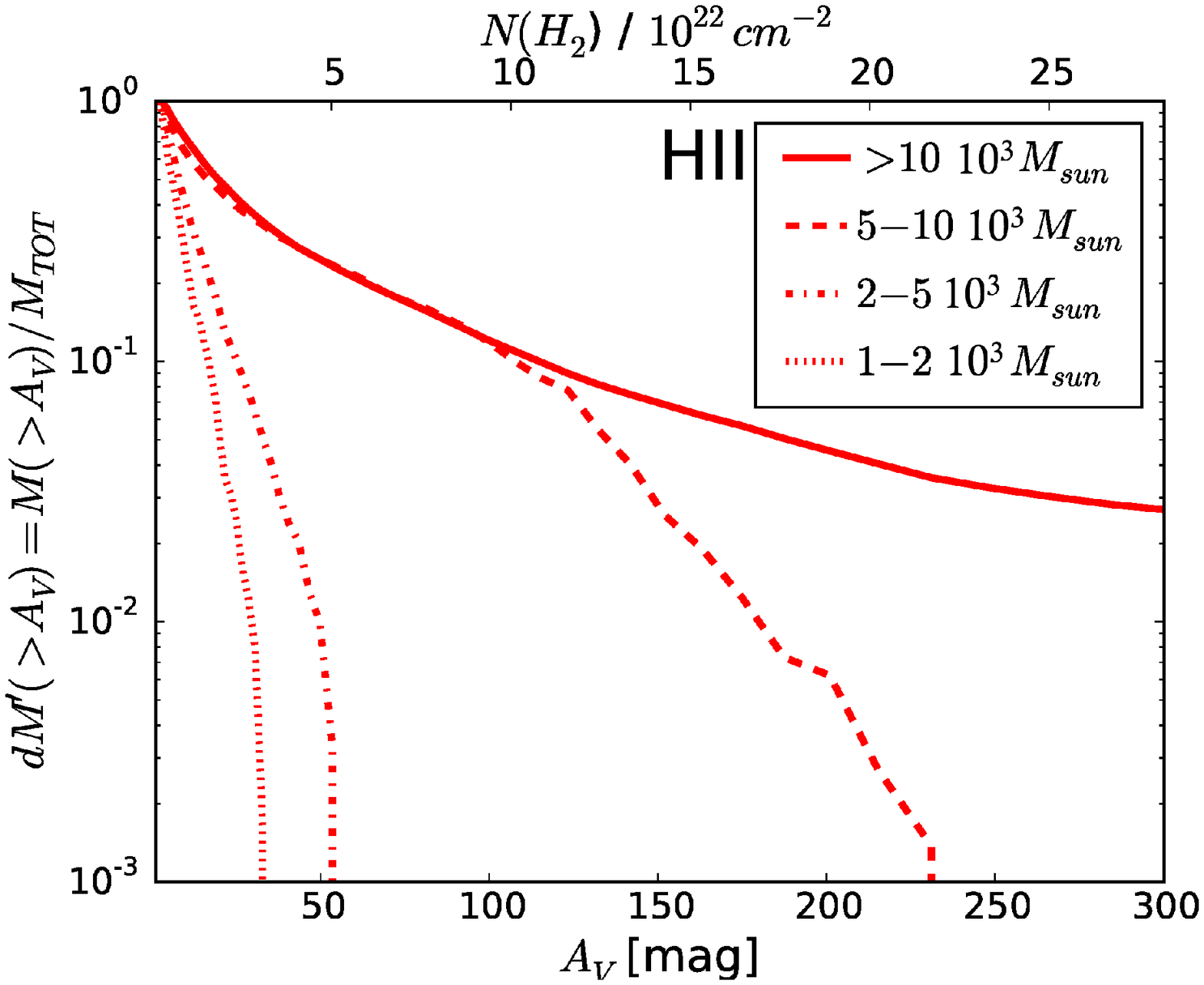}
\includegraphics[width=0.33\textwidth]{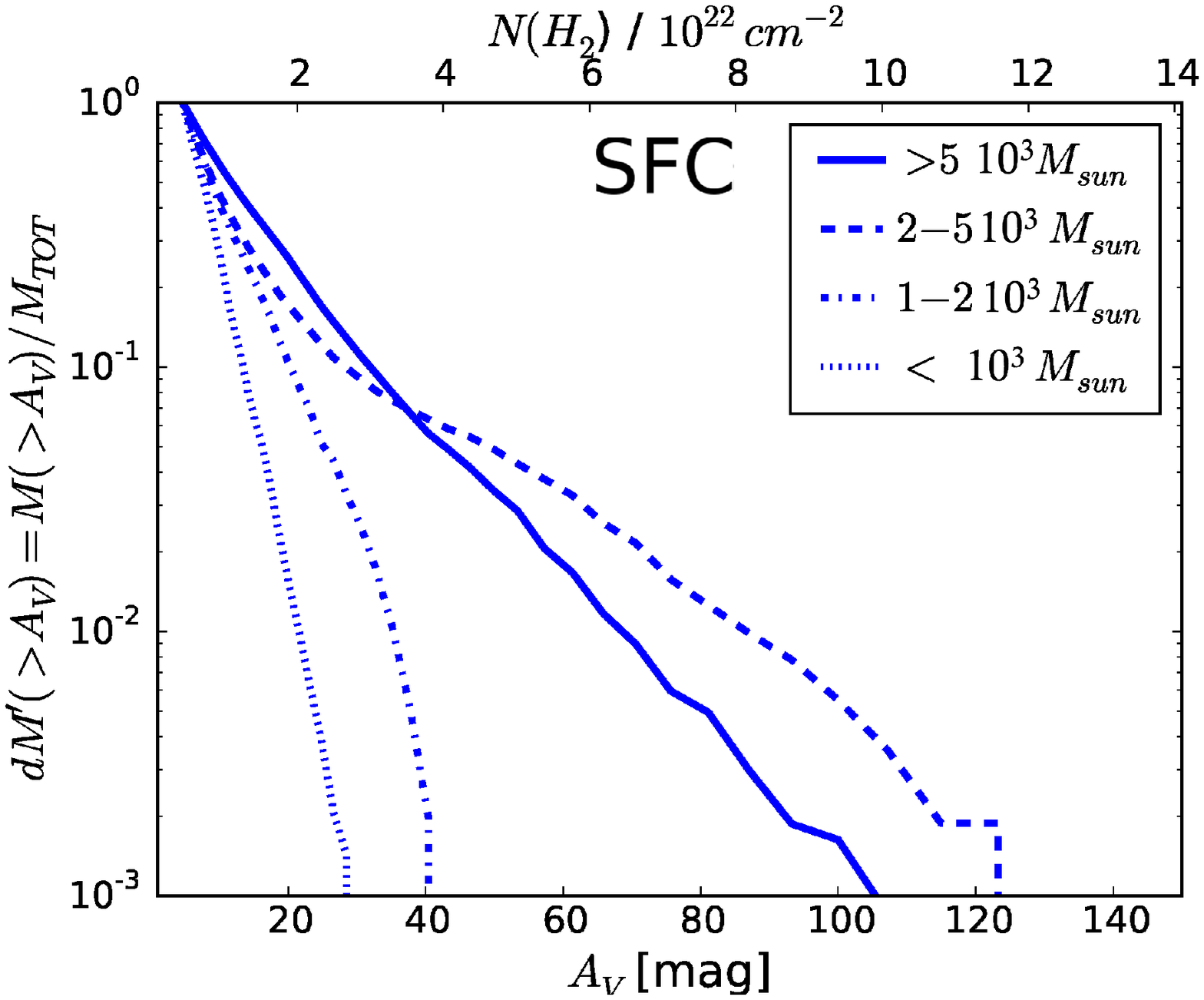}
\includegraphics[width=0.33\textwidth]{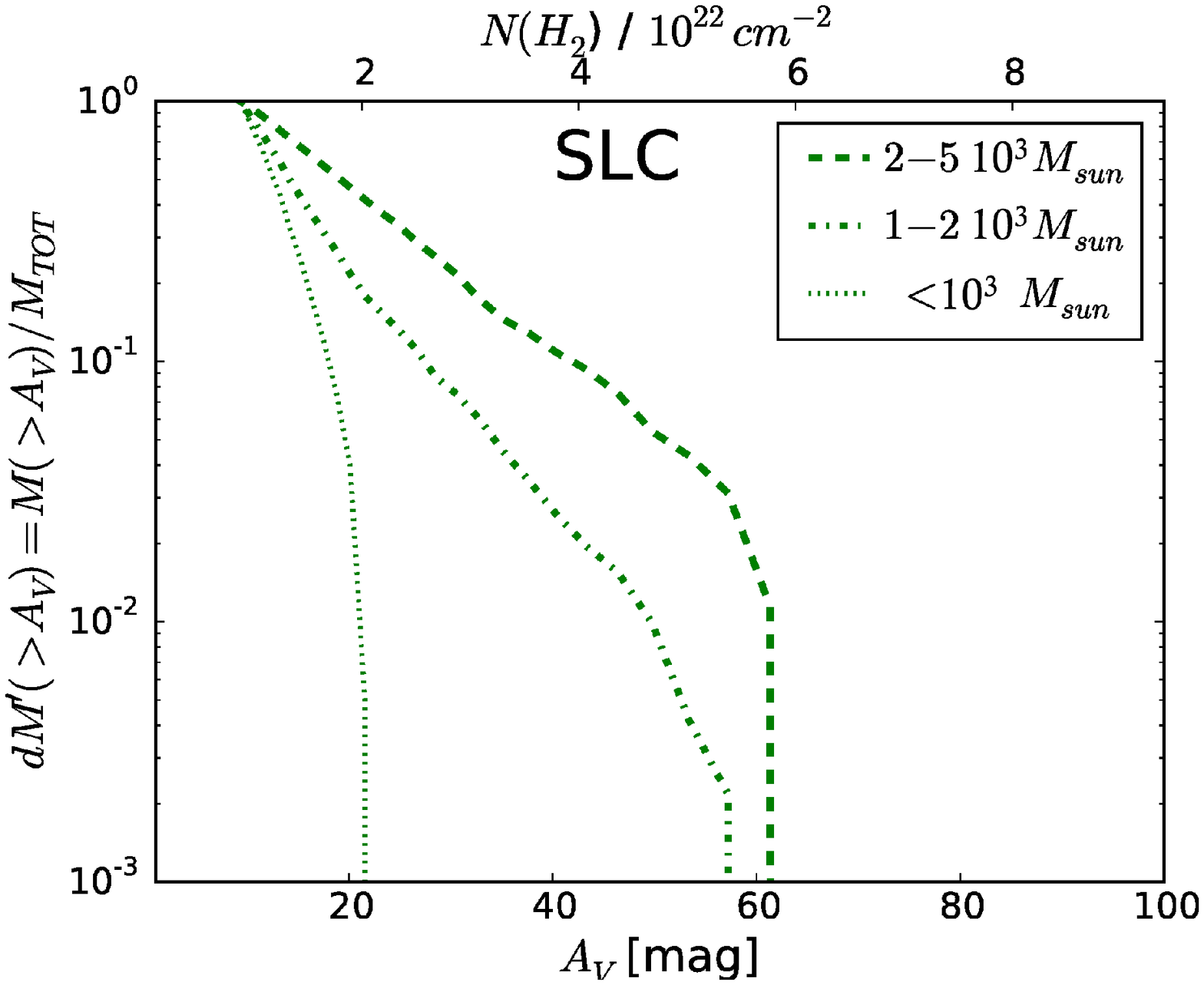}
\caption{Mass-binned average DGMFs for each evolutionary class.  Each line
shows the DGMF for each of the mass intervals listed in the
corresponding panel and defined in Table~\ref{tab:dgmf-mass}.  Dotted lines,
dotted-dashed lines, dashed lines, and solid lines progress from less
to most massive bins, respectively. The DGMFs were normalized
following the procedure described in Section~\ref{sec:dgmf} and shown in
Fig.~\ref{fig:dgmf-mean}.
}
\label{fig:dgmf-mass}
\end{figure*}

Figure~\ref{fig:dgmf-mass} shows the mean DGMFs of each mass
interval for the three evolutionary classes.
In all evolutionary classes, most massive regions
have shallower DGMFs
than those of less massive regions.
We fit the mean DGMFs with exponential and
power-law functions as described in Sect.~\ref{sec:dgmf}.
Most DGMFs could not be fitted well with the combination of both
functions over their entire column density range.
Only the DGMFs of the most massive SFCs and \HII regions
required two component functions; DGMFs of less massive regions are well
described by an exponential alone. Exponents derived from
this analysis are shown in Table~\ref{tab:dgmf-mass-exp}.

\begin{figure}[h!!]
\centering
\includegraphics[scale = 0.38]{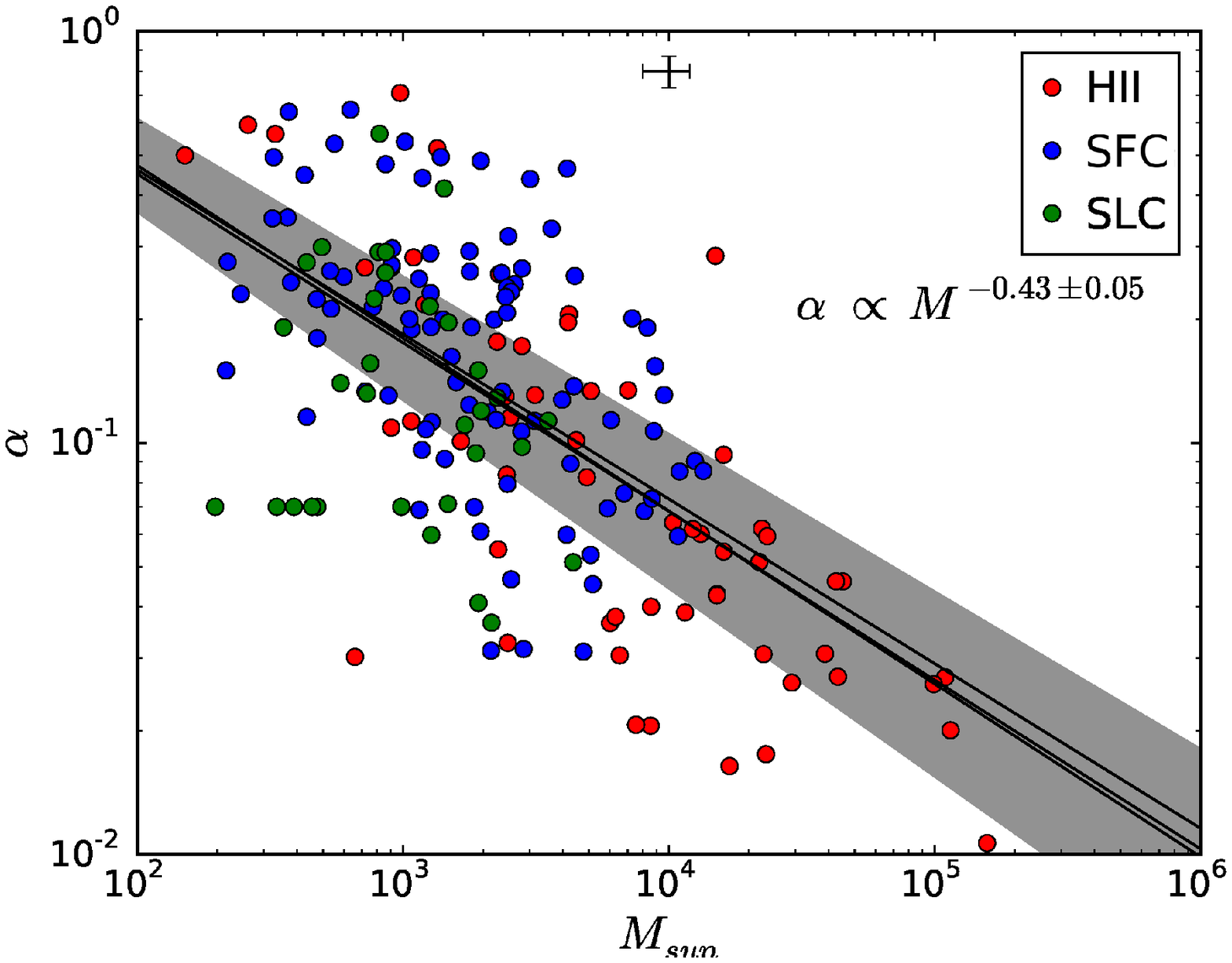}\\
\includegraphics[scale = 0.38]{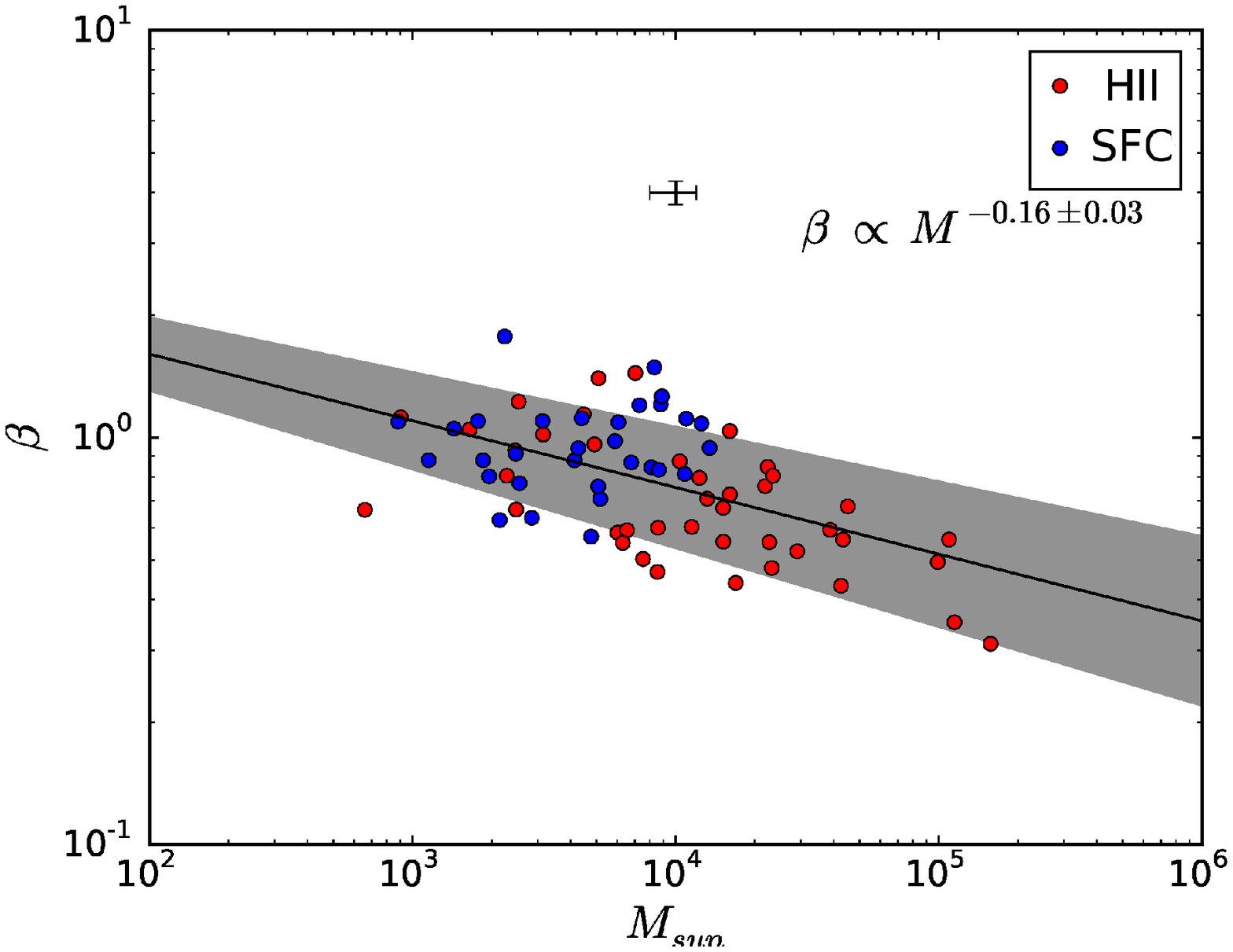}\\
\includegraphics[scale = 0.38]{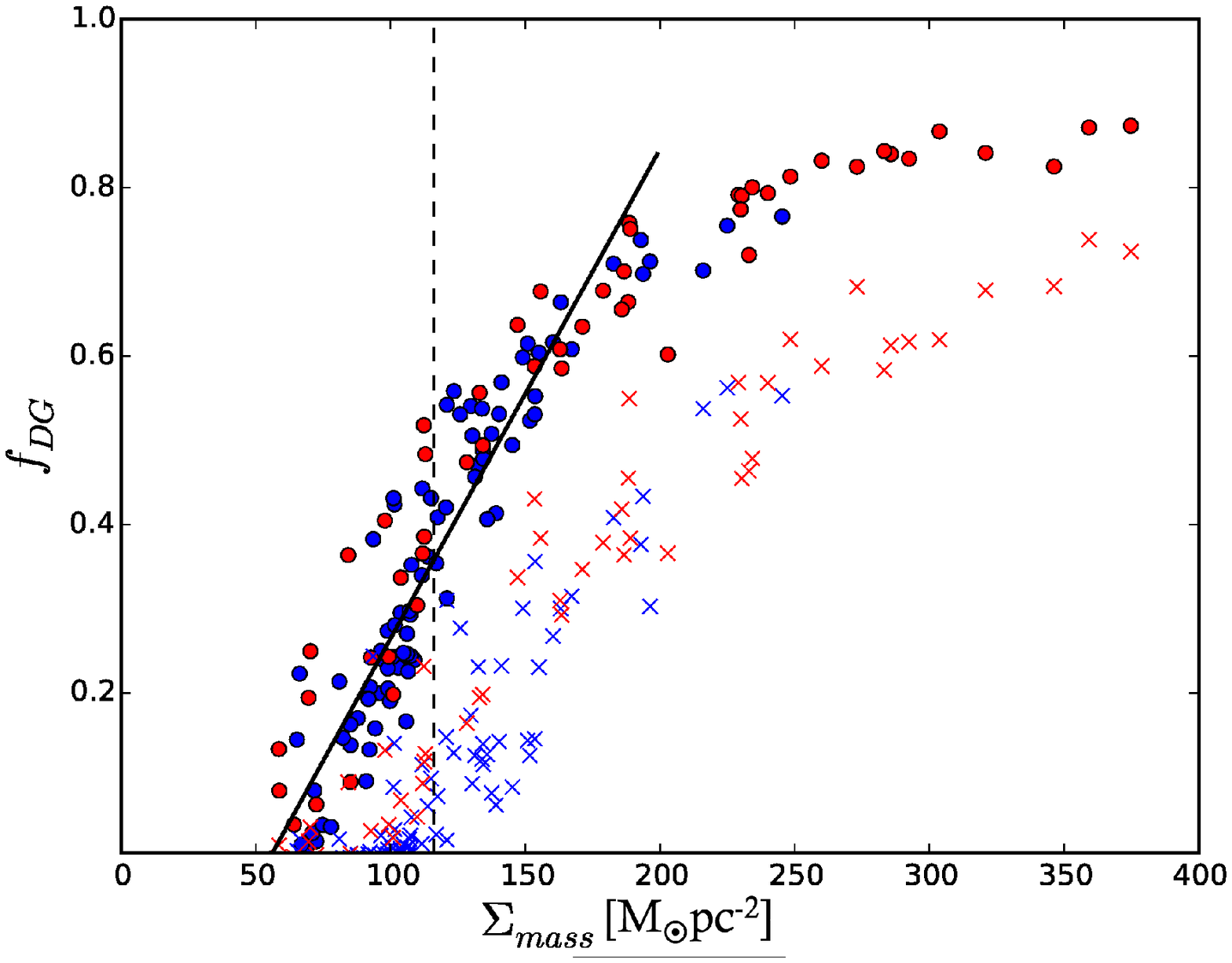}
\caption{\emph{Top:} relationship between the mass of the
regions [$M_{\sun}$] and
the exponent of the exponential fit to the DGMFs ($\alpha_{exp}$).
Black solid line shows the best fit to the data
and the shaded region shows its $\sigma$ error. The dotted and
dashed lines show the best fit when the  most massive 
\HII regions are removed. Colors indicate the
evolutionary class of each point as indicated. \emph{Middle:} relationship
between
mass of the regions and the
slope of the power-law range of the DGMFs.
Black line shows the best fit to the data
and the shaded region shows its $\sigma$ error.
\emph{Bottom: }relationship between
the mean gas mass surface density of the MCs, $\Sigma_{\mathrm{mass}}$,
and the dense gas mass fraction of gas,
$f_{\mathrm{DG}}=\frac{M(A_{V}>7.0\,\mathrm{mag})}{M_{\mathrm{tot}}}$.
The crosses show the $f_{\mathrm{DG}}$ obtained 
integrating the exponential regime of the DGMFs of each region
in the range $A_{V}=0-7$\,mag (see third paragraph in Sect.~\ref{sec:4.2}). 
Black line shows a linear fit to the data in the range
$\Sigma_{\mathrm{mass}}=50-200\,\mathrm{M_{\sun}pc^{-2}}$.
Vertical dashed line at
$\Sigma_{\mathrm{mass}}=116\,\mathrm{M_{\sun}pc^{-2}}$~\citep{2010ApJ...724..687L,2012ApJ...745..190L}
indicates the threshold for the dense gas.}
\label{fig:mass-dgmf-fits}
\end{figure}

In all evolutionary classes, the exponent of the
exponential function, $\alpha$, increases
with mass (see Table~\ref{tab:dgmf-mass-exp}).
In order to further investigate this correlation, we repeated the
same fitting procedure for each individual region. Results are shown in
Fig.~\ref{fig:mass-dgmf-fits}.
Top panel of Fig.~\ref{fig:mass-dgmf-fits} shows
the relationship between $\alpha$ and 
the mass of each region: $\alpha\propto M^{-0.43\pm0.05}$,
that has a correlation coefficient $r=0.64$ and
a significance value $p=0.18$. 
The fit parameters and their errors were obtained  
from a Monte-Carlo simulation of 10$^{6}$ cycles. On each cycle
we selected a random sample of points and fitted the resulting data set.
We adopt the average of the best fit parameters obtained on each cycle
as the best fit values and their standard deviation as the
error of the fit.
It could be argued that the correlation between $\alpha$ and 
mass is dominated by the most massive ($M>3\times10^{4}M_{\sun}$) \HII regions. 
To establish whether the correlation strength depends 
strongly on these few massive clouds we also explored 
this correlation without those extreme points.
The correlation coefficient is somewhat lower in this case $r=0.56$
and a significant value $p=0.21$.
However, there is no significant difference in the 
resulting fit ($\alpha\propto M^{-0.40\pm0.08}$). 
Power-law exponents of DGMFs also exhibit
a correlation with mass: $\beta\propto M^{-0.16\pm0.03}$ (see middle panel
of Fig.~\ref{fig:mass-dgmf-fits}).
The larger scatter seen in the data from the exponential fits
relative to that seen in power-law fits may indicate
that the power-law regimes of DGMFs are much better constrained
than the exponential regimes.

\begin{table}
\caption{Slopes of the exponential and power-law fits to DGMFs, $\alpha$,
$\beta$
for the mass ranges presented in Table~\ref{tab:dgmf-mass}.}
\centering 
\begin{tabular}{c c c c c c} 
\hline\hline 
& $M_{i.\,1}$ & $M_{i.\,2}$ & $M_{i.\,3}$ & $M_{i.\,4}$ & $M_{i.\,4}$\\
\hline 
$\alpha$(\HII) & ---   & -0.25 & -0.22 & -0.06 & -0.04  \\
$\alpha$(SFCs) & -0.29 & -0.20 & -0.18 & -0.10 & ---    \\
$\alpha$(SLC)  & -0.32 & -0.19 & -0.09 & ---   & ---    \\
$\beta$(\HII)  & ---   & ---   & ---   & -0.59 & -0.61  \\
$\beta$(SFCs)  & ---   & ---   & -1.03 & ---   & ---    \\
\hline
\label{tab:dgmf-mass-exp}
\end{tabular}
\end{table}

\section{Discussion}\label{disc}

\subsection{\textit{N}-PDFs as a measure of the evolutionary stage of objects.}\label{sec:pdf-evol}

The total \textit{N}-PDFs of different evolutionary classes exhibit clear differences; 
these differences can be linked to differences in the 
mechanisms that drive the evolution of objects within the various classes.
The \textit{N}-PDF of SLCs is well described by a single log-normal function
(see Fig. ~\ref{fig:total-pdfs}). This agrees with previous observations of
starless low-mass clouds~\citep{2009A&A...508L..35K} or starless regions of
star-forming clouds~\citep{2012A&A...540L..11S,2013ApJ...766L..17S,2013A&A...554A..42R}.
In particular, this simple log-normal form agrees with predictions for
turbulence-dominated media from numerical
simulations~\citep{1997ApJ...474..730P,2001ApJ...557..727V}.

In contrast, the total \textit{N}-PDFs of star-forming clouds, i.e.,
\HII regions and SFCs, show two components that can be described by
log-normal and power-law functions. 
The  power-law components of the \textit{N}-PDFs  of the \HII regions
are shallower than those of the SFCs. Previous studies have found
that the power-law slopes are within $p=[-1.5, -3.3]$, with shallower
slopes related to most active star-forming regions.
In the only study with a resolution similar to
ours,~\citet{2013A&A...554A..42R}
 found a non-star-forming region in NGC\,6334 to have a steep \textit{N}-PDF
slope\footnote{~\citet{2013A&A...554A..42R} quote the equivalent radial
density profile ($\kappa$), which can be related with the slope of
the power-law tail of the \textit{N}-PDF via $p = -2/(\kappa-1)$.} 
($p=-5.7$), moderately star-forming regions to have shallower slopes
($p=-3.3, -3.0$), and an \HII region to have the shallowest slope ($p=-1.5$).
Their trend to have shallower \textit{N}-PDF in a cloud region that 
contains an \HII region is similar to what we find in our work.

Theories and simulations that consider turbulent gas under the influence
of gravity predict power-law-like tails for \textit{N}-PDFs with exponents
comparable to what is observed~\citep{2011ApJ...727L..20K,2013ApJ...763...51F},
possibly featuring flattening of the power-law over time-scales relevant for
star formation~\citep{2011MNRAS.416.1436B,2013ApJ...763...51F,2014ApJ...781...91G}.
~\citet{2011ApJ...727L..20K} showed that a collapsing spherical
cloud with a power-law density distribution, $\rho \propto r^{-\kappa}$,
will have a power-law \textit{N}-PDF with a slope of $p = -2/(\kappa-1)$.
The power-law slopes that we observe (see Table~\ref{tab:fit-results})
indicate $\kappa=1.9$ and 1.5 for the \HII regions and SFCs, respectively.
The former is very close to the value $\kappa=2$ of a collapsing isothermal
sphere~\citep{1987ARA&A..25...23S}, suggesting that the density
distribution of \HII regions may be dominated by self-gravity.
The value of $\kappa=1.5$ we find for SFCs can also be
indicative of a collapse slowed down by turbulence-driving
effects~\citep{2014ApJ...781...91G}.
We note a caveat in this analysis. Our SFCs and \HII regions are unlikely
to be close to spheres and their large sizes make them unlikely to be
under general free-fall collapse. 
However, these regions are composed of numerous smaller ATLASGAL clumps (see
Fig.~\ref{fig:galPlane}) that may
be closer to spherical symmetry, and we are averaging the emission of all
these smaller clumps. Indeed, the density profile exponent of our SFCs is
similar to that found by~\citet{2002ApJ...566..945B} in a sample of small high mass
star-forming objects, which correspond to our definition of SFCs.

Recent works based on \emph{Herschel} observations have
explored possible effects of other processes (e.g. ionising radiation or shock
compression) on the \textit{N}-PDFs of \HII regions~\citep{2012A&A...540L..11S,2013ApJ...766L..17S,2014A&A...564A.106T}.
~\citet{2014A&A...564A.106T} report \textit{N}-PDFs with two log-normal components. They relate 
the log-normal component at low column-densities to the
turbulent motions of the gas and
the component at high-column densities to ionization pressure.
They also suggest that the presence of these double-peaked \textit{N}-PDFs
depends on the relative importance of ionizing and turbulent pressures. 
The total \textit{N}-PDFs of our \HII regions, composed of 60
individual regions, does not exhibit
such behavior. This could originate from a combination 
of several factors: \textit{i)} the low-column density component detected by
~\citet{2014A&A...564A.106T} is at column densities of $A_{v}\lesssim6$\, mag. 
These column densities are generally filtered out from 
the ATLASGAL data; \textit{ii)} the size-scales of the molecular clouds
studied in~\citet{2014A&A...564A.106T} and this work
are different and it may happen that the ionisation front
of the \HII regions is not spatially resolved in our observations.

%
%
%

The above models offer an attractive possibility to link the observed
\textit{N}-PDF tails to self-gravitating gas in molecular clouds. However, it has not yet
been shown observationally that the power-law parts would be definitely caused by
self-gravity; an alternative interpretation has been proposed
by~\citet{2011A&A...536A..48K} who suggested that the overall pressure conditions in the clouds may play
a role in producing the observed power-law-like behavior in low-mass molecular
clouds.

\subsection{Dense Gas Mass Fraction in molecular clouds}\label{sec:4.2}

With our cloud sample, we are able to study the DGMFs
of molecular clouds over a relatively wide dynamic range of 
column densities and separately in various evolutionary classes.
The continuous DGMF functions (Eq.~\ref{dgmf}) allow for a more complete
census of the dense gas in the
clouds than the analysis of the ratios of two tracers, e.g., of
CO emission and dust emission.
We find that the DGMFs of \HII regions are shallower
than those of SFCs and SLCs.
This suggests a direct relation between
the star-forming activity of molecular clouds
and their relative dense gas mass fraction.
Similar results have also
been previously found in 
nearby regions~\citep{2009ApJ...703...52L,2010ApJ...724..687L,2009A&A...508L..35K,2013A&A...549A..53K}
and filmentary clouds~\citep{2010A&A...518L.102A}.

We detect a clear correlation between the DGMF slope
and cloud mass (see Fig.~\ref{fig:dgmf-mass}).
Previously,~\citet{2014ApJ...780..173B}
found no correlation between molecular cloud mass and the dense
gas fraction in a large sample of molecular clouds.
They defined the dense gas fraction as the ratio of dust
emission-derived mass, traced with 
1\,mm flux, above $A_{V}=9.5$\,mag to
CO-derived mass: $f_{\mathrm{DG}}=M_{\mathrm{dust}}/M^{\mathrm{CO}}_{GMC}$.
Their result imply that there is no correlation between the
mass of CO-traced gas ($A_{V}\sim3-8$\,mag) and
the mass of gas at column densities $A_{V}>9$\,mag.
Unfortunately, we do not measure the CO mass
of our MCs and therefore we cannot directly compare
our results with those of~\citet{2014ApJ...780..173B}.
The correlation we find between the molecular cloud masses
and the slope of DGMFs suggests
that the dense gas fraction depends on the mass of moderately dense gas
($A_{V}\gtrsim10$\,mag) rather than the CO mass of the clouds.

~\citet{2010ApJ...724..687L,2012ApJ...745..190L} suggested that
star formation rates depend linearly on the amount of dense gas in
molecular clouds: $\Sigma_{\mathrm{SFR}}\propto f_{\mathrm{DG}}\Sigma_{\mathrm{mass}}$,
with $f_\mathrm{DG} = M(A_\mathrm{V} > 7 \ \mathrm{mag}) / M_\mathrm{tot}$.
Combining this relation with ~\citet{2011ApJ...739...84G},
who derived the relation $\Sigma_{\mathrm{SFR}}\propto\Sigma^{2}_{\mathrm{mass}}$,
suggests $f_{\mathrm{DG}}\propto\Sigma_{\mathrm{mass}}$.
We find that this correlation indeed exists in the
range $\Sigma_{\mathrm{mass}}=50-200\,\mathrm{M_{\sun}pc^{-2}}$
(see Fig.~\ref{fig:mass-dgmf-fits}).
At higher surface densities the relationship flattens at $f_{\mathrm{DG}}\cong0.8$,
suggesting that the maximum amount of dense gas that a MC can harbor is
around 80\% of its total mass. Consequently, the maximum $\Sigma_{\mathrm{SFR}}$ of a molecular
cloud is reached at $f_{\mathrm{DG}}\cong0.8$. This value depends on the definition
of the column density threshold ($A_{V}^{\mathrm{th}}$) of the dense gas
becoming lower for higher values of $A_{V}^{\mathrm{th}}$.
The spatial filtering of ATLASGAL data (see Appendix~\ref{sec:herComp})
results in overestimated $f_{\mathrm{DG}}$ values. We therefore propose
$f_{\mathrm{DG}}\cong0.8$ as an upper limit to the actual maximum $f_{\mathrm{DG}}$ of a MC.
The overestimation of the $f_{\mathrm{DG}}$ values derived
above can be studied using DGMFs. 
In general, DGMFs have been shown to follow an exponential function,
$\propto e^{\alpha A_{V}}$, down to low column densities~\citep{2009A&A...508L..35K,2013A&A...549A..53K}.
We adopted the $\alpha$ values calculated in Section~\ref{sec:massDgmf}
and integrated the exponential DGMF in the range
$A_{V}=0-7$\,mag to obtain an estimate of $f_{\mathrm{DG}}$.
The result is shown with crosses in bottom panel of Fig~\ref{fig:mass-dgmf-fits}.
The mean overestimation of $f_{\mathrm{DG}}$ in SFCs and \HII regions
is 2 and $\sim$1.3 respectively. 
We did not include the SLCs in this experiment
because their reliability limit is $A_{V}=9$\,mag
and therefore they have $f_{\mathrm{DG}}=1$ (i.e.
all its mass is enclosed in regions $A_{V}>7$\,mag).

\begin{figure}[h!!]
\centering
\includegraphics[width = 0.5\textwidth]{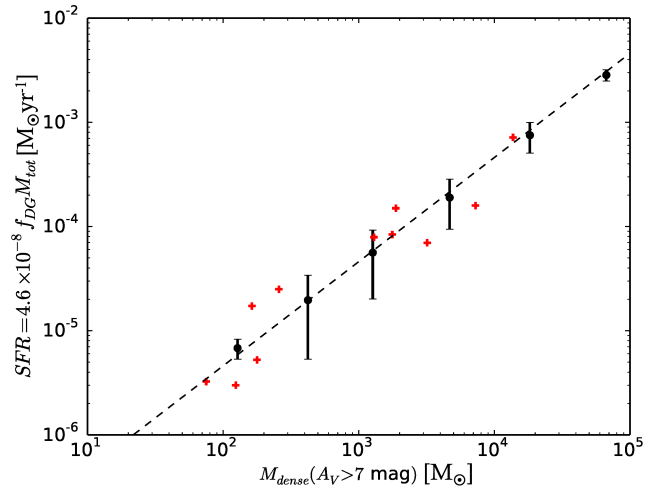}
\caption{SFR as defined in~\citet{2012ApJ...745..190L}
for different mass ranges of SFCs and \HII regions.
Red crosses show data from~\citet{2012ApJ...745..190L}.
Solid black vertical lines show the standard deviation, $\sigma$, for
each mass bin for our study. Black dashed line shows the constant value $f_{\mathrm{DG}}=1$.}
\label{fig:fDGmass}
\end{figure}

Our data can also help to understand the $SFR$ - dense gas mass relation
suggested by~\citet{2012ApJ...745..190L}. The $SFR$ - dense gas mass relation
shows significant scatter of star formation rates for a given dense gas mass,
about 0.6\,dex (see Fig. 2~\citet{2012ApJ...745..190L} and
Fig.~\ref{fig:fDGmass}).
This scatter shows that not all clouds with the same amount of dense gas
form stars with the same rate. To gain insight into this, we calculated
the dense gas mass fractions and star formation rates for our regions as
defined by~\citet{2012ApJ...745..190L},
i.e., $SFR = 4.6 \times 10^{-8} f_\mathrm{DG} M_\mathrm{tot}$\, M$_{sun}$yr$^{-1}$.
Figure~\ref{fig:fDGmass} shows the $SFR$ - dense gas mass relation with
data points
from~\citet{2010ApJ...724..687L}. The figure also shows the mean SFR of
our regions in six mass bins, with error bars showing the relative standard
deviation of $f_\mathrm{DG}$. The standard deviations are also listed in
Table~\ref{tab:scatter}.
The relative standard deviation of $f_\mathrm{DG}$ over the entire mass range
of our regions is 0.71, which is slightly higher than the relative scatter of
SFR in~\citet{2012ApJ...745..190L}, $f_\mathrm{DG}=0.56$. We conclude that 
the scatter in star formation rates for a given dense gas mass can
originate
from differences in dense gas fractions, i.e., in the total masses of
clouds for a given
dense gas mass. This, in turn, suggests that the dense gas mass is not the
only ingredient
affecting the star formation rate, but the lower-density envelope of the cloud
also plays
a significant role. However, we note the caveat that ATLASGAL filters out
low-column
densities , which may make the dense gas fractions we derive not comparable
with those in~\citet{2012ApJ...745..190L}, derived using dust extinction data.

\begin{table}
\caption{Statistics of $f_{\mathrm{DG}}$ in this work and
in~\citep{2012ApJ...745..190L}.}
\centering 
\begin{tabular}{c c c c} 
\hline\hline 
$M_{\mathrm{tot}}$ & $\overline{f_{\mathrm{DG}}}$ & $\sigma/(\overline{f_{\mathrm{DG}}})$ & \# of
regions \\
\hline 
 This work & & & \\\hline
 $<0.8\times10^{3}$             & $0.24$ & $0.22$ & 13 \\
 $0.8-2.2\times10^{3}$          & $0.29$ & $0.73$ & 42 \\
 $2.2-6.0\times10^{3}$          & $0.31$ & $0.64$ & 39 \\
 $6.0-17\times10^{3}$           & $0.41$ & $0.50$ & 27 \\
 $17-46\times10^{3}$            & $0.58$ & $0.33$ & 10 \\
 $>46\times10^{3}$              & $0.83$ & $0.16$ & 4  \\
 \textbf{Entire range} &  \textbf{0.39 }&  \textbf{0.71} & \textbf{135}\\
 \hline
 \citep{2012ApJ...745..190L} & & & \\\hline
 $0.8-100\times10^{3}$          & $0.11$ & $0.56$ & 11\\
\hline
\label{tab:scatter}
\end{tabular}
\end{table}

\subsection{Evolutionary time-scales of the evolutionary classes as indicated by their \textit{N}-PDFs}

If \textit{N}-PDFs evolve during the lives of molecular clouds, could they give us 
information about the evolutionary timescales of the clouds in the three classes we have defined?
\citet{2014ApJ...781...91G} have developed an analytical model which
predicts the evolution of the \textit{$\rho$}-PDFs of a system in
free-fall collapse.
They estimate the relative evolution time-scale, $t_{\mathrm{E}}$,
from the free-fall time at the mean
density, $\overline{\rho}$, of the molecular cloud,
$t_{\mathrm{ff}}(\overline{\rho})$,
and the density at which the \textit{$\rho$}-PDFs begin to show a
power-law shape, $\rho_{\mathrm{tail}}$
\begin{equation}\label{giri}
t_{\mathrm{E}} = \sqrt{0.2\frac{\overline{\rho}}{\rho_{\mathrm{tail}}}}t_{\mathrm{ff}}(\overline{\rho}).
\end{equation}
The model also predicts the mass fraction
of gas in regions with $\rho > \rho_{\mathrm{tail}}$.
We denote this mass as $M_{\mathrm{dense}}$.

Since our work is based on column densities 
instead of volume densities, we need to write
Eq.~\ref{giri} in terms of column
densities. To this aim, we assume a ratio between 2D and
3D variances, $R=\sigma^{2}_{N/<N>}/\sigma^{2}_{\rho/\overline{\rho}}$.
This relation is also valid for the ratios
$\overline{\rho}/\rho_{\mathrm{tail}}$ and $\overline{A}_{V}/A_{V}^{\mathrm{tail}}=e^{-s_{t}}$,
where $A_{V}^{\mathrm{tail}}$ is the column density value at
which the \textit{N}-PDF becomes a power-law and $s_{t}$ is
the mean normalized $A_{V}^{\mathrm{tail}}$. Then, Eq.~\ref{giri} can be written as
\begin{equation}\label{giri2}
t_{\mathrm{E}} =
\sqrt{\frac{0.2}{\sqrt{R}}e^{-s_{t}}}t_{\mathrm{ff}}(\overline{A}_{V}).
\end{equation}
This equation allows us to estimate
the evolutionary time-scale of a molecular
cloud using two observable quantities, namely $\overline{A}_{V}$ and
$A_{V}^{\mathrm{tail}}$.
The factor $R$ is still not well constrained.
~\citet{2014Sci...344..183K} obtained observationally a value of
$R\sim0.4$ while~\citet{2010MNRAS.403.1507B} obtained
$R=[0.03,0.15]$ in MHD turbulence simulations without gravity. In the following we use
the observationally derived value, $R=0.4$, to estimate
the time-scales of our three evolutionary classes.
We estimate the uncertainty in the time-scales as
the relative uncertainty between $R=0.4$ and $R=0.15$,
which is roughly 30\%.

We find that our \HII and SFCs classes
have evolutionary time-scales $t_{E,\HII}=t_{\mathrm{E,SFC}}=0.4\pm0.1t_{\mathrm{ff}}$
and their relative mass of gas in regions
with $s>s_{t}$ are $M_{\mathrm{dense,\HII}}\sim30\pm0.05\%$
and $M_{\mathrm{dense,SFC}}\sim10\pm0.06\%$ where the uncertainties were obtained
from the 1-$\sigma$ uncertainties in $s_{t}$ (see Section~\ref{sec:pdf}).
Since the \textit{N}-PDF of the SLCs has no power-law tail, we calculated
an upper limit of their evolutionary time-scale by using the
largest extinction in their \textit{N}-PDFs as a lower
limit, $s_{t}>1.2$. We obtained $t_{E,SLC}<0.3\pm0.1t_{\mathrm{ff}}$.

\begin{table}
\caption{Evolutionary time-scales}
\centering 
\begin{tabular}{c c c c c} 
\hline\hline 
& $t_E$ [$\mathrm{t_{\mathrm{ff}}}$] & $t_E$ [Myr] & $M_{\mathrm{dense}}$ [\%] &
$\overline{\rho}$ [cm$^{-3}$]\\
\hline 
\HII  & $0.4\pm0.1$    & $0.7\pm0.2$              & $30^{+0.05}_{-0.06}$ 
& $0.3\times10^{3}$ \\
SFCs  & $0.4\pm0.1$    & $0.3\pm0.1$              & $10^{+0.06}_{-0.04}$ 
& $1.5\times10^{3}$ \\
SLCs  & $0.3\pm0.1$    & $\lesssim0.1\pm0.03$     & ---                  
& $4.7\times10^{3}$  \\
\hline
\label{tab:giriTab}
\end{tabular}
\end{table}

The above relative time-scales can be used to estimate absolute
time-scales if the free-fall time is known. We estimate the free-fall
time of each evolutionary class as $t_{\mathrm{ff}}=\sqrt{3\pi/32G\overline{\rho}}$.
The mean density of each class was estimated
using their mean masses and effective radii\footnote{We define the
effective radius as the radius of a
circle with the same area as the projected area of a given cloud.}
and assuming spherical symmetry.
We find that the mean evolutionary
time-scale for our \HII regions is $t_{E,\HII}=0.7\pm0.2$\,Myr,
and the time-scale of SFCs is
$t_{E,\mathrm{SFC}}=0.3\pm0.1$\,Myr.
SLCs have the shortest time-scales,
$t_{E,\mathrm{SLC}}<0.1\pm0.03$\,Myr.
We note that the absolute time-scales are measured
using the onset of the gravitational collapse in the molecular cloud as $t=0$ 
and that they were specifically estimated independently for each of the three 
classes of clouds defined in this work.

Do the above results agree with previous time-scale estimations?
The evolutionary time-scale of our SLC
sample is within the range of collapse life-times
derived in other studies.
For example,~\citet{2012A&A...540A.113T} derived a life-time of
$6\times10^{4}$\,yr and~\citet{2013A&A...559A..79R}, $7-17\times10^{4}$\,yr
for the starless core phase. 
Furthermore,~\citet{csengeri-2014} found a time-scale
of $7.5\pm2.5\times10^4$\,yr for massive starless clumps in the Galaxy using
ATLASGAL data. In all these studies, as well as in the present paper,
the starless clumps are massive enough to be able to harbor high-mass
star-forming activity. Similar evolutionary time-scales have also been found
in regions that are more likely to only form low-mass stars, 
e.g., in Perseus~\citep{2014AAS...22345433W}.
The SFC evolutionary time-scale
is close to recent age estimates
of Class 0+1 protostars,
$\sim0.5-0.4$\,Myr~\citep[Table~1]{2014prpl.conf..195D}.
\HII regions are subject to other physical processes
apart from gravity, such as Rayleigh-Taylor (RT) instabilities involved in
the expansion
of the \HII regions and shocks due to stellar feedback.
These processes make this simple evolutionary
model hardly applicable to them and we therefore do not
discuss the evolutionary time obtained for \HII regions further.

Finally, we mention several caveats associated with
the time-scales derived above.
The mean column density used in Eq.~\ref{giri2}
corresponds only to the mean observed column
density and not necessarily to the actual
mean column density that should be used in Eq.~\ref{giri2}.
In addition, the factor, $R$, relating 2D and 3D
variances of mean normalized densities is still not well constrained.
Furthermore, this model assumes a single cloud undergoing 
free-fall collapse. While this assumption can be true
for the SLCs, it is unlikely to be the case in SFCs.
As mentioned in Section~\ref{sec:pdf-evol},
we assume that the smaller ATLASGAL clumps which compose
each SFC region are close to spherical symmetry.
With these caveats, we only aim to study
the evolutionary time-scales in terms of orders of magnitude.
Considering these caveats, we conclude that the method of
estimating evolutionary time-scale presented agrees
with independently derived typical ages for SLCs and SFCs.

\section{Conclusions}\label{conc}

We have used ATLASGAL 870\,$\mu$m dust continuum data
to study the column density distribution of
330 molecular clouds molecular clouds that we divide in three 
evolutionary classes: starless clumps (SLCs), star-forming clouds (SFCs),
and \HII regions.
Our large sample of molecular clouds allows us
to study their column density distributions at
Galactic scale for the first time.
We study the column density distributions
of the clouds over a wide dynamic range
$A_{V}\sim3-1000$\,mag, spanning a wide range of
cloud masses ($10^{2}-10^{5}\,\mathrm{M_{\sun}}$).
In the following we summarize the main results obtained.

\begin{itemize}

\item The total \textit{N}-PDFs of SLCs is well described by a log-normal
function with a width of about $\sigma_{s}\sim0.5$. The total
\textit{N}-PDF of SFCs and \HII
regions show power-law tails at high column
densities, with \HII regions having a shallower slope.
These observations agree with a picture in which the density distribution
of SLCs is dominated by turbulent motions.
The SFCs are significantly affected by gravity,
although turbulence may still play a role in structuring the clouds.
The density distributions of \HII regions are consistent with
gravity-dominated media.
Our statistical sample shows that this picture, earlier observed in clouds of
the Solar neighborhood, is relevant also at Galactic scale.

\item DGMFs of SLCs are well described by exponential functions
with exponent $\alpha_{exp}=-0.1$.
The DGMFs of \HII regions and SFCs are better described by power-laws
with exponents of $\beta=-1.0$ and $\beta=-2.1$ respectively.
The DGMF shape depends on cloud mass, being shallower for
the most massive clouds and steeper for the less massive clouds.
This dependence exists in all evolutionary classes.

\item We find an approximately linear correlation
$f_{\mathrm{DG}}\propto\Sigma_{\mathrm{mass}}$ for
$\Sigma_{\mathrm{mass}}=50-200\,\mathrm{M_{\sun}pc^{-2}}$, valid for all
evolutionary classes. This relation flatens at $f_{\mathrm{DG}}\cong0.8$ in MCs,
suggesting that the maximum
star-forming activity in MCs is reached at $f_{\mathrm{DG}}\cong0.8$.
We also find that the intrinsic scatter of $f_{\mathrm{DG}}$ is ($\sim$0.7\,dex)
is similar to the scatter seen in the relation
SFR - dense gas mass of~\citep{2010ApJ...724..687L,2012ApJ...745..190L}.
This suggests that both, the dense gas mass
and the lower-density envelope of the cloud, play
a significant role in affecting the star formation rate.

\item We estimate the evolutionary time-scales
of our three classes using an analytical model which
predicts the evolution of the PDF of a cloud in free-fall
collapse~\citep{2014ApJ...781...91G}.
We found $t_{\mathrm{E}}\lesssim0.1$\,Myr,
$t_{\mathrm{E}}\sim0.3$\,Myr, and $t_{\mathrm{E}}\lesssim0.7$\,Myr for SLCs, SFCs, and \HII
regions, respectively.
Both time-scales agree with previous,
independent age estimates of corresponding objects,
suggesting that molecular cloud evolution
may indeed be imprinted into the observable \textit{N}-PDF functions.
\HII regions show a complexity of physical processes that
make this model hard to apply to them. 
\end{itemize}

\begin{acknowledgements}
The work of J.A. is supported by the Sonderforschungsbereich (SFB) 881
\textquotedblleft The Milky Way System\textquotedblright \,and
the International Max-Planck Research School (IMPRS) at Heidelberg University.
The work of J.K. and A.S. was supported by the
Deutsche Forschungsgemeinschaft priority
program 1573 (\textquotedblleft Physics of the Interstellar
Medium\textquotedblright).
This research has made use of the SIMBAD
database, operated at CDS, Strasbourg, France.
\end{acknowledgements}

\bibliographystyle{aa} 

\begin{thebibliography}{89}
\expandafter\ifx\csname natexlab\endcsname\relax\def\natexlab#1{#1}\fi

\bibitem[{{Aguirre} {et~al.}(2011){Aguirre}, {Ginsburg}, {Dunham}, {Drosback},
  {Bally}, {Battersby}, {Bradley}, {Cyganowski}, {Dowell}, {Evans}, {Glenn},
  {Harvey}, {Rosolowsky}, {Stringfellow}, {Walawender}, \& {Williams}}]{bgps}
{Aguirre}, J.~E., {Ginsburg}, A.~G., {Dunham}, M.~K., {et~al.} 2011, \apjs,
  192, 4

\bibitem[{{Alves} {et~al.}(2014){Alves}, {Lombardi}, \& {Lada}}]{alves-2014}
{Alves}, J., {Lombardi}, M., \& {Lada}, C.~J. 2014, \aap, 565, A18

\bibitem[{{Anderson} {et~al.}(2012){Anderson}, {Zavagno}, {Deharveng},
  {Abergel}, {Motte}, {Andr{\'e}}, {Bernard}, {Bontemps}, {Hennemann}, {Hill},
  {Rod{\'o}n}, {Roussel}, \& {Russeil}}]{anderson-12}
{Anderson}, L.~D., {Zavagno}, A., {Deharveng}, L., {et~al.} 2012, \aap, 542,
  A10

\bibitem[{{Andr{\'e}} {et~al.}(2010){Andr{\'e}}, {Men'shchikov}, {Bontemps},
  {K{\"o}nyves}, {Motte}, {Schneider}, {Didelon}, {Minier}, {Saraceno},
  {Ward-Thompson}, {di Francesco}, {White}, {Molinari}, {Testi}, {Abergel},
  {Griffin}, {Henning}, {Royer}, {Mer{\'{\i}}n}, {Vavrek}, {Attard},
  {Arzoumanian}, {Wilson}, {Ade}, {Aussel}, {Baluteau}, {Benedettini},
  {Bernard}, {Blommaert}, {Cambr{\'e}sy}, {Cox}, {di Giorgio}, {Hargrave},
  {Hennemann}, {Huang}, {Kirk}, {Krause}, {Launhardt}, {Leeks}, {Le Pennec},
  {Li}, {Martin}, {Maury}, {Olofsson}, {Omont}, {Peretto}, {Pezzuto}, {Prusti},
  {Roussel}, {Russeil}, {Sauvage}, {Sibthorpe}, {Sicilia-Aguilar}, {Spinoglio},
  {Waelkens}, {Woodcraft}, \& {Zavagno}}]{2010A&A...518L.102A}
{Andr{\'e}}, P., {Men'shchikov}, A., {Bontemps}, S., {et~al.} 2010, \aap, 518,
  L102

\bibitem[{{Ballesteros-Paredes} {et~al.}(2011){Ballesteros-Paredes},
  {V{\'a}zquez-Semadeni}, {Gazol}, {Hartmann}, {Heitsch}, \&
  {Col{\'{\i}}n}}]{2011MNRAS.416.1436B}
{Ballesteros-Paredes}, J., {V{\'a}zquez-Semadeni}, E., {Gazol}, A., {et~al.}
  2011, \mnras, 416, 1436

\bibitem[{{Battisti} \& {Heyer}(2014)}]{2014ApJ...780..173B}
{Battisti}, A.~J. \& {Heyer}, M.~H. 2014, \apj, 780, 173

\bibitem[{{Beuther} {et~al.}(2002){Beuther}, {Schilke}, {Menten}, {Motte},
  {Sridharan}, \& {Wyrowski}}]{2002ApJ...566..945B}
{Beuther}, H., {Schilke}, P., {Menten}, K.~M., {et~al.} 2002, \apj, 566, 945

\bibitem[{{Bohlin} {et~al.}(1978){Bohlin}, {Savage}, \&
  {Drake}}]{1978ApJ...224..132B}
{Bohlin}, R.~C., {Savage}, B.~D., \& {Drake}, J.~F. 1978, \apj, 224, 132

\bibitem[{{Bronfman} {et~al.}(1996){Bronfman}, {Nyman}, \&
  {May}}]{1996A&AS..115...81B}
{Bronfman}, L., {Nyman}, L.-A., \& {May}, J. 1996, \aaps, 115, 81

\bibitem[{{Brunt} {et~al.}(2010){Brunt}, {Federrath}, \&
  {Price}}]{2010MNRAS.403.1507B}
{Brunt}, C.~M., {Federrath}, C., \& {Price}, D.~J. 2010, \mnras, 403, 1507

\bibitem[{{Csengeri} {et~al.}(2014){Csengeri}, {Urquhart}, {Schuller}, {Motte},
  {Bontemps}, {Wyrowski}, {Menten}, {Bronfman}, {Beuther}, {Henning}, {Testi},
  {Zavagno}, \& {Walmsley}}]{csengeri-2014}
{Csengeri}, T., {Urquhart}, J.~S., {Schuller}, F., {et~al.} 2014, \aap, 565,
  A75

\bibitem[{{Cyganowski} {et~al.}(2008){Cyganowski}, {Whitney}, {Holden},
  {Braden}, {Brogan}, {Churchwell}, {Indebetouw}, {Watson}, {Babler},
  {Benjamin}, {Gomez}, {Meade}, {Povich}, {Robitaille}, \&
  {Watson}}]{cyganowski-2008}
{Cyganowski}, C.~J., {Whitney}, B.~A., {Holden}, E., {et~al.} 2008, \aj, 136,
  2391

\bibitem[{{Dobbs} \& {Burkert}(2012)}]{2012MNRAS.421.2940D}
{Dobbs}, C.~L. \& {Burkert}, A. 2012, \mnras, 421, 2940

\bibitem[{{Draine}(2011)}]{2011piim.book.....D}
{Draine}, B.~T. 2011, {Physics of the Interstellar and Intergalactic
  Medium.~Princeton University Press, 2011.~ISBN: 978-0-691-12214-4}

\bibitem[{{Dunham} {et~al.}(2011){Dunham}, {Robitaille}, {Evans}, {Schlingman},
  {Cyganowski}, \& {Urquhart}}]{2011ApJ...731...90D}
{Dunham}, M.~K., {Robitaille}, T.~P., {Evans}, II, N.~J., {et~al.} 2011, \apj,
  731, 90

\bibitem[{{Dunham} {et~al.}(2014){Dunham}, {Stutz}, {Allen}, {Evans},
  {Fischer}, {Megeath}, {Myers}, {Offner}, {Poteet}, {Tobin}, \&
  {Vorobyov}}]{2014prpl.conf..195D}
{Dunham}, M.~M., {Stutz}, A.~M., {Allen}, L.~E., {et~al.} 2014, Protostars and
  Planets VI, 195

\bibitem[{{Ellsworth-Bowers} {et~al.}(2013){Ellsworth-Bowers}, {Glenn},
  {Rosolowsky}, {Mairs}, {Evans}, {Battersby}, {Ginsburg}, {Shirley}, \&
  {Bally}}]{2013ApJ...770...39E}
{Ellsworth-Bowers}, T.~P., {Glenn}, J., {Rosolowsky}, E., {et~al.} 2013, \apj,
  770, 39

\bibitem[{{Federrath} \& {Klessen}(2013)}]{2013ApJ...763...51F}
{Federrath}, C. \& {Klessen}, R.~S. 2013, \apj, 763, 51

\bibitem[{{Forster} \& {Caswell}(2000)}]{2000ApJ...530..371F}
{Forster}, J.~R. \& {Caswell}, J.~L. 2000, \apj, 530, 371

\bibitem[{{Garay} {et~al.}(1993){Garay}, {Rodriguez}, {Moran}, \&
  {Churchwell}}]{1993ApJ...418..368G}
{Garay}, G., {Rodriguez}, L.~F., {Moran}, J.~M., \& {Churchwell}, E. 1993,
  \apj, 418, 368

\bibitem[{{Girichidis} {et~al.}(2014){Girichidis}, {Konstandin}, {Whitworth},
  \& {Klessen}}]{2014ApJ...781...91G}
{Girichidis}, P., {Konstandin}, L., {Whitworth}, A.~P., \& {Klessen}, R.~S.
  2014, \apj, 781, 91

\bibitem[{{Goodman} {et~al.}(2009){Goodman}, {Pineda}, \&
  {Schnee}}]{2009ApJ...692...91G}
{Goodman}, A.~A., {Pineda}, J.~E., \& {Schnee}, S.~L. 2009, \apj, 692, 91

\bibitem[{{Griffin} {et~al.}(2010){Griffin}, {Abergel}, {Abreu}, {Ade},
  {Andr{\'e}}, {Augueres}, {Babbedge}, {Bae}, {Baillie}, {Baluteau}, {Barlow},
  {Bendo}, {Benielli}, {Bock}, {Bonhomme}, {Brisbin}, {Brockley-Blatt},
  {Caldwell}, {Cara}, {Castro-Rodriguez}, {Cerulli}, {Chanial}, {Chen},
  {Clark}, {Clements}, {Clerc}, {Coker}, {Communal}, {Conversi}, {Cox},
  {Crumb}, {Cunningham}, {Daly}, {Davis}, {de Antoni}, {Delderfield}, {Devin},
  {di Giorgio}, {Didschuns}, {Dohlen}, {Donati}, {Dowell}, {Dowell}, {Duband},
  {Dumaye}, {Emery}, {Ferlet}, {Ferrand}, {Fontignie}, {Fox}, {Franceschini},
  {Frerking}, {Fulton}, {Garcia}, {Gastaud}, {Gear}, {Glenn}, {Goizel},
  {Griffin}, {Grundy}, {Guest}, {Guillemet}, {Hargrave}, {Harwit}, {Hastings},
  {Hatziminaoglou}, {Herman}, {Hinde}, {Hristov}, {Huang}, {Imhof}, {Isaak},
  {Israelsson}, {Ivison}, {Jennings}, {Kiernan}, {King}, {Lange}, {Latter},
  {Laurent}, {Laurent}, {Leeks}, {Lellouch}, {Levenson}, {Li}, {Li},
  {Lilienthal}, {Lim}, {Liu}, {Lu}, {Madden}, {Mainetti}, {Marliani}, {McKay},
  {Mercier}, {Molinari}, {Morris}, {Moseley}, {Mulder}, {Mur}, {Naylor},
  {Nguyen}, {O'Halloran}, {Oliver}, {Olofsson}, {Olofsson}, {Orfei}, {Page},
  {Pain}, {Panuzzo}, {Papageorgiou}, {Parks}, {Parr-Burman}, {Pearce},
  {Pearson}, {P{\'e}rez-Fournon}, {Pinsard}, {Pisano}, {Podosek}, {Pohlen},
  {Polehampton}, {Pouliquen}, {Rigopoulou}, {Rizzo}, {Roseboom}, {Roussel},
  {Rowan-Robinson}, {Rownd}, {Saraceno}, {Sauvage}, {Savage}, {Savini},
  {Sawyer}, {Scharmberg}, {Schmitt}, {Schneider}, {Schulz}, {Schwartz},
  {Shafer}, {Shupe}, {Sibthorpe}, {Sidher}, {Smith}, {Smith}, {Smith},
  {Spencer}, {Stobie}, {Sudiwala}, {Sukhatme}, {Surace}, {Stevens}, {Swinyard},
  {Trichas}, {Tourette}, {Triou}, {Tseng}, {Tucker}, {Turner}, {Vaccari},
  {Valtchanov}, {Vigroux}, {Virique}, {Voellmer}, {Walker}, {Ward}, {Waskett},
  {Weilert}, {Wesson}, {White}, {Whitehouse}, {Wilson}, {Winter}, {Woodcraft},
  {Wright}, {Xu}, {Zavagno}, {Zemcov}, {Zhang}, \&
  {Zonca}}]{2010A&A...518L...3G}
{Griffin}, M.~J., {Abergel}, A., {Abreu}, A., {et~al.} 2010, \aap, 518, L3

\bibitem[{{Gutermuth} {et~al.}(2011){Gutermuth}, {Pipher}, {Megeath}, {Myers},
  {Allen}, \& {Allen}}]{2011ApJ...739...84G}
{Gutermuth}, R.~A., {Pipher}, J.~L., {Megeath}, S.~T., {et~al.} 2011, \apj,
  739, 84

\bibitem[{{Heiderman} {et~al.}(2010){Heiderman}, {Evans}, {Allen}, {Huard}, \&
  {Heyer}}]{2010ApJ...723.1019H}
{Heiderman}, A., {Evans}, II, N.~J., {Allen}, L.~E., {Huard}, T., \& {Heyer},
  M. 2010, \apj, 723, 1019

\bibitem[{{Hennebelle} \& {Falgarone}(2012)}]{2012A&ARv..20...55H}
{Hennebelle}, P. \& {Falgarone}, E. 2012, \aapr, 20, 55

\bibitem[{{Henning} {et~al.}(2010){Henning}, {Linz}, {Krause}, {Ragan},
  {Beuther}, {Launhardt}, {Nielbock}, \& {Vasyunina}}]{2010A&A...518L..95H}
{Henning}, T., {Linz}, H., {Krause}, O., {et~al.} 2010, \aap, 518, L95

\bibitem[{{Kainulainen} {et~al.}(2011{\natexlab{a}}){Kainulainen}, {Alves},
  {Beuther}, {Henning}, \& {Schuller}}]{2011A&A...536A..48K}
{Kainulainen}, J., {Alves}, J., {Beuther}, H., {Henning}, T., \& {Schuller}, F.
  2011{\natexlab{a}}, \aap, 536, A48

\bibitem[{{Kainulainen} {et~al.}(2011{\natexlab{b}}){Kainulainen}, {Beuther},
  {Banerjee}, {Federrath}, \& {Henning}}]{2011A&A...530A..64K}
{Kainulainen}, J., {Beuther}, H., {Banerjee}, R., {Federrath}, C., \&
  {Henning}, T. 2011{\natexlab{b}}, \aap, 530, A64

\bibitem[{{Kainulainen} {et~al.}(2009){Kainulainen}, {Beuther}, {Henning}, \&
  {Plume}}]{2009A&A...508L..35K}
{Kainulainen}, J., {Beuther}, H., {Henning}, T., \& {Plume}, R. 2009, \aap,
  508, L35

\bibitem[{{Kainulainen} {et~al.}(2013{\natexlab{a}}){Kainulainen}, {Federrath},
  \& {Henning}}]{2013A&A...553L...8K}
{Kainulainen}, J., {Federrath}, C., \& {Henning}, T. 2013{\natexlab{a}}, \aap,
  553, L8

\bibitem[{{Kainulainen} {et~al.}(2014){Kainulainen}, {Federrath}, \&
  {Henning}}]{2014Sci...344..183K}
{Kainulainen}, J., {Federrath}, C., \& {Henning}, T. 2014, Science, 344, 183

\bibitem[{{Kainulainen} {et~al.}(2013{\natexlab{b}}){Kainulainen}, {Ragan},
  {Henning}, \& {Stutz}}]{2013A&A...557A.120K}
{Kainulainen}, J., {Ragan}, S.~E., {Henning}, T., \& {Stutz}, A.
  2013{\natexlab{b}}, \aap, 557, A120

\bibitem[{{Kainulainen} \& {Tan}(2013)}]{2013A&A...549A..53K}
{Kainulainen}, J. \& {Tan}, J.~C. 2013, \aap, 549, A53

\bibitem[{{Klessen}(2000)}]{2000ApJ...535..869K}
{Klessen}, R.~S. 2000, \apj, 535, 869

\bibitem[{{Klessen} \& {Burkert}(2000)}]{2000ApJS..128..287K}
{Klessen}, R.~S. \& {Burkert}, A. 2000, \apjs, 128, 287

\bibitem[{{Kritsuk} {et~al.}(2011){Kritsuk}, {Norman}, \&
  {Wagner}}]{2011ApJ...727L..20K}
{Kritsuk}, A.~G., {Norman}, M.~L., \& {Wagner}, R. 2011, \apjl, 727, L20

\bibitem[{{Lada} {et~al.}(2012){Lada}, {Forbrich}, {Lombardi}, \&
  {Alves}}]{2012ApJ...745..190L}
{Lada}, C.~J., {Forbrich}, J., {Lombardi}, M., \& {Alves}, J.~F. 2012, \apj,
  745, 190

\bibitem[{{Lada} {et~al.}(2009){Lada}, {Lombardi}, \&
  {Alves}}]{2009ApJ...703...52L}
{Lada}, C.~J., {Lombardi}, M., \& {Alves}, J.~F. 2009, \apj, 703, 52

\bibitem[{{Lada} {et~al.}(2010){Lada}, {Lombardi}, \&
  {Alves}}]{2010ApJ...724..687L}
{Lada}, C.~J., {Lombardi}, M., \& {Alves}, J.~F. 2010, \apj, 724, 687

\bibitem[{{Launhardt} {et~al.}(2013){Launhardt}, {Stutz}, {Schmiedeke},
  {Henning}, {Krause}, {Balog}, {Beuther}, {Birkmann}, {Hennemann},
  {Kainulainen}, {Khanzadyan}, {Linz}, {Lippok}, {Nielbock}, {Pitann}, {Ragan},
  {Risacher}, {Schmalzl}, {Shirley}, {Stecklum}, {Steinacker}, \&
  {Tackenberg}}]{2013A&A...551A..98L}
{Launhardt}, R., {Stutz}, A.~M., {Schmiedeke}, A., {et~al.} 2013, \aap, 551,
  A98

\bibitem[{{Lockman}(1989)}]{1989ApJS...71..469L}
{Lockman}, F.~J. 1989, \apjs, 71, 469

\bibitem[{{Lockman} {et~al.}(1996){Lockman}, {Pisano}, \&
  {Howard}}]{1996ApJ...472..173L}
{Lockman}, F.~J., {Pisano}, D.~J., \& {Howard}, G.~J. 1996, \apj, 472, 173

\bibitem[{{Lombardi} \& {Alves}(2001)}]{2001A&A...377.1023L}
{Lombardi}, M. \& {Alves}, J. 2001, \aap, 377, 1023

\bibitem[{{Lombardi} {et~al.}(2015){Lombardi}, {Alves}, \&
  {Lada}}]{lombardi-15}
{Lombardi}, M., {Alves}, J., \& {Lada}, C.~J. 2015, \aap, 576, L1

\bibitem[{{Lumsden} {et~al.}(2013){Lumsden}, {Hoare}, {Urquhart}, {Oudmaijer},
  {Davies}, {Mottram}, {Cooper}, \& {Moore}}]{lumsden-2013}
{Lumsden}, S.~L., {Hoare}, M.~G., {Urquhart}, J.~S., {et~al.} 2013, \apjs, 208,
  11

\bibitem[{{Marshall} {et~al.}(2009){Marshall}, {Joncas}, \&
  {Jones}}]{2009ApJ...706..727M}
{Marshall}, D.~J., {Joncas}, G., \& {Jones}, A.~P. 2009, \apj, 706, 727

\bibitem[{{Molinari} {et~al.}(2010){Molinari}, {Swinyard}, {Bally}, {Barlow},
  {Bernard}, {Martin}, {Moore}, {Noriega-Crespo}, {Plume}, {Testi}, {Zavagno},
  {Abergel}, {Ali}, {Andr{\'e}}, {Baluteau}, {Benedettini}, {Bern{\'e}},
  {Billot}, {Blommaert}, {Bontemps}, {Boulanger}, {Brand}, {Brunt}, {Burton},
  {Campeggio}, {Carey}, {Caselli}, {Cesaroni}, {Cernicharo}, {Chakrabarti},
  {Chrysostomou}, {Codella}, {Cohen}, {Compiegne}, {Davis}, {de Bernardis}, {de
  Gasperis}, {Di Francesco}, {di Giorgio}, {Elia}, {Faustini}, {Fischera},
  {Fukui}, {Fuller}, {Ganga}, {Garcia-Lario}, {Giard}, {Giardino}, {Glenn},
  {Goldsmith}, {Griffin}, {Hoare}, {Huang}, {Jiang}, {Joblin}, {Joncas},
  {Juvela}, {Kirk}, {Lagache}, {Li}, {Lim}, {Lord}, {Lucas}, {Maiolo},
  {Marengo}, {Marshall}, {Masi}, {Massi}, {Matsuura}, {Meny}, {Minier},
  {Miville-Desch{\^e}nes}, {Montier}, {Motte}, {M{\"u}ller}, {Natoli}, {Neves},
  {Olmi}, {Paladini}, {Paradis}, {Pestalozzi}, {Pezzuto}, {Piacentini},
  {Pomar{\`e}s}, {Popescu}, {Reach}, {Richer}, {Ristorcelli}, {Roy}, {Royer},
  {Russeil}, {Saraceno}, {Sauvage}, {Schilke}, {Schneider-Bontemps},
  {Schuller}, {Schultz}, {Shepherd}, {Sibthorpe}, {Smith}, {Smith},
  {Spinoglio}, {Stamatellos}, {Strafella}, {Stringfellow}, {Sturm}, {Taylor},
  {Thompson}, {Tuffs}, {Umana}, {Valenziano}, {Vavrek}, {Viti}, {Waelkens},
  {Ward-Thompson}, {White}, {Wyrowski}, {Yorke}, \&
  {Zhang}}]{2010PASP..122..314M}
{Molinari}, S., {Swinyard}, B., {Bally}, J., {et~al.} 2010, \pasp, 122, 314

\bibitem[{{Motte} \& {Hennebelle}(2009)}]{2009EAS....34..195M}
{Motte}, F. \& {Hennebelle}, P. 2009, in EAS Publications Series, Vol.~34, EAS
  Publications Series, ed. L.~{Pagani} \& M.~{Gerin}, 195--211

\bibitem[{{Ossenkopf} \& {Henning}(1994)}]{1994A&A...291..943O}
{Ossenkopf}, V. \& {Henning}, T. 1994, \aap, 291, 943

\bibitem[{{Ostriker} {et~al.}(2001){Ostriker}, {Stone}, \&
  {Gammie}}]{2001ApJ...546..980O}
{Ostriker}, E.~C., {Stone}, J.~M., \& {Gammie}, C.~F. 2001, \apj, 546, 980

\bibitem[{{Ott}(2010)}]{2010ASPC..434..139O}
{Ott}, S. 2010, in Astronomical Society of the Pacific Conference Series, Vol.
  434, Astronomical Data Analysis Software and Systems XIX, ed. Y.~{Mizumoto},
  K.-I. {Morita}, \& M.~{Ohishi}, 139

\bibitem[{{Padoan} {et~al.}(1997){Padoan}, {Jones}, \&
  {Nordlund}}]{1997ApJ...474..730P}
{Padoan}, P., {Jones}, B.~J.~T., \& {Nordlund}, A.~P. 1997, \apj, 474, 730

\bibitem[{{Padoan} \& {Nordlund}(2011)}]{2011ApJ...730...40P}
{Padoan}, P. \& {Nordlund}, {\AA}. 2011, \apj, 730, 40

\bibitem[{{Peretto} \& {Fuller}(2010)}]{2010ApJ...723..555P}
{Peretto}, N. \& {Fuller}, G.~A. 2010, \apj, 723, 555

\bibitem[{{Pilbratt} {et~al.}(2010){Pilbratt}, {Riedinger}, {Passvogel},
  {Crone}, {Doyle}, {Gageur}, {Heras}, {Jewell}, {Metcalfe}, {Ott}, \&
  {Schmidt}}]{2010A&A...518L...1P}
{Pilbratt}, G.~L., {Riedinger}, J.~R., {Passvogel}, T., {et~al.} 2010, \aap,
  518, L1

\bibitem[{{Poglitsch} {et~al.}(2010){Poglitsch}, {Waelkens}, {Geis},
  {Feuchtgruber}, {Vandenbussche}, {Rodriguez}, {Krause}, {Renotte}, {van
  Hoof}, {Saraceno}, {Cepa}, {Kerschbaum}, {Agn{\`e}se}, {Ali}, {Altieri},
  {Andreani}, {Augueres}, {Balog}, {Barl}, {Bauer}, {Belbachir}, {Benedettini},
  {Billot}, {Boulade}, {Bischof}, {Blommaert}, {Callut}, {Cara}, {Cerulli},
  {Cesarsky}, {Contursi}, {Creten}, {De Meester}, {Doublier}, {Doumayrou},
  {Duband}, {Exter}, {Genzel}, {Gillis}, {Gr{\"o}zinger}, {Henning},
  {Herreros}, {Huygen}, {Inguscio}, {Jakob}, {Jamar}, {Jean}, {de Jong},
  {Katterloher}, {Kiss}, {Klaas}, {Lemke}, {Lutz}, {Madden}, {Marquet},
  {Martignac}, {Mazy}, {Merken}, {Montfort}, {Morbidelli}, {M{\"u}ller},
  {Nielbock}, {Okumura}, {Orfei}, {Ottensamer}, {Pezzuto}, {Popesso},
  {Putzeys}, {Regibo}, {Reveret}, {Royer}, {Sauvage}, {Schreiber}, {Stegmaier},
  {Schmitt}, {Schubert}, {Sturm}, {Thiel}, {Tofani}, {Vavrek}, {Wetzstein},
  {Wieprecht}, \& {Wiezorrek}}]{2010A&A...518L...2P}
{Poglitsch}, A., {Waelkens}, C., {Geis}, N., {et~al.} 2010, \aap, 518, L2

\bibitem[{{Ragan} {et~al.}(2012){Ragan}, {Henning}, {Krause}, {Pitann},
  {Beuther}, {Linz}, {Tackenberg}, {Balog}, {Hennemann}, {Launhardt}, {Lippok},
  {Nielbock}, {Schmiedeke}, {Schuller}, {Steinacker}, {Stutz}, \&
  {Vasyunina}}]{2012A&A...547A..49R}
{Ragan}, S., {Henning}, T., {Krause}, O., {et~al.} 2012, \aap, 547, A49

\bibitem[{{Ragan} {et~al.}(2013){Ragan}, {Henning}, \&
  {Beuther}}]{2013A&A...559A..79R}
{Ragan}, S.~E., {Henning}, T., \& {Beuther}, H. 2013, \aap, 559, A79

\bibitem[{{Robitaille} {et~al.}(2008){Robitaille}, {Meade}, {Babler},
  {Whitney}, {Johnston}, {Indebetouw}, {Cohen}, {Povich}, {Sewilo}, {Benjamin},
  \& {Churchwell}}]{robitaille-2008}
{Robitaille}, T.~P., {Meade}, M.~R., {Babler}, B.~L., {et~al.} 2008, \aj, 136,
  2413

\bibitem[{{Roman-Duval} {et~al.}(2009){Roman-Duval}, {Jackson}, {Heyer},
  {Johnson}, {Rathborne}, {Shah}, \& {Simon}}]{2009ApJ...699.1153R}
{Roman-Duval}, J., {Jackson}, J.~M., {Heyer}, M., {et~al.} 2009, \apj, 699,
  1153

\bibitem[{{Roussel}(2013)}]{2013PASP..125.1126R}
{Roussel}, H. 2013, \pasp, 125, 1126

\bibitem[{{Rudolph} {et~al.}(2006){Rudolph}, {Fich}, {Bell}, {Norsen},
  {Simpson}, {Haas}, \& {Erickson}}]{2006ApJS..162..346R}
{Rudolph}, A.~L., {Fich}, M., {Bell}, G.~R., {et~al.} 2006, \apjs, 162, 346

\bibitem[{{Russeil} {et~al.}(2013){Russeil}, {Schneider}, {Anderson},
  {Zavagno}, {Molinari}, {Persi}, {Bontemps}, {Motte}, {Ossenkopf},
  {Andr{\'e}}, {Arzoumanian}, {Bernard}, {Deharveng}, {Didelon}, {Di
  Francesco}, {Elia}, {Hennemann}, {Hill}, {K{\"o}nyves}, {Li}, {Martin},
  {Nguyen Luong}, {Peretto}, {Pezzuto}, {Polychroni}, {Roussel}, {Rygl},
  {Spinoglio}, {Testi}, {Tig{\'e}}, {Vavrek}, {Ward-Thompson}, \&
  {White}}]{2013A&A...554A..42R}
{Russeil}, D., {Schneider}, N., {Anderson}, L.~D., {et~al.} 2013, \aap, 554,
  A42

\bibitem[{{Sadavoy} {et~al.}(2014){Sadavoy}, {Di Francesco}, {Andre},
  {Pezzuto}, {Bernard}, {Maury}, {Men'shchikov}, {Motte}, {Nguyn-Luong},
  {Schneider}, {Arzoumanian}, {Benedettini}, {Bontemps}, {Elia}, {Hennemann},
  {Hill}, {Konyves}, {Louvet}, {Peretto}, {Roy}, \&
  {White}}]{2014ApJ...787L..18S}
{Sadavoy}, S.~I., {Di Francesco}, J., {Andre}, P., {et~al.} 2014, \apjl, 787,
  L18

\bibitem[{{Scalo} {et~al.}(1998){Scalo}, {Vazquez-Semadeni}, {Chappell}, \&
  {Passot}}]{1998ApJ...504..835S}
{Scalo}, J., {Vazquez-Semadeni}, E., {Chappell}, D., \& {Passot}, T. 1998,
  \apj, 504, 835

\bibitem[{{Schneider} {et~al.}(2013){Schneider}, {Andr{\'e}}, {K{\"o}nyves},
  {Bontemps}, {Motte}, {Federrath}, {Ward-Thompson}, {Arzoumanian},
  {Benedettini}, {Bressert}, {Didelon}, {Di Francesco}, {Griffin}, {Hennemann},
  {Hill}, {Palmeirim}, {Pezzuto}, {Peretto}, {Roy}, {Rygl}, {Spinoglio}, \&
  {White}}]{2013ApJ...766L..17S}
{Schneider}, N., {Andr{\'e}}, P., {K{\"o}nyves}, V., {et~al.} 2013, \apjl, 766,
  L17

\bibitem[{{Schneider} {et~al.}(2012){Schneider}, {Csengeri}, {Hennemann},
  {Motte}, {Didelon}, {Federrath}, {Bontemps}, {Di Francesco}, {Arzoumanian},
  {Minier}, {Andr{\'e}}, {Hill}, {Zavagno}, {Nguyen-Luong}, {Attard},
  {Bernard}, {Elia}, {Fallscheer}, {Griffin}, {Kirk}, {Klessen}, {K{\"o}nyves},
  {Martin}, {Men'shchikov}, {Palmeirim}, {Peretto}, {Pestalozzi}, {Russeil},
  {Sadavoy}, {Sousbie}, {Testi}, {Tremblin}, {Ward-Thompson}, \&
  {White}}]{2012A&A...540L..11S}
{Schneider}, N., {Csengeri}, T., {Hennemann}, M., {et~al.} 2012, \aap, 540, L11

\bibitem[{{Schneider} {et~al.}(2014){Schneider}, {Csengeri}, {Klessen},
  {Tremblin}, {Ossenkopf}, {Peretto}, {Simon}, {Bontemps}, \&
  {Federrath}}]{schneider-14}
{Schneider}, N., {Csengeri}, T., {Klessen}, R.~S., {et~al.} 2014, ArXiv
  e-prints

\bibitem[{{Schneider} {et~al.}(2015){Schneider}, {Ossenkopf}, {Csengeri},
  {Klessen}, {Federrath}, {Tremblin}, {Girichidis}, {Bontemps}, \&
  {Andr{\'e}}}]{schneider-15}
{Schneider}, N., {Ossenkopf}, V., {Csengeri}, T., {et~al.} 2015, \aap, 575, A79

\bibitem[{{Schuller} {et~al.}(2009){Schuller}, {Menten}, {Contreras},
  {Wyrowski}, {Schilke}, {Bronfman}, {Henning}, {Walmsley}, {Beuther},
  {Bontemps}, {Cesaroni}, {Deharveng}, {Garay}, {Herpin}, {Lefloch}, {Linz},
  {Mardones}, {Minier}, {Molinari}, {Motte}, {Nyman}, {Reveret}, {Risacher},
  {Russeil}, {Schneider}, {Testi}, {Troost}, {Vasyunina}, {Wienen}, {Zavagno},
  {Kovacs}, {Kreysa}, {Siringo}, \& {Wei{\ss}}}]{2009A&A...504..415S}
{Schuller}, F., {Menten}, K.~M., {Contreras}, Y., {et~al.} 2009, \aap, 504, 415

\bibitem[{{Shirley} {et~al.}(2011){Shirley}, {Huard}, {Pontoppidan}, {Wilner},
  {Stutz}, {Bieging}, \& {Evans}}]{2011ApJ...728..143S}
{Shirley}, Y.~L., {Huard}, T.~L., {Pontoppidan}, K.~M., {et~al.} 2011, \apj,
  728, 143

\bibitem[{{Shirley} {et~al.}(2005){Shirley}, {Nordhaus}, {Grcevich}, {Evans},
  {Rawlings}, \& {Tatematsu}}]{2005ApJ...632..982S}
{Shirley}, Y.~L., {Nordhaus}, M.~K., {Grcevich}, J.~M., {et~al.} 2005, \apj,
  632, 982

\bibitem[{{Shu} {et~al.}(1987){Shu}, {Adams}, \&
  {Lizano}}]{1987ARA&A..25...23S}
{Shu}, F.~H., {Adams}, F.~C., \& {Lizano}, S. 1987, \araa, 25, 23

\bibitem[{{Simon} {et~al.}(2006){Simon}, {Jackson}, {Rathborne}, \&
  {Chambers}}]{2006ApJ...639..227S}
{Simon}, R., {Jackson}, J.~M., {Rathborne}, J.~M., \& {Chambers}, E.~T. 2006,
  \apj, 639, 227

\bibitem[{{Solomon} \& {Rivolo}(1989)}]{1989ApJ...339..919S}
{Solomon}, P.~M. \& {Rivolo}, A.~R. 1989, \apj, 339, 919

\bibitem[{{Stutz} {et~al.}(2010){Stutz}, {Launhardt}, {Linz}, {Krause},
  {Henning}, {Kainulainen}, {Nielbock}, {Steinacker}, \&
  {Andr{\'e}}}]{2010A&A...518L..87S}
{Stutz}, A., {Launhardt}, R., {Linz}, H., {et~al.} 2010, \aap, 518, L87

\bibitem[{{Stutz} \& {Kainulainen}(2015)}]{2015arXiv150405188S}
{Stutz}, A.~M. \& {Kainulainen}, J. 2015, ArXiv:1504.05188

\bibitem[{{Tackenberg} {et~al.}(2012){Tackenberg}, {Beuther}, {Henning},
  {Schuller}, {Wienen}, {Motte}, {Wyrowski}, {Bontemps}, {Bronfman}, {Menten},
  {Testi}, \& {Lefloch}}]{2012A&A...540A.113T}
{Tackenberg}, J., {Beuther}, H., {Henning}, T., {et~al.} 2012, \aap, 540, A113

\bibitem[{{Tassis} {et~al.}(2010){Tassis}, {Christie}, {Urban}, {Pineda},
  {Mouschovias}, {Yorke}, \& {Martel}}]{2010MNRAS.408.1089T}
{Tassis}, K., {Christie}, D.~A., {Urban}, A., {et~al.} 2010, \mnras, 408, 1089

\bibitem[{{T{\'o}th} {et~al.}(2014){T{\'o}th}, {Marton}, {Zahorecz},
  {Bal{\'a}zs}, {Ueno}, {Tamura}, {Kawamura}, {Kiss}, \&
  {Kitamura}}]{2014PASJ...66...17T}
{T{\'o}th}, L.~V., {Marton}, G., {Zahorecz}, S., {et~al.} 2014, \pasj, 66, 17

\bibitem[{{Tremblin} {et~al.}(2014){Tremblin}, {Schneider}, {Minier},
  {Didelon}, {Hill}, {Anderson}, {Motte}, {Zavagno}, {Andr{\'e}},
  {Arzoumanian}, {Audit}, {Benedettini}, {Bontemps}, {Csengeri}, {Di
  Francesco}, {Giannini}, {Hennemann}, {Nguyen Luong}, {Marston}, {Peretto},
  {Rivera-Ingraham}, {Russeil}, {Rygl}, {Spinoglio}, \&
  {White}}]{2014A&A...564A.106T}
{Tremblin}, P., {Schneider}, N., {Minier}, V., {et~al.} 2014, \aap, 564, A106

\bibitem[{{Urquhart} {et~al.}(2013){Urquhart}, {Thompson}, {Moore}, {Purcell},
  {Hoare}, {Schuller}, {Wyrowski}, {Csengeri}, {Menten}, {Lumsden}, {Kurtz},
  {Walmsley}, {Bronfman}, {Morgan}, {Eden}, \& {Russeil}}]{2013MNRAS.435..400U}
{Urquhart}, J.~S., {Thompson}, M.~A., {Moore}, T.~J.~T., {et~al.} 2013, \mnras,
  435, 400

\bibitem[{{Vazquez-Semadeni}(1994)}]{1994ApJ...423..681V}
{Vazquez-Semadeni}, E. 1994, \apj, 423, 681

\bibitem[{{V{\'a}zquez-Semadeni} \&
  {Garc{\'{\i}}a}(2001)}]{2001ApJ...557..727V}
{V{\'a}zquez-Semadeni}, E. \& {Garc{\'{\i}}a}, N. 2001, \apj, 557, 727

\bibitem[{{Walker-LaFollette} {et~al.}(2014){Walker-LaFollette}, {Shirley},
  {Amaya}, {Becker}, {Biddle}, {Lichtenberger}, {Nieberding}, {Raphael},
  {Romine}, {Small}, {Stanford-Jones}, {Smith}, {Thompson}, {Towner}, {Turner},
  {Watson}, {Cates}, {McGraw}, {Pearson}, {Robertson}, \&
  {Tombleson}}]{2014AAS...22345433W}
{Walker-LaFollette}, A., {Shirley}, Y.~L., {Amaya}, H., {et~al.} 2014, in
  American Astronomical Society Meeting Abstracts, Vol. 223, American
  Astronomical Society Meeting Abstracts, 454

\bibitem[{{Walsh} {et~al.}(1997){Walsh}, {Hyland}, {Robinson}, \&
  {Burton}}]{1997MNRAS.291..261W}
{Walsh}, A.~J., {Hyland}, A.~R., {Robinson}, G., \& {Burton}, M.~G. 1997,
  \mnras, 291, 261

\bibitem[{{Wienen} {et~al.}(2012){Wienen}, {Wyrowski}, {Schuller}, {Menten},
  {Walmsley}, {Bronfman}, \& {Motte}}]{2012A&A...544A.146W}
{Wienen}, M., {Wyrowski}, F., {Schuller}, F., {et~al.} 2012, \aap, 544, A146

\bibitem[{{Wood} \& {Churchwell}(1989)}]{1989ApJS...69..831W}
{Wood}, D.~O.~S. \& {Churchwell}, E. 1989, \apjs, 69, 831

\end{thebibliography}

\appendix

\section{Comparing ATLASGAL and Herschel}\label{sec:herComp}

Every observational technique to estimate
\textit{N}-PDFs has its own limitations. The ATLASGAL data
reduction process filters out extended
emission from the maps in scales larger than 2.5$\arcmin$~\citep{2009A&A...504..415S}.
The FIR emission observed by \textit{Herschel}~\citep{2010A&A...518L...1P}
is very likely to be contaminated by emission from dust
unrelated to the cloud of interest~\citep{schneider-15}.
We explore now how the \textit{N}-PDFs derived with
\textit{Herschel} and ATLASGAL differ. We do this for 
one example object of each evolutionary class
using the \HII region M17 (\#248), the SFC IRDC G11.11-0.12 (\#54)
and the SLC (\#54c).

The \textit{Herschel} data of the SFC  
the SLC were taken as part of the \textit{Herschel} guaranteed time key program 
Earliest Phases of Star formation~\citep[EPOS]{2010A&A...518L..95H,2012A&A...547A..49R}.
The data of M17 was obtained from the \textit{Herschel} program Hi-GAL~\citep{2010PASP..122..314M}.
We used the three SPIRE~\citep{2010A&A...518L...3G} wavelengths 
(250\,$\mu$m, 350\,$\mu$m and 500\,$\mu$m), reduced using
\texttt{scanamorphos \,v23}~\citep{2013PASP..125.1126R}, and PACS\,160\,$\mu$m~\citep{2010A&A...518L...2P},
reduced using \texttt{HIPE\,v12}~\citep{2010ASPC..434..139O}. 
In the flux calibration process, the Planck zero-point correction 
was applied only to the M17 data.

\begin{figure*}[h!!!]
\centering
\includegraphics[width = 0.3\textwidth]{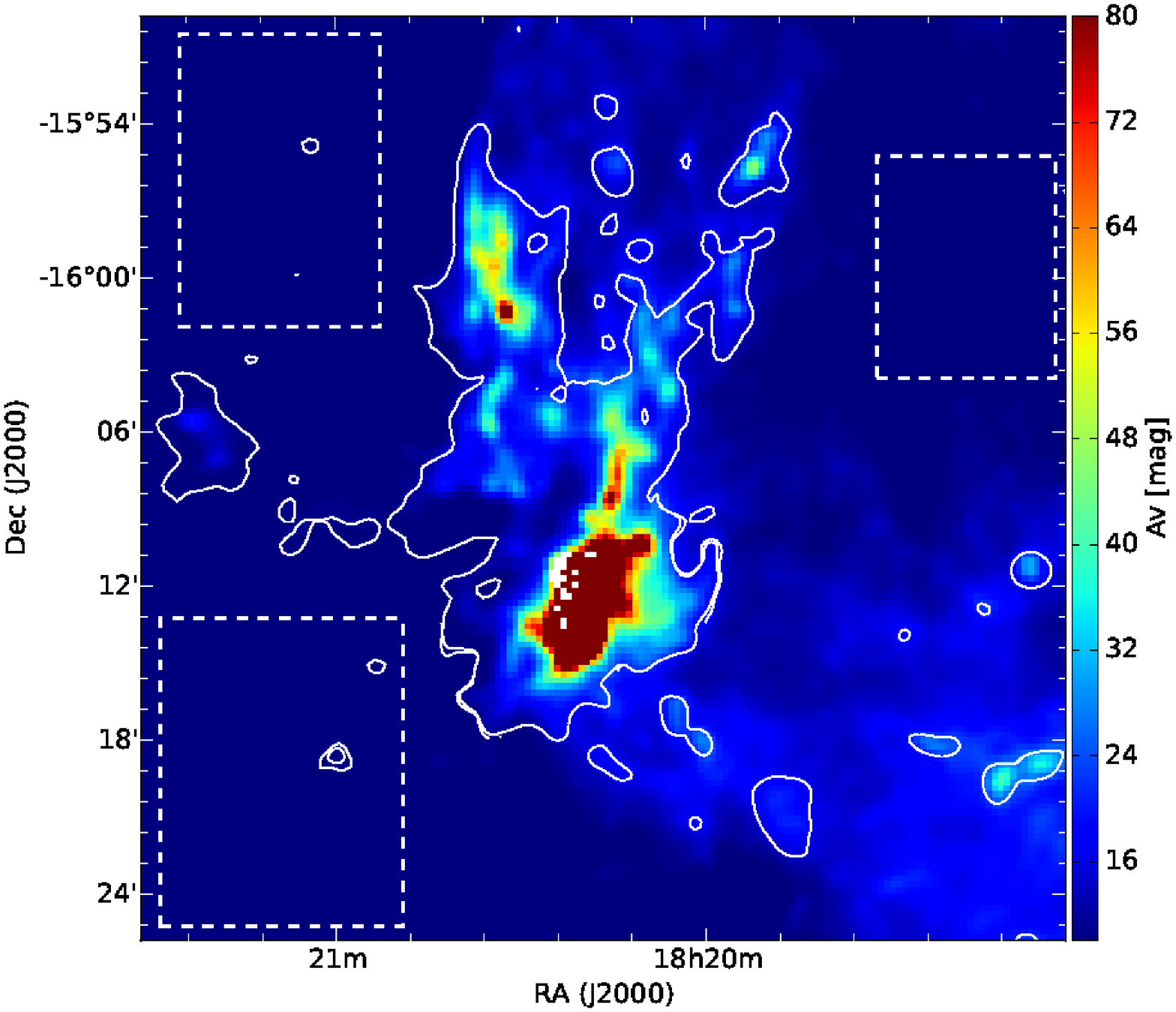}
\includegraphics[width = 0.3\textwidth]{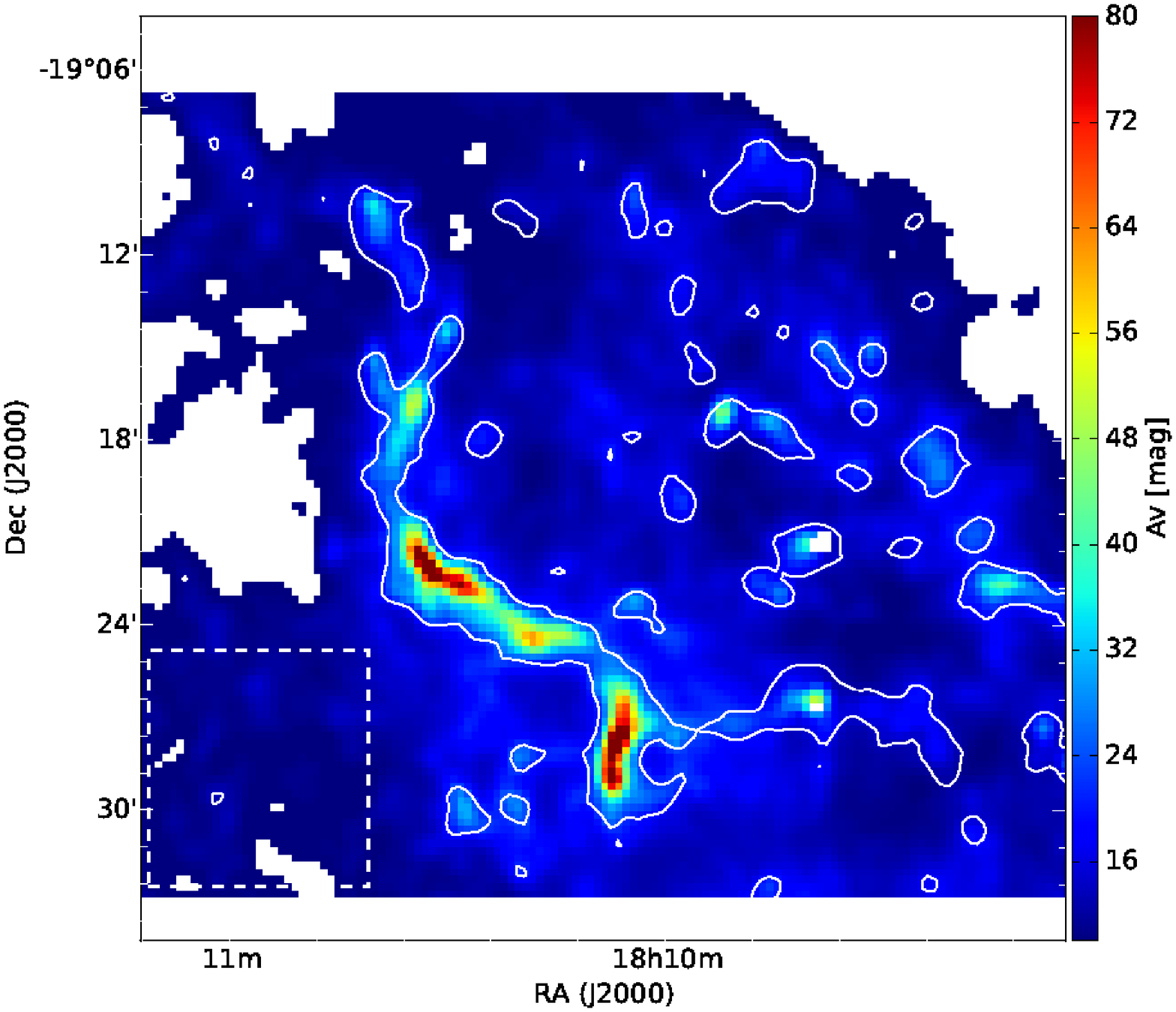}
\includegraphics[width = 0.3\textwidth]{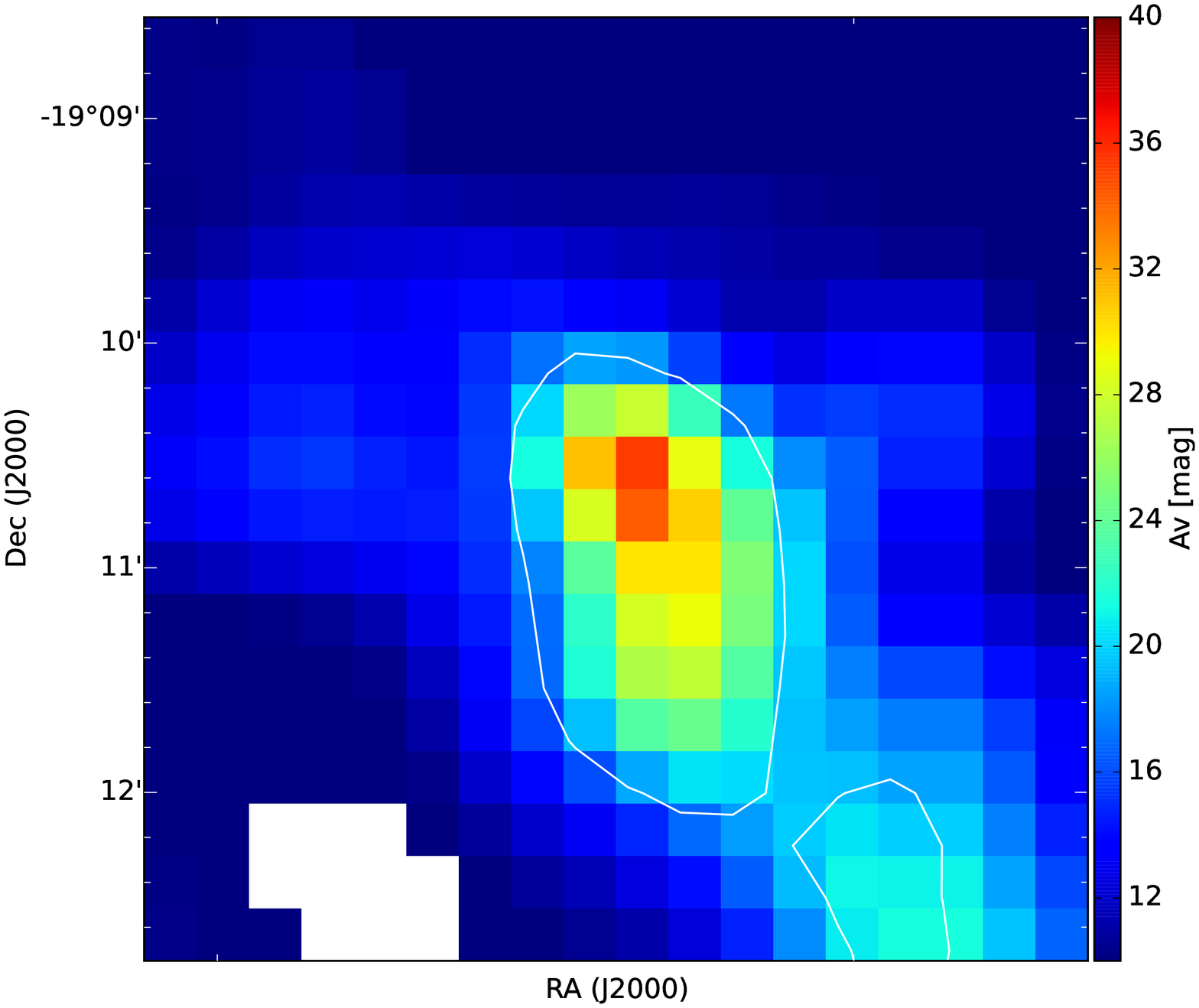}\\
\includegraphics[width = 0.3\textwidth]{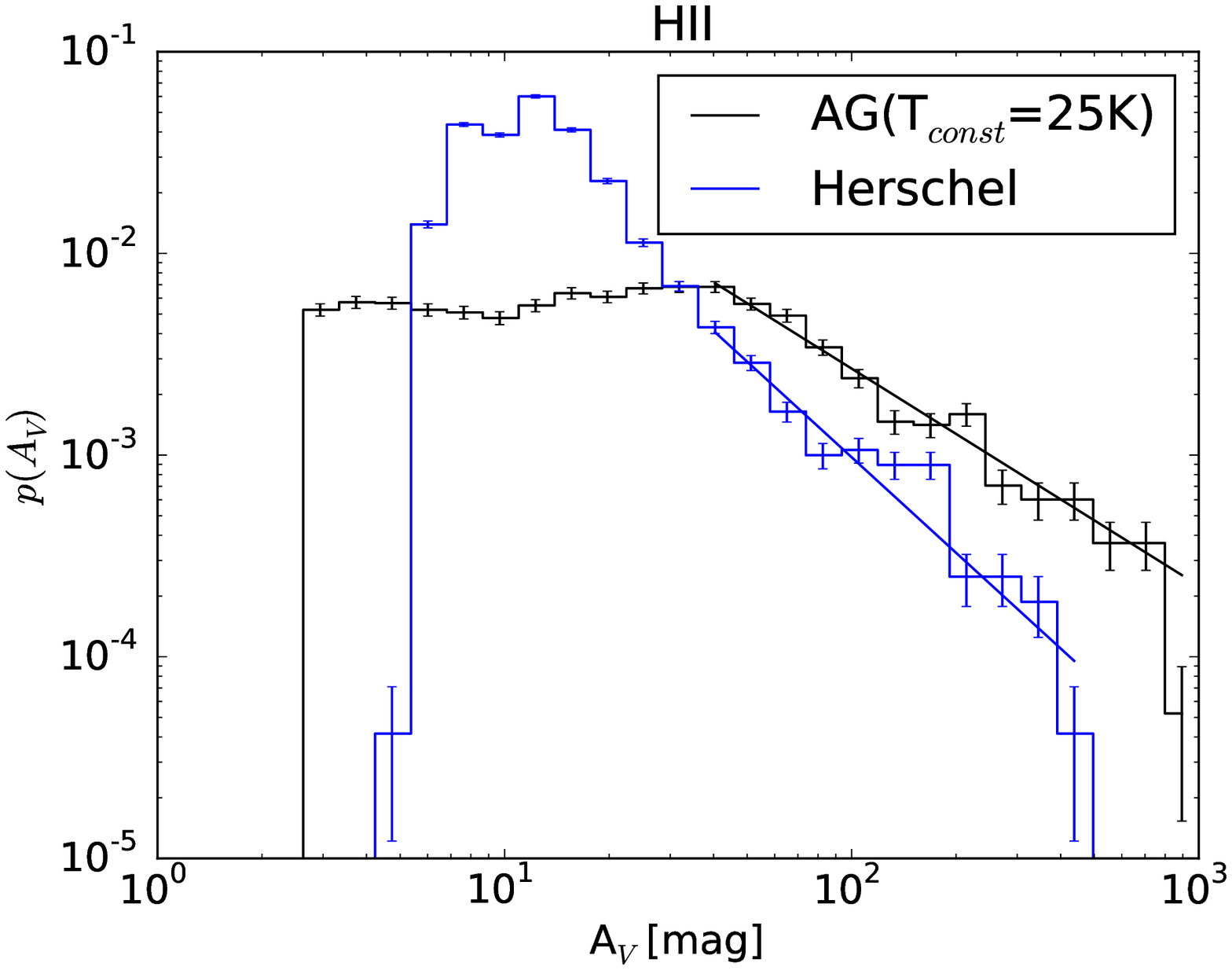}
\includegraphics[width = 0.3\textwidth]{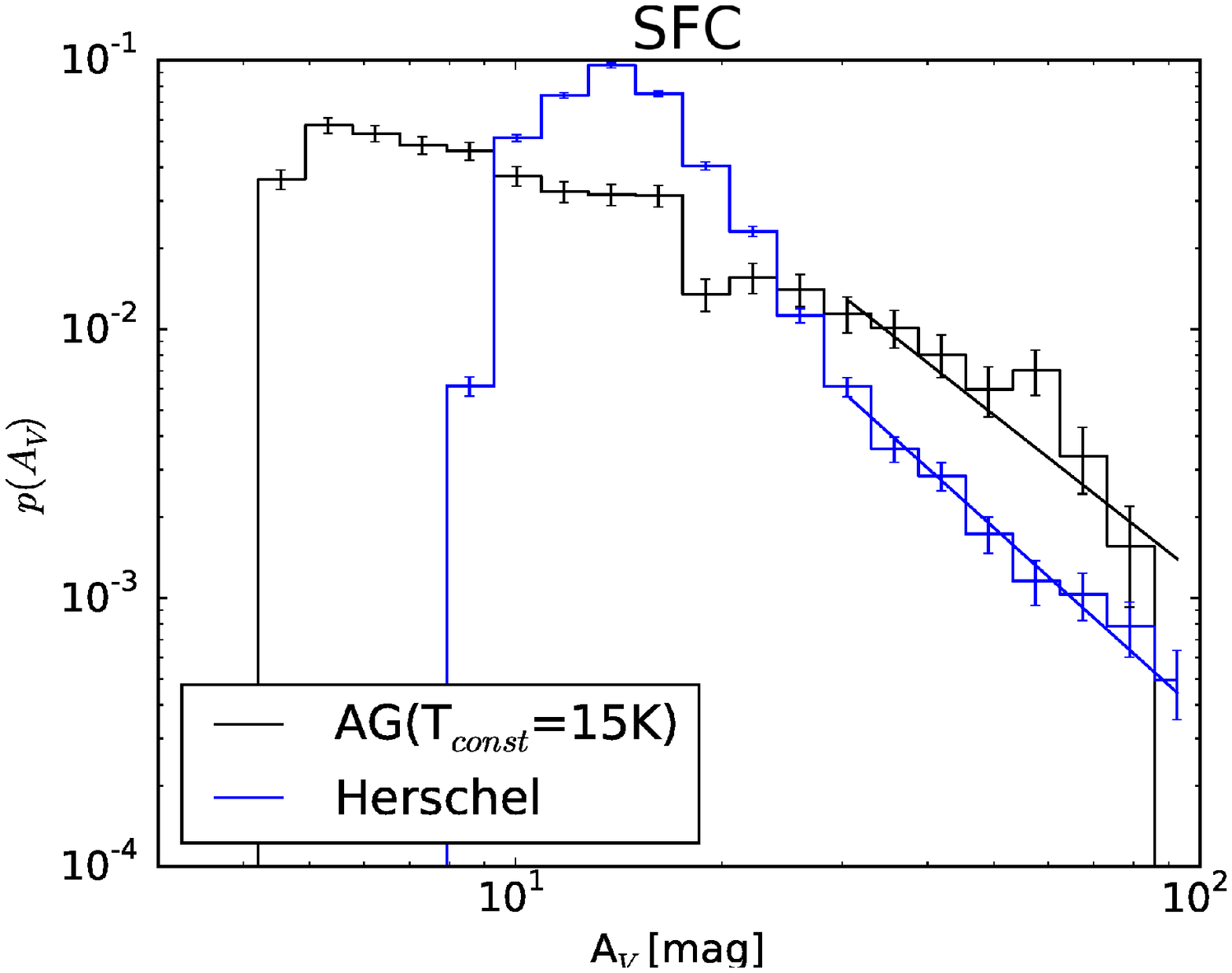}
\includegraphics[width = 0.3\textwidth]{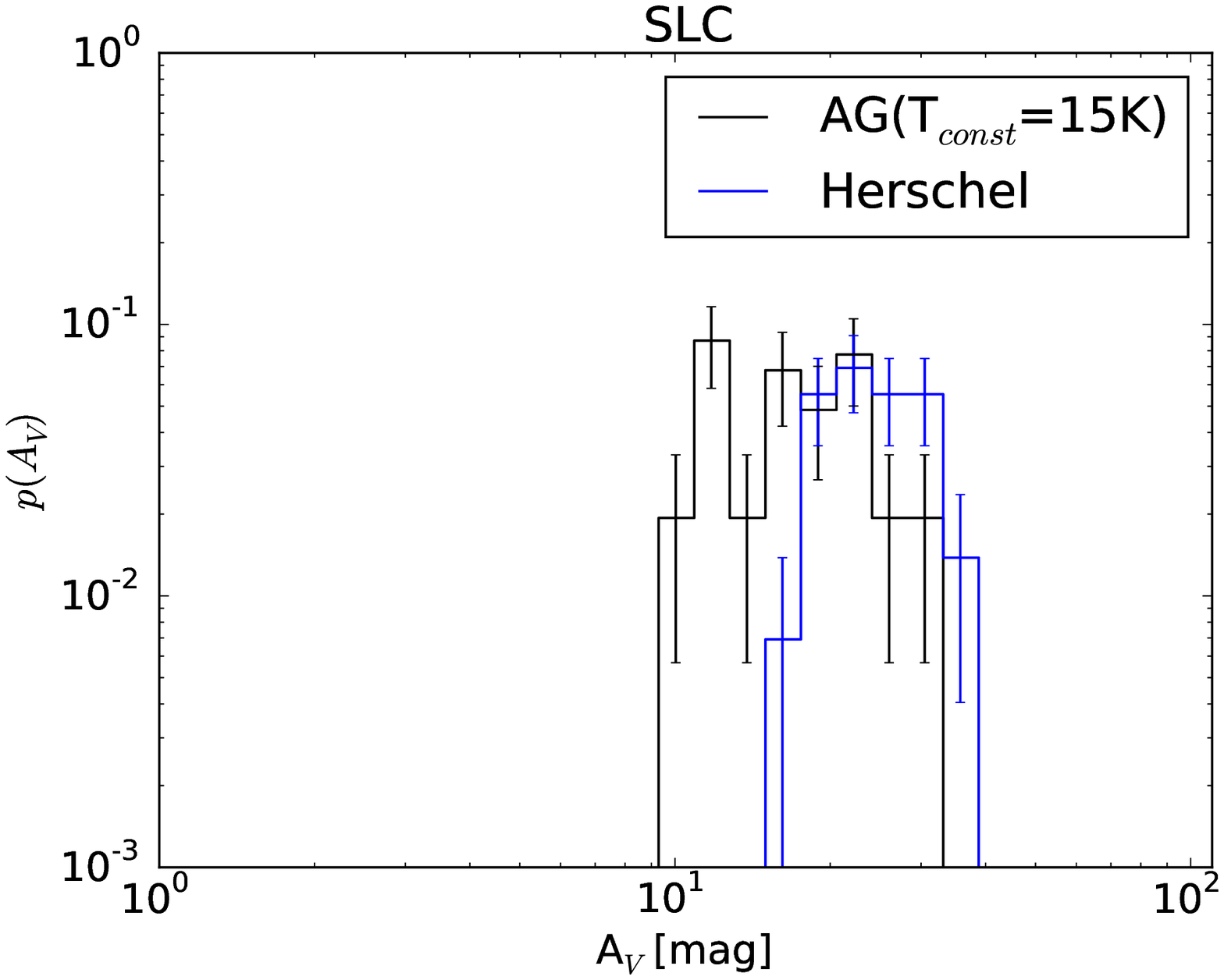}\\
\includegraphics[width = 0.3\textwidth]{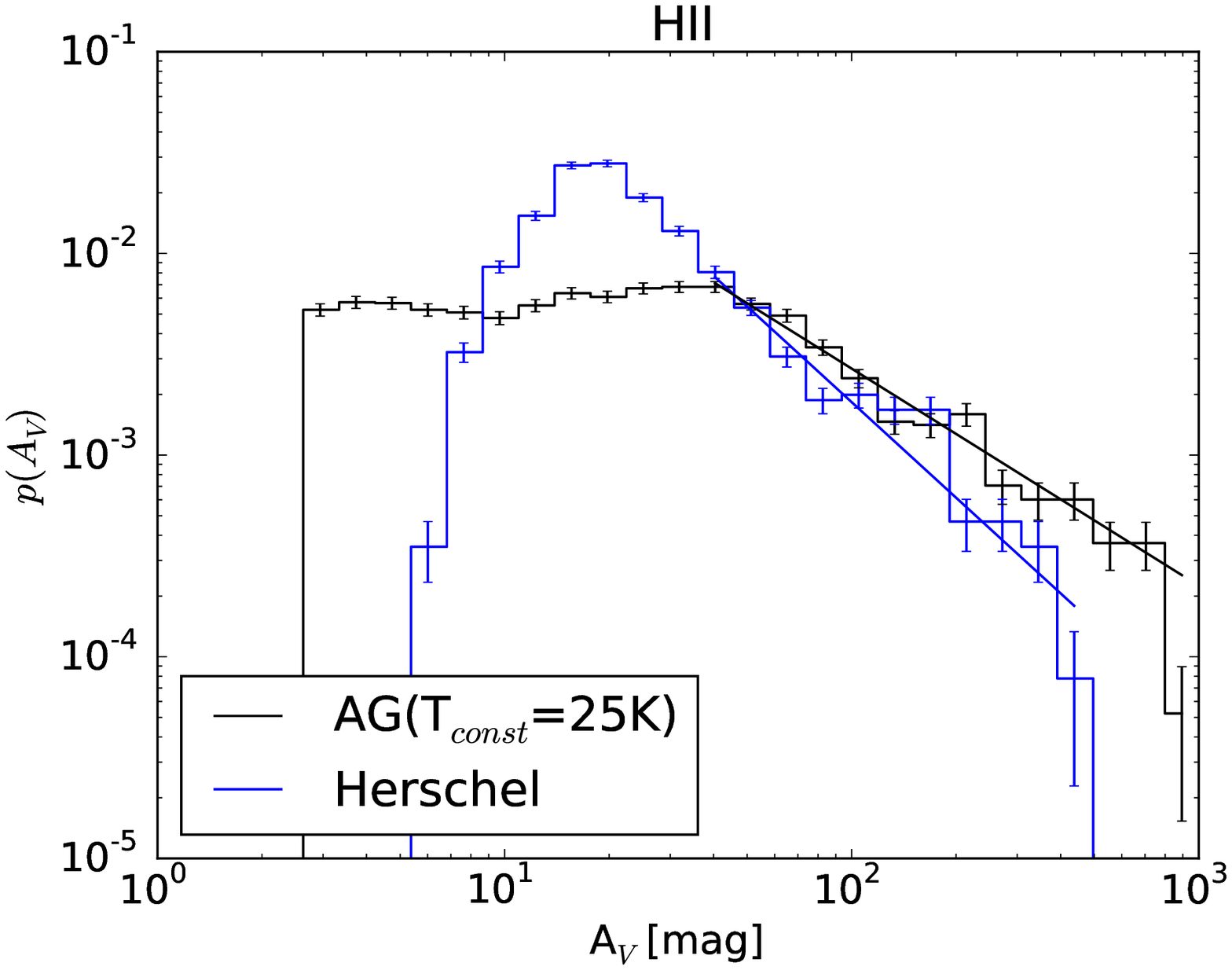}
\includegraphics[width = 0.3\textwidth]{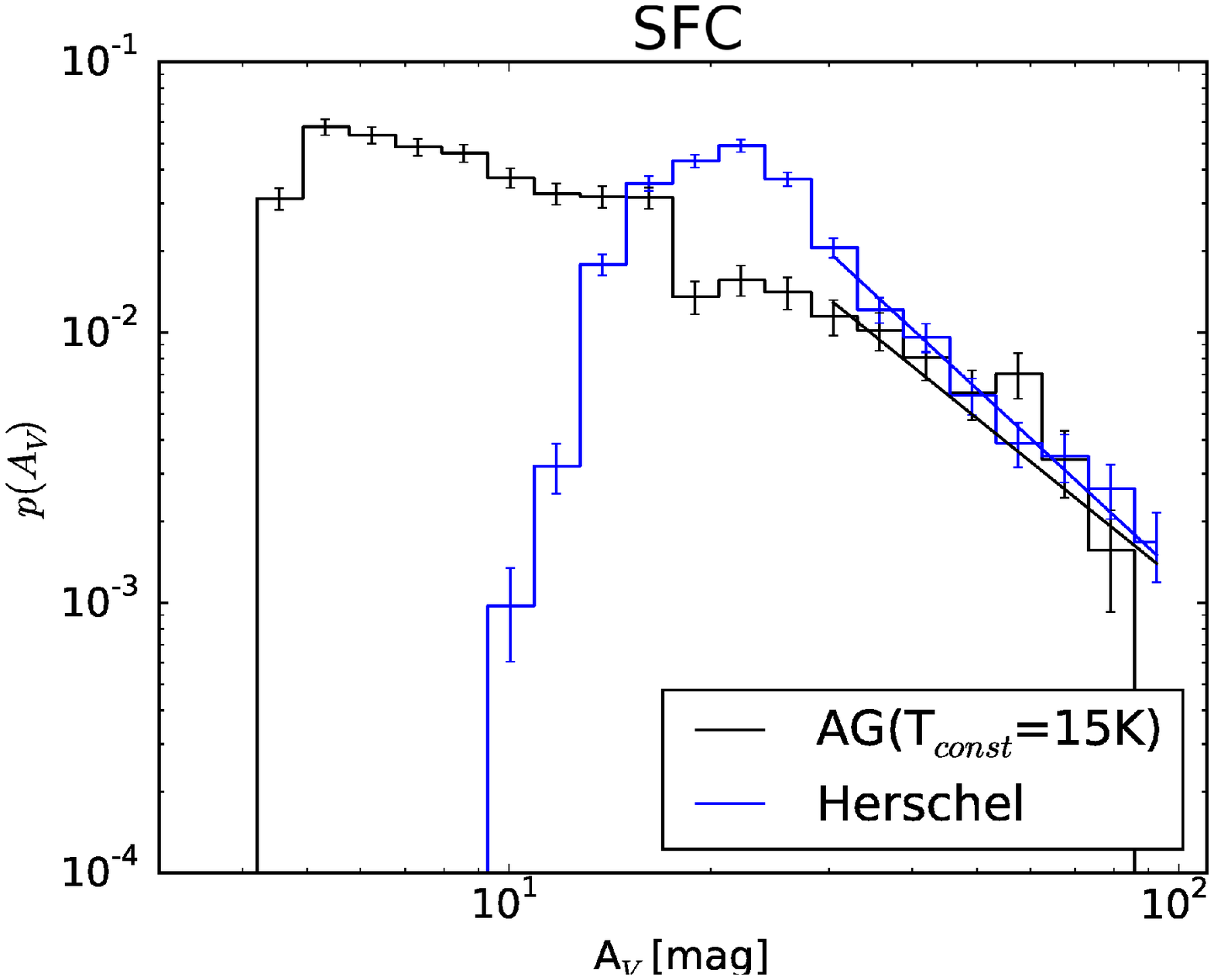}
\includegraphics[width = 0.3\textwidth]{./fig/SLC_PDFs_AgHer_eqArea.eps}\\
\includegraphics[width = 0.3\textwidth]{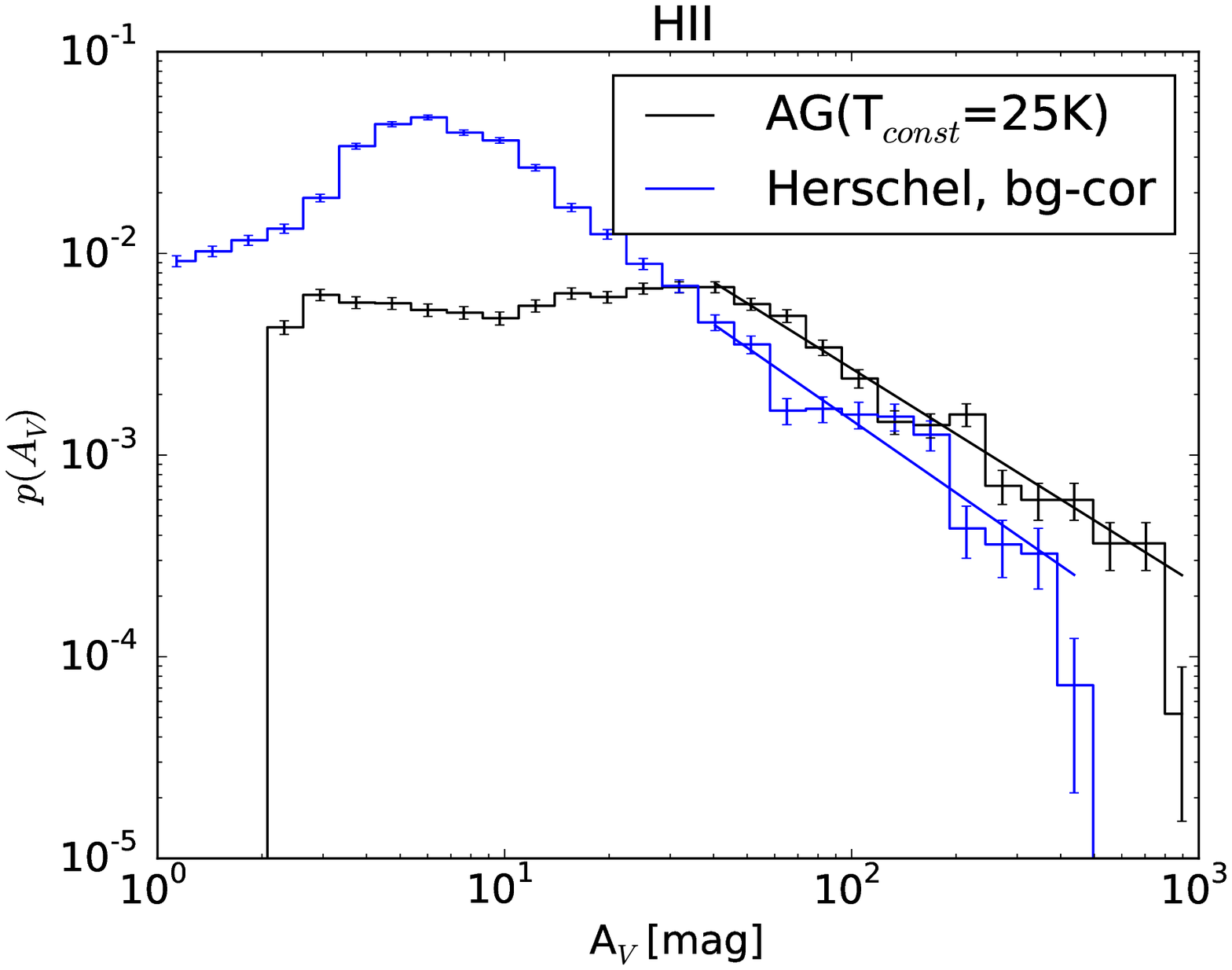}
\includegraphics[width = 0.3\textwidth]{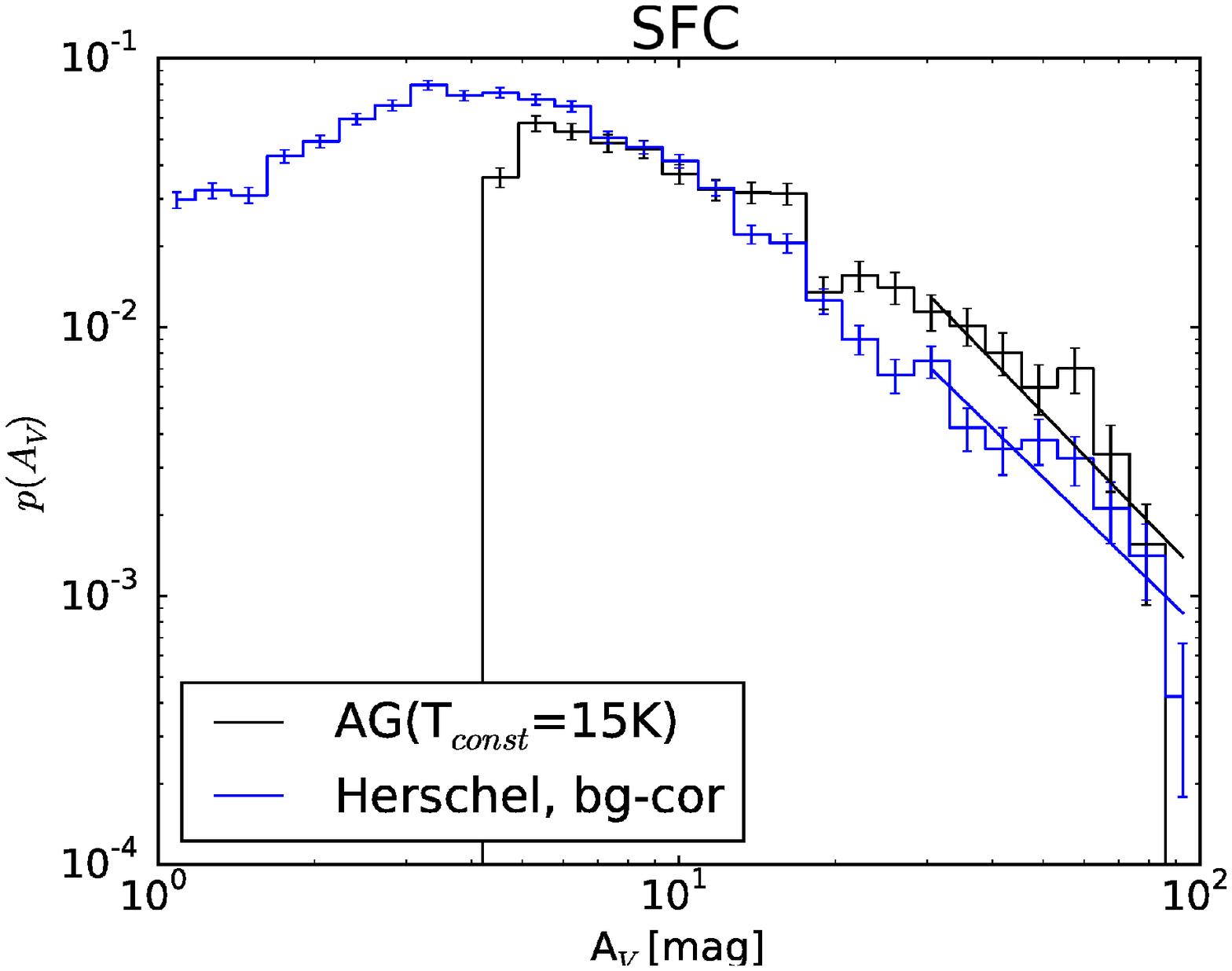}
\includegraphics[width = 0.3\textwidth]{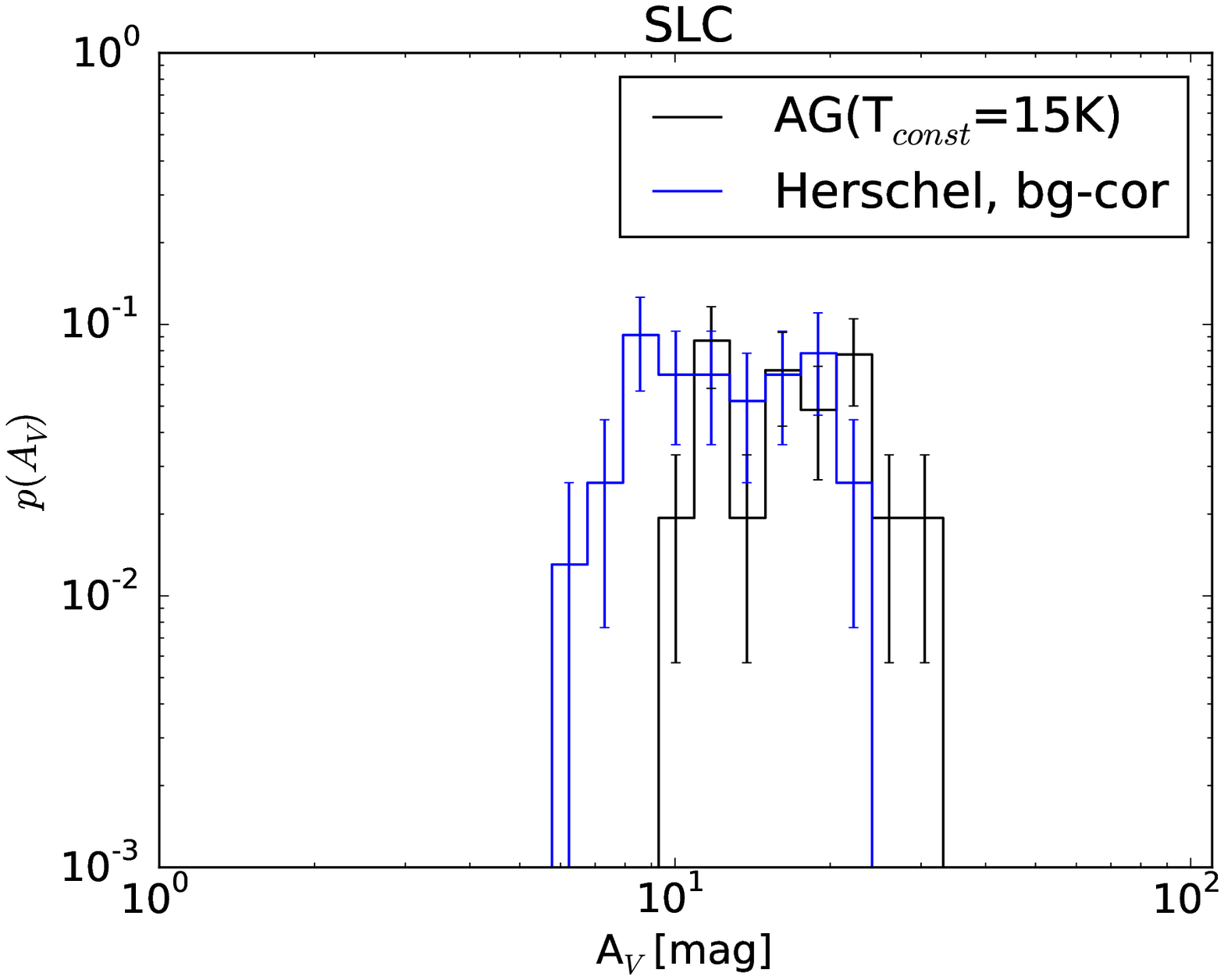}\\
\includegraphics[width = 0.3\textwidth]{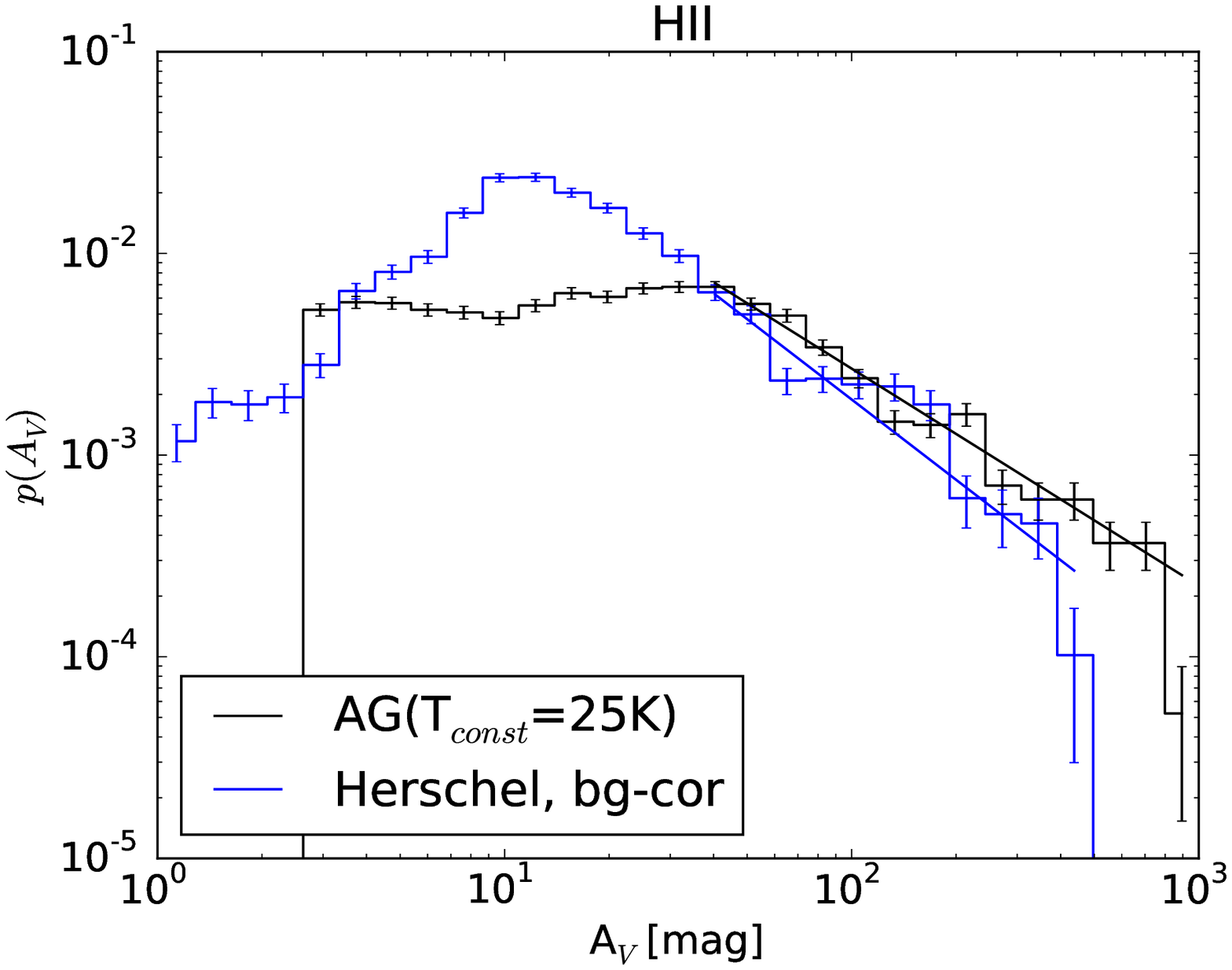}
\includegraphics[width = 0.3\textwidth]{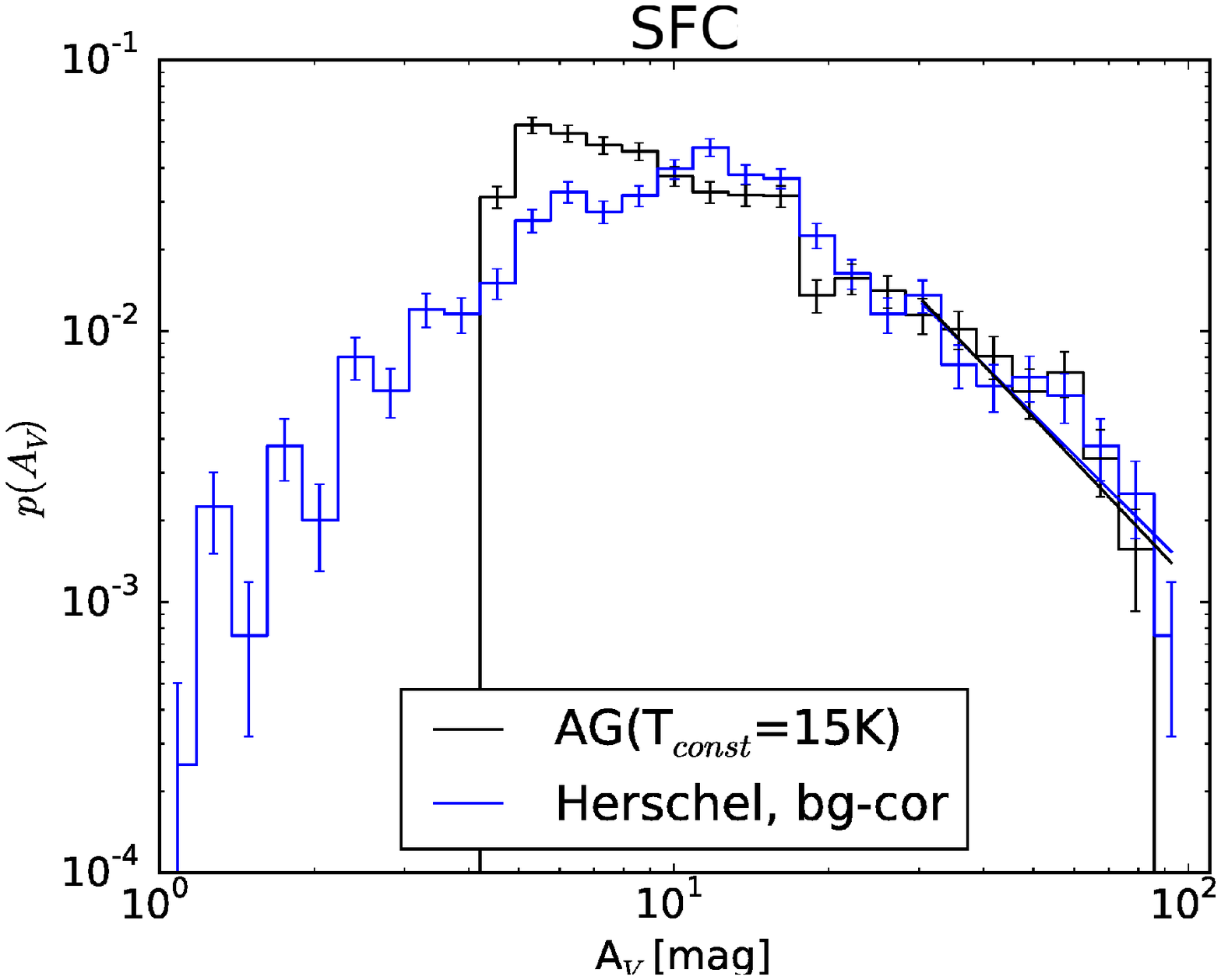}
\includegraphics[width = 0.3\textwidth]{./fig/SLC_PDFs_AgHer_eqArea_noBg.eps}
\caption{\emph{Top row:} \textit{Herschel}-derived 
column density maps of M17 (\HII region), G11 (SFC) and \#53c (SLC), in units of $A_{V}$. 
The white contours show the dense gas area ($A_{V}> $2\,mag, 4.5\,mag and 9\,mag for the \HII region,
SFC and SLC respectively). The white dashed boxes show the regions where
the background contamination of \textit{Herschel} has been calculated.
\emph{Second row: } \textit{N}-PDFs as seen by \textit{Herschel} (blue)
and ATLASGAL (black) in the maps shown in the top row.
The vertical error bars show the Poison standard deviation.
The solid lines show the best fit to the power-law tail. 
\emph{Third row: } ATLASGAL-derived and \textit{Herschel}-derived
\textit{N}-PDFs in the dense gas area.
\emph{Fourth row: } Background corrected ATLASGAL-derived and \textit{Herschel}-derived
\textit{N}-PDFs in the whole map area of the top row. The background emission
was estimated as the mean column density in the dashed boxes
of the first row, seen by \textit{Herschel}.
\emph{Bottom row: } Background corrected \textit{N}-PDFs
evaluated in the dense gas area.}
\label{fig:CompHerAG}
\end{figure*}

We derived the column density and
temperature maps for each of the three selected regions
through a pixel-to-pixel modified greybody fit 
to the four \textit{Herschel} continuum maps, all of them 
smoothed to a resolution of 36$\arcsec$.
For consistency with the ATLASGAL data analysis, we
adopted the dust opacity by interpolation of
the~\citet{1994A&A...291..943O} dust
model of grains with thin ice mantles and a mean density of
$n=10^{6}\mathrm{\,cm^{-3}}$. The mean uncertainty obtained 
in our greybody fitting technique for the 
temperature maps is $\sim2.5$\,K. The relative uncertainty of the 
column density maps is $\sim40\%$. The column density maps
are shown in Fig.~\ref{fig:CompHerAG}.

The area over which a molecular cloud 
shows significant emission is different in
ATLASGAL and \textit{Herschel} column density maps. 
We face this issue by comparing the \textit{N}-PDFs 
over two different areas: the area over which ATLASGAL shows significant emission,
which we will refer to as dense gas area
(white contours in Fig.~\ref{fig:CompHerAG} and panels in the mid row in the same figure).
We also compare the \textit{N}-PDFs derived from the entire
areas shown in Fig.~\ref{fig:CompHerAG}.

We now describe how the \textit{N}-PDFs derived from 
\textit{Herschel} and ATLASGAL data look like. In the SFC and the \HII region
\textit{Herschel}-derived \textit{N}-PDFs show a clear log-normal and
power-law combination. This combination is seen in both cases of area selection: dense gas area 
and whole map. The ATLASGAL-derived \textit{N}-PDFs of the SFC  and the \HII region also
have power-law tails at high column densities but they do not
show a log-normal distribution at low column densities. In both cases, 
\textit{Herschel}-derived \textit{N}-PDFs do not probe regions with $A_{V}\lesssim10$\,mag.
Both, the ATLASGAL-derived and \textit{Herschel}-derived \textit{N}-PDFs of the SLC are unfortunately dominated by noise,
making a comparison impossible.
At 36$\arcsec$ of resolution the SLCs do not have enough pixels for an analysis.

The absence of column densities $A_{V}\lesssim10$\,mag in the 
\textit{Herschel} data-set is very likely related to the line-of-sight 
contamination. To compare \textit{N}-PDFs of the datasets without 
this contamination, we subtracted the background emission
from the \textit{Herschel} column density maps.
We estimated the magnitude of the line-of-sight contamination
averaging the \textit{Herschel}-derived column densities inside the
white boxes shown in the top row of Fig.~\ref{fig:CompHerAG}.
We found $A_{V}^{\mathrm{bg}}=11.2\pm1.2$\,mag in the SFC and SLC regions and
$A_{V}^{\mathrm{bg}}=7.0\pm1.6$\,mag in the \HII region. The background-subtracted
\textit{N}-PDFs are shown in the fourth and fifth rows of Fig.~\ref{fig:CompHerAG}
for the whole map and the dense gas area.
The background subtraction significantly widens the \textit{Herschel}-derived
\textit{N}-PDFs in the low column density regime. It has, however, 
very small effect in the high column-density regime,
which is slightly flattened. 

To estimate the difference between the fitted
parameters in the datasets we fitted the power-law 
tails of the \textit{N}-PDFs, following the same procedure as
in Section~\ref{sec:pdf}. The fits were performed 
in the column density regimes $A_{V}>30$\,mag and
$A_{V}>40$\,mag in the SFC and \HII region respectively.
In all cases, the power-law portion of ATLASGAL is somewhat shallower ($p_{\mathrm{HII,AG}}=-1.2$, $p_{\mathrm{SFC,AG}}=-2.0$)
than that obtained for \textit{Herschel} ($p_{\mathrm{HII,H}}=-1.6$, $p_{\mathrm{SFC,H}}=-2.3$).
The power-law slopes obtained in this work for the SFC are 
shallower than those reported by~\citet{schneider-14}.
The difference is probably caused by the different 
column density ranges used to fit the power-law in
the works.
When the background component of \textit{Herschel} is removed,
the power-law tails flatten and become much more similar to those 
observed by ATLASGAL ($p^{\mathrm{bg}}_{HII,H}=-1.2$, $p^{\mathrm{bg}}_{SFC,H}=-1.9$). 
Using only the ATLASGAL emission area (middle row of Fig.~\ref{fig:CompHerAG})
or the whole map (bottom row of Fig.~\ref{fig:CompHerAG}) makes no significant difference
in the slope of the power-law tails obtained.
We conclude that the high-column density power-law parts of the
\textit{N}-PDFs are in good agreement 
between ATLASGAL and \textit{Herschel}. The agreement is even
better when the background contamination component of \textit{Herschel} is removed.
Note that a background correction to the \textit{Herschel} column densities is 
usually necessary, as the diffuse Galactic dust component is 
significant at the Galactic plane. Therefore, one should consider
the background subtracted \textit{N}-PDF as a better estimate
of the \textit{N}-PDF of the cloud.

We identify the absence of log-normal 
components in the ATLASGAL-derived \textit{N}-PDFs 
as an effect associated to the spatial filtering in the data reduction
process. This effect is significant in both \HII regions and SFCs, being
less important in denser regions of molecular clouds where SLC objects lie.
Spatial filtering is clearly seen at column densities $A_{V}\sim10-20$\,mag
in Fig.~\ref{fig:CompHerAG}, where the \emph{Herschel}-derived \textit{N}-PDFs shows a clear excess
compared to the ATLASGAL-derived \textit{N}-PDFs.  
Despite the significant differences shown by the \textit{N}-PDFs
derived at low column density regimes, the power-law tails at high column 
densities are in good agreement, showing the ATLASGAL-derived \textit{N}-PDFs
marginally flatter distributions than the \textit{Herschel}-derived \textit{N}-PDFs.
Similar results are obtained when the ATLASGAL-derived and \textit{Herschel}-derived
DGMFs are compared.

\section{Effects of the isothermal assumption on the \textit{N}-PDFs}\label{app:tcomp}

Obtaining column densities via dust emission maps at sub-mm
wavelengths requires the use of the dust temperature (see Eq.~\ref{eq:NH2}).
When only one wavelength is available, as in the case of this paper,
the most simple assumption is that the dust is isothermal.
However, molecular clouds are not isothermal
and the isothermal assumption can therefore generate artificial
features in the column density distributions of the maps derived with this method.
When several wavelengths are available, as in the case of \textit{Herschel} observations,
the line-of-sight averaged temperature and column density distributions can be simultaneously obtained
via modified blackbody fitting to the FIR/sub-mm spectral energy distribution.

\begin{figure*}[h!!!]
\centering
\includegraphics[width = 0.3\textwidth]{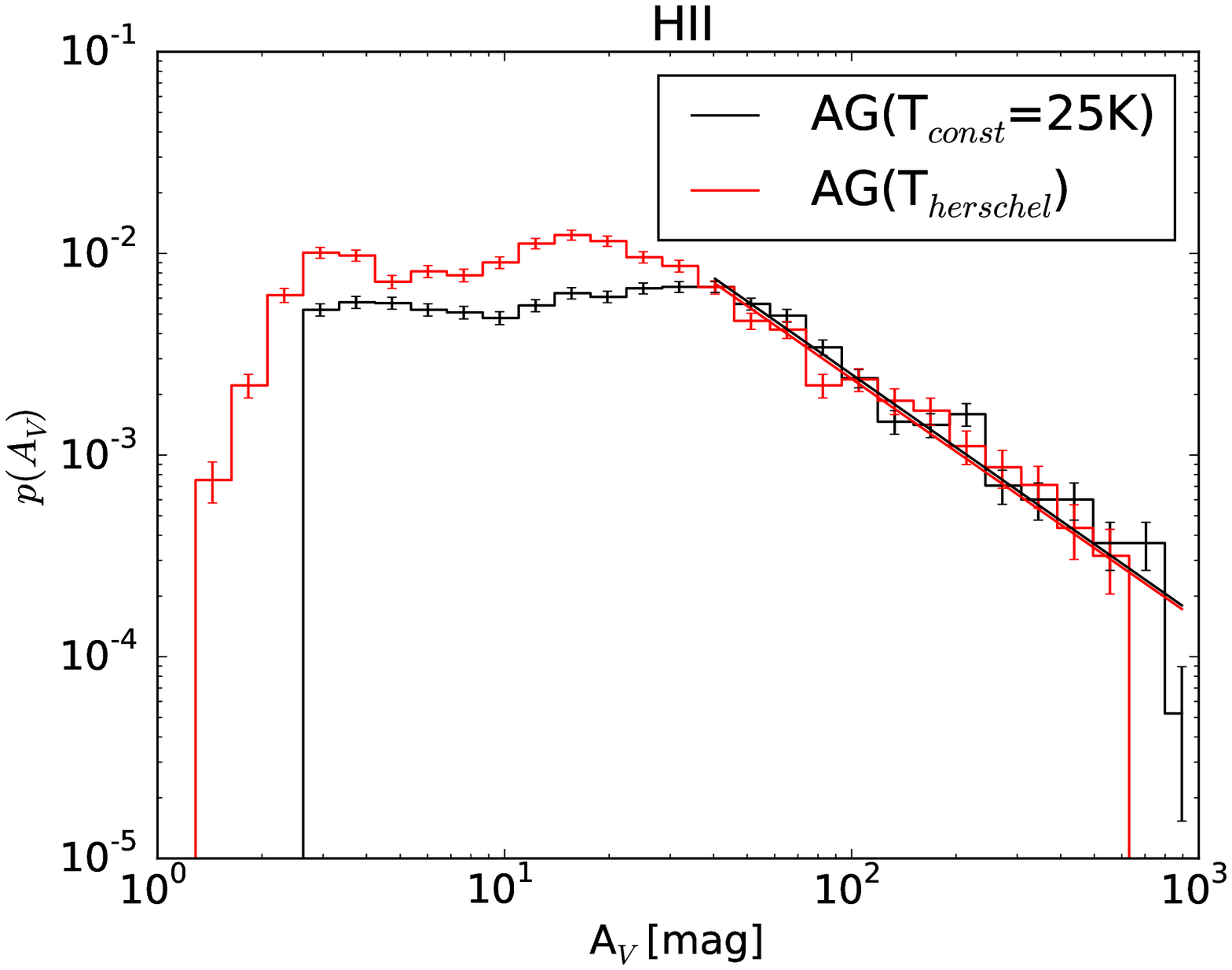}
\includegraphics[width = 0.3\textwidth]{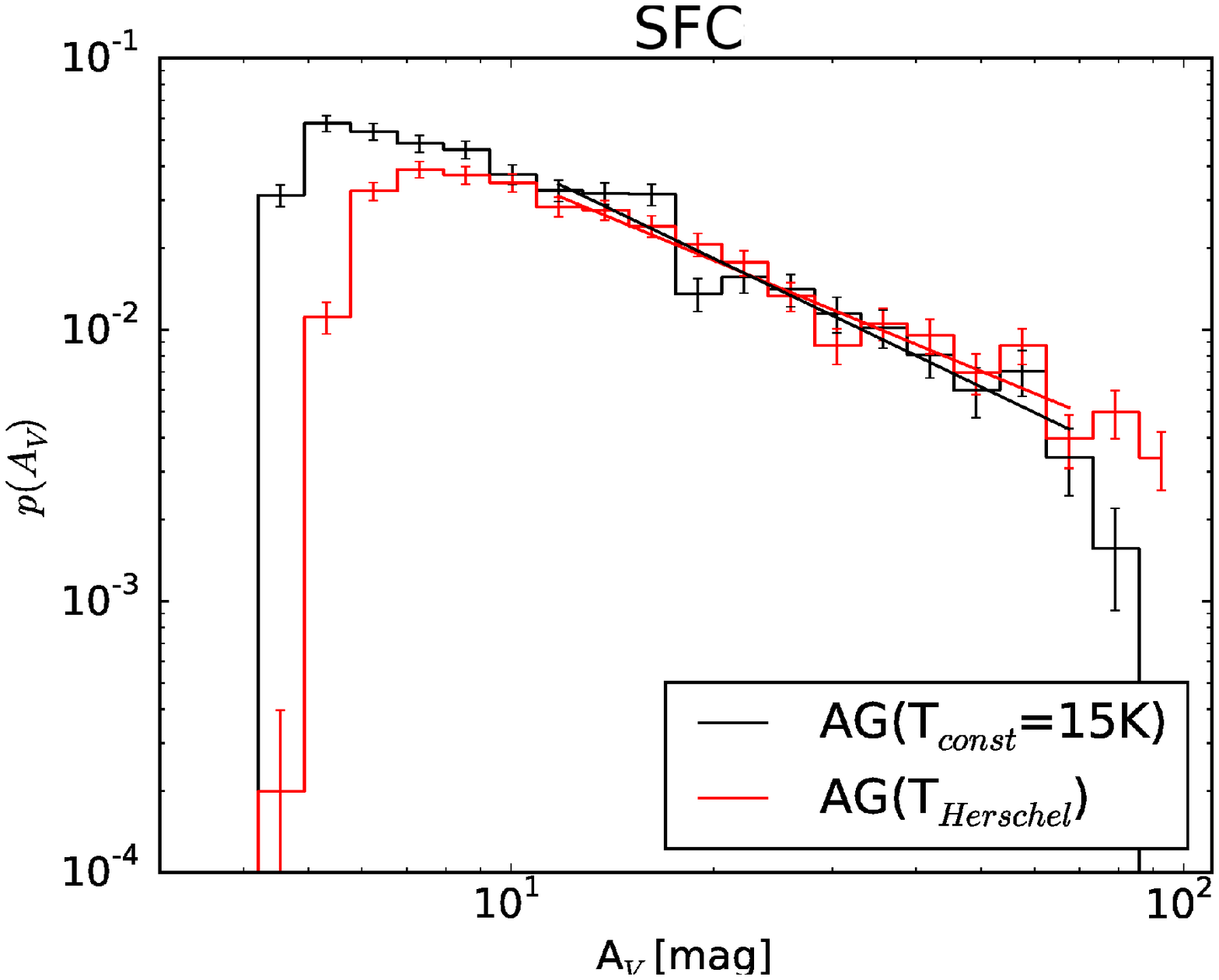}
\includegraphics[width = 0.3\textwidth]{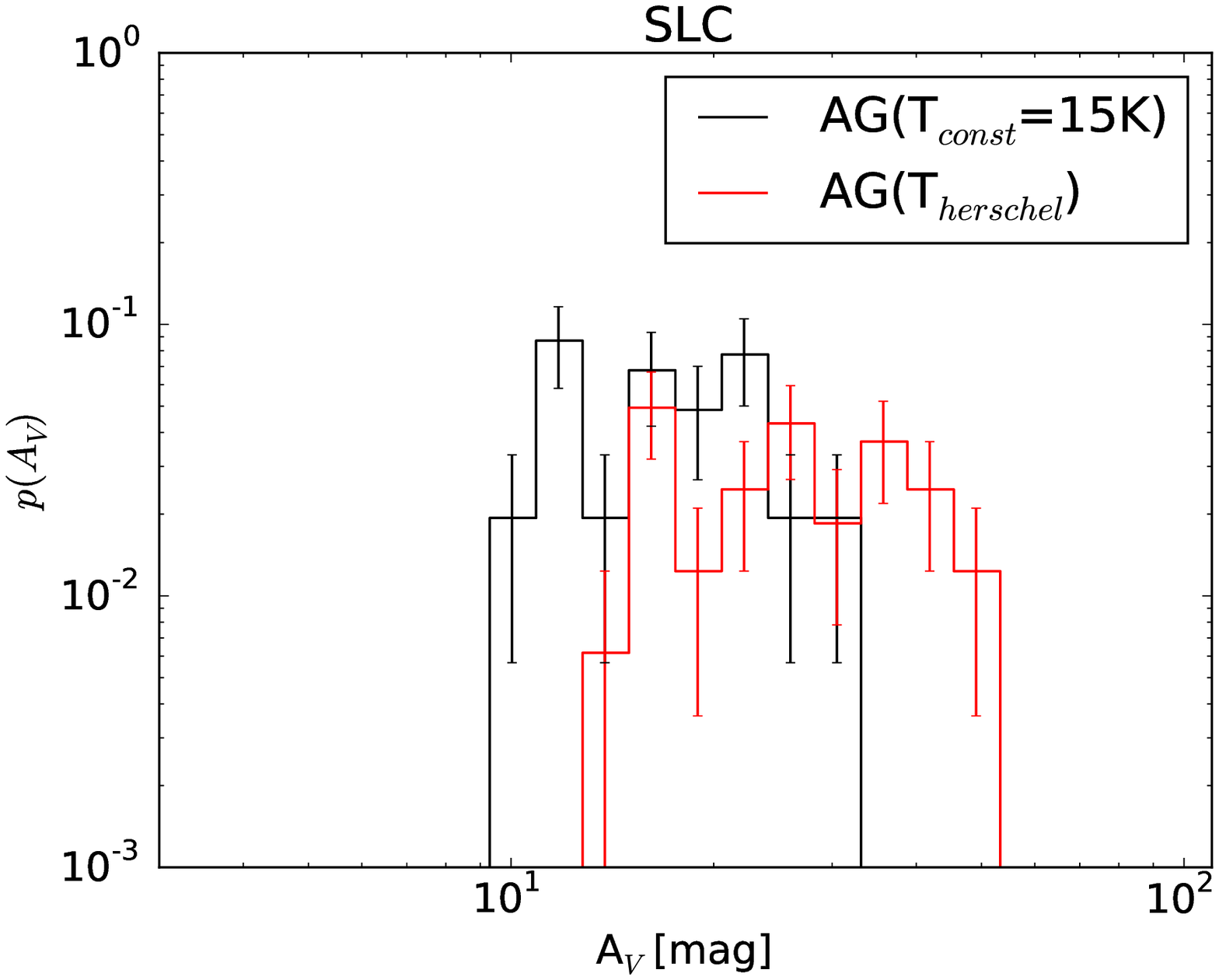}

\caption{Isothermal ATLASGAL-derived \textit{N}-PDFs (black)
and \textit{N}-PDFs derived using ATLASGAL emission maps 
together with \textit{Herschel}-derived temperature maps (red).
From left to right, \HII region, SFC and SLC.
The vertical error bars show the Poison standard deviation.
The solid lines show the power-law fit to the data in the column
density range covered by the lines. }
\label{fig:CompAGtemp}
\end{figure*}

To study the temperature effects on the resulting \textit{N}-PDFs
we used the \textit{Herschel} derived temperature distributions in the previous
section to reconstruct the ATLASGAL column density maps of the same three regions.
The results of this experiment are shown in Fig.~\ref{fig:CompAGtemp}.
In the \HII region, the isothermal assumption underestimates
the low column density regimes of the \textit{N}-PDF, which remain
practically unaffected at $A_{V}>40$\,mag. 
The isothermal \textit{N}-PDF of the SFC overestimates the
low column density regime and remain similar to the
\textit{N}-PDF of the \textit{Herschel}-derived temperature distribution
at $A_{V}=10-90$\,mag. 
The isothermal \textit{N}-PDF in the SLC is shifted to lower
column densities.

The isothermal assumption is therefore valid
in the high column density regime (i.e. in the power-law tail)
of the \HII region and the SFC examples shown here. 
We note that the relative temperature uncertainties are
larger in the coldest regions (T$\sim12-15$\,K) of molecular clouds (i.e. in the
densest regions) and these uncertainties could also result in the
underestimate of the \textit{N}-PDF observed at $A_{V}>90$\,mag in the SFC.
Unfortunately, we cannot quantify the possible differences in the shape of
the isothermal and the \textit{Herschel}-derived temperature distribution
\textit{N}-PDFs of the SLC.
The isothermal \textit{N}-PDF therefore offers a more accurate
reproduction of the \textit{Herschel}-derived temperature distribution \textit{N}-PDF
in the column density regime of the power-law tail than at low-column densities.

\section{MIPSGAL 24$\mu$m maps with ATLASGAL contours and
regions}\label{ap:maps}

Please find this appendix in the journal version

\section{ATLASGAL maps from starless clumps}\label{app:TSL}

Please find this appendix in the journal version

\section{\HII regions}\label{app:HII}

Please find this appendix in the journal version

\section{star-forming clouds}\label{app:SFC}

Please find this appendix in the journal version

\end{document}